\documentclass[11pt,a4paper]{article} 
\usepackage[english,german]{babel}
\usepackage[latin1]{inputenc}
\usepackage[T1]{fontenc}
\usepackage{ae,aecompl}
\usepackage{times}
\usepackage{amsmath}
\usepackage{amsfonts,bbm,theorem}

\ifx\pdftexversion\undefined
  \usepackage[dvips]{graphicx}
\else
  \usepackage[pdftex]{graphicx}
\fi

\usepackage{epsfig,url}
\newcommand{\myfigure}[2]{ \includegraphics*[#1]{#2} }

\numberwithin{equation}{section}

\newtheorem{theorem}{Theorem}[section]
\newtheorem{definition}[theorem]{Definition}
\newtheorem{corollary}[theorem]{Corollary}
\newtheorem{assumption}[theorem]{Assumption}
\newtheorem{lemma}[theorem]{Lemma}

\newtheorem{proposition}[theorem]{Proposition}
\newenvironment{proof}{\begin{trivlist}\item[]{\em Proof:}\/}{%
\hfill\mbox{$\Box$}\end{trivlist}}
\newenvironment{proofof}[1]{\begin{trivlist}\item[]{\em Proof of #1:}\/}{%
\hfill\mbox{$\Box$}\end{trivlist}}

\newcounter{jlisti}
\newenvironment{jlist}[1][(\thejlisti)]{\begin{list}{{\rm #1}\ \ }{ %
      \usecounter{jlisti} %
    \setlength{\itemsep}{0pt} 
    \setlength{\parsep}{0pt}  %
    \setlength{\leftmargin}{0pt} %
    \setlength{\labelwidth}{0pt} %
    \setlength{\labelsep}{0pt} %
}}{\end{list}}
\newcounter{printapu}

\newcommand{\sabs}[1]{\langle #1\rangle}
\newcommand{\Bigsabs}[1]{\Bigl\langle #1\Bigr\rangle}

\newcommand{\vep}{\varepsilon}
\newcommand{\qand}{\qquad\text{and}\qquad}

\newcommand{\defset}[2]{ \left\{ #1\left|\, #2 %
      \makebox[0cm]{$\displaystyle\phantom{#1}$}\right.\!\right\} }
\newcommand{\set}[1]{\{#1\}}
\newcommand{\mean}[1]{\langle #1\rangle}

\newcommand{\ci}{{\rm i}}
\newcommand{\rmd}{{\rm d}}
\newcommand{\rme}{{\rm e}}
\newcommand{\1}{\mathbbm{1}}
\newcommand{\re}{{\rm Re\,}}
\newcommand{\im}{{\rm Im\,}}
\newcommand{\Z}{{\mathbb Z}}
\newcommand{\R}{{\mathbb R}}
\newcommand{\T}{{\mathbb T}}
\newcommand{\C}{{\mathbb C\hspace{0.05 ex}}}
\newcommand{\E}{{\mathbb E}}
\newcommand{\N}{{\mathbb N}}
\newcommand{\M}{{\mathbb M}}

\newcommand{\frechet}{Fr\'{e}chet}

\newcommand{\norm}[1]{\Vert #1\Vert}

\newcommand{\order}[1]{{\cal O}\!\left( #1\right)}

\newcommand{\braket}[2]{ \left\langle #1\! \left| #2 %
      \makebox[0cm]{$\displaystyle\phantom{#1}$}\right.\!\right\rangle}
\newcommand{\wvep}{W^{\vep}}
\newcommand{\cwvep}{\mathcal{W}^{\vep}}
\newcommand{\owvep}{\overline{W}^{\vep}}
\newcommand{\owvept}{\overline{W}^{\vep}\!(t)}
\newcommand{\owl}{\overline{W}}
\newcommand{\ulr}{\underline{r}}
\newcommand{\tmacro}{\bar{t}}
\newcommand{\nmacro}{\bar{n}}
\newcommand{\pmacro}{\bar{p}}
\newcommand{\Kpart}{\mathcal{K}}
\newcommand{\Kpamp}{\mathcal{K}^{(\text{amp})}}
\newcommand{\Kpmain}{\mathcal{K}^{(\text{main})}}
\newcommand{\ommax}{\omega_{\text{max}}}
\newcommand{\ommin}{\omega_{\text{min}}}
\newcommand{\totalcs}{\sigma}
\newcommand{\csmax}{\sigma_{\text{max}}}
\newcommand{\ho}{H}
\newcommand{\Emax}{E_{0}}

\newcommand{\VR}{V^{R}}

\newcommand{\Edens}{\mathcal{E}}
\newcommand{\hilb}{\mathcal{H}}
\newcommand{\banach}{\mathcal{B}}
\newcommand{\cals}{\mathcal{S}}

\newcommand{\Vy}{V^{(y)}}
\newcommand{\Wy}{W^{(y)}}

\newcommand{\gpath}{{\Gamma_\beta}}

\newcommand{\setm}{\!\setminus }

\newcommand{\FT}[1]{\smash{\widehat{#1}}}
\newcommand{\IFT}[1]{\smash{\widetilde{#1}}}
\newcommand{\psierr}{\psi_{\text{err}}}
\newcommand{\psimain}{\psi_{\text{main}}}
\newcommand{\Fmain}{F^\vep_{\text{main}}}
\newcommand{\comega}{c_{\omega}}

\begin{document}
\selectlanguage{english}

\newcommand{\email}[1]{E-mail: \tt #1}
\newcommand{\emailjani}{\email{jlukkari@ma.tum.de}}
\newcommand{\addressjani}{\em Zentrum Mathematik, 
Technische Universit\"at M\"unchen, \\
\em Boltzmannstr. 3, D-85747 Garching, Germany}

\newcommand{\emailherbert}{\email{spohn@ma.tum.de}}

\title{Kinetic Limit for Wave Propagation in a Random Medium}
\author{Jani Lukkarinen\thanks{\emailjani}, 
  Herbert Spohn\thanks{\emailherbert}\\[1em]
\addressjani }

\maketitle

\begin{abstract}
We study crystal dynamics in the harmonic approximation. The atom\-ic
masses are weakly disordered, in the sense that their deviation from
uniformity is of order $\sqrt{\vep}$.  The dispersion relation is assumed
to be a Morse function and to suppress crossed recollisions.
We then prove that in the limit $\vep\to 0$
the disorder averaged Wigner function on the kinetic scale, time and
space of order $\vep^{-1}$, is governed by a linear Boltzmann equation.
\end{abstract}

\tableofcontents

\section{Introduction}

When investigating 
the propagation of waves, one has to deal with the fact that the
supporting medium  often is not perfectly homogeneous, but suffers from
irregularities.  A standard method is then to assume that
the material coefficients characterizing the
medium are random, being homogeneous only in average.
Examples abound: Shallow water waves travelling in a canal with uneven bottom,
radar waves propagating through turbulent air, elastic waves dispersing in a
random compound of two materials.   The arguably simplest prototype is the
scalar wave equation
\begin{align}\label{eq:randomwaveeq}
n^2 \partial_t^2 u = c^2 \Delta u
\end{align}
with a random index of refraction $n$.  We will be interested in the case
where the randomness is frozen in, or at most varies slowly on the time scale
of the wave propagation.  To say, we assume $x\mapsto n(x)$ to be a stationary
stochastic process with short range correlations.

An important special case is a random medium with 
a small variance of $n(x)$, which one can write as
\begin{align}\label{eq:smalldisord} 
n(x) = (1 + \sqrt{\vep} \xi(x))^{-1}
\end{align}
with $\xi(x)$ order $1$ and $\vep\ll 1$.  As argued many times, ranging
from isotope disordered harmonic crystals to seismic waves propagating in
the crust of the Earth, for such weak disorder a
kinetic description becomes possible and offers a valuable approximation to
the complete equation (\ref{eq:randomwaveeq}) -- we refer to the highly
instructive survey by Ryzhik, Keller and Papanicolaou \cite{ryzhik96} for
details. 
In the kinetic limit one considers times of order $\vep^{-1}$ and spatial
distances of order $\vep^{-1}$.  On that scale, the Wigner function $W$
associated to the solution $u$ of (\ref{eq:randomwaveeq}) is, in a good
approximation, governed by the Boltzmann type transport equation
\begin{align}\label{eq:genBtransporteq}
 & \partial_t W(x,k,t) + \nabla \omega(k)\cdot
 \nabla_x W(x,k,t) \nonumber \\
 &\quad = \int \! \rmd k' \left(r(k',k) W(x,k',t) - 
   r(k,k') W(x,k,t) \right) .
\end{align}
Here $x\in \R^3$, the physical space, and $k$ denotes the 
wave number.  $\omega$ is the dispersion relation,
$\omega(k)=c|k|$ with $k\in \R^3$ for (\ref{eq:randomwaveeq}).
Note that the left hand side of (\ref{eq:genBtransporteq}) is
the semiclassical approximation to (\ref{eq:randomwaveeq}) with $n(x)=1$.
The collision operator on the right hand side of (\ref{eq:genBtransporteq})
describes the scattering from the inhomogeneities with a rate kernel 
$r(k,k')\rmd k'$ which depends on the particular model under consideration.

Despite the wide use of the kinetic approximation (\ref{eq:genBtransporteq}),
there is no complete mathematical justification for the step from 
microscopic equations like
(\ref{eq:randomwaveeq}), together with (\ref{eq:smalldisord}),
to (\ref{eq:genBtransporteq}) apart from one exception: Erd\H{o}s 
and Yau \cite{erdyau99} (see also \cite{chen03,chen04,eng04,erdos02,erdyau04}) 
investigate the random Schr\"{o}dinger equation
\begin{align}\label{eq:randomSch}
 \ci \partial_t \psi(x,t) =(-\Delta + \sqrt{\vep} V)\psi(x,t),
\end{align}
where $\psi$ is the $\C$-valued wave function. This equation can be 
thought of as a two component wave equation for our purposes.  
In \cite{erdyau99} it is established
that (\ref{eq:genBtransporteq}) becomes valid on the kinetic scale.
Of course,
the proof exploits special properties of the Schr\"{o}dinger equation.
For us one motivation leading to the present investigation was to understand
whether the techniques developed in \cite{erdyau99} carry over to standard
wave equations such as (\ref{eq:randomwaveeq}).  In fact, with the proper
adjustments they do, and we are quite confident that also other wave equations
with small random coefficients, as e.g.\
discussed in \cite{ryzhik96}, can be treated in
the same way.  Due to the intricate nature of the estimates, we do
not claim this to be an easy exercise, but
there is a blue-print which now can be followed.

Even restricting to the scalar wave equation (\ref{eq:randomwaveeq}) there are
choices to be made. One could add dispersion as 
$c^2(\Delta u-u)$ or
the randomness could sit in the Laplacian as 
$\nabla\cdot(c(x)^2\nabla u)$ with $c(x)$ random and $n(x)=1$. 
To have a model of physical relevance, in our
contribution we will
consider a dielectric crystal in the harmonic approximation.  If, for
simplicity, the crystal structure is simple cubic, then $u_y$, $y\in\Z^3$,
are the displacements of the atoms from their equilibrium position. Their
movement is governed by Newton's equations of motion
\begin{align}\label{eq:cubicNewt}
 m_y \frac{\rmd^2}{\rmd t^2} u_y = (\Delta u)_y, \qquad y\in\Z^3.
\end{align}
Here $\Delta$ is the lattice Laplacian, which corresponds to an elastic
coupling between nearest neighbour atoms, and $m_y$ is the mass of the atom at
$y$.  (\ref{eq:cubicNewt}) can be regarded as the space discretized version
of (\ref{eq:randomwaveeq}).
Real crystals come as isotope mixtures.  For instance,
natural silicon consists in 
92.23\% of ${}^{28}$Si, 4.68\% of ${}^{29}$Si, and 3.09\% of ${}^{30}$Si.
Thus $\text{Var}(m_x)/\text{Av}(m_x)^2 \approx 10^{-4}$ and, 
in the appropriate units, we set
\begin{equation}
  \label{eq:defmi}
m_y = (1+\sqrt{\vep}\, \xi_y)^{-2}, \qquad \vep \ll 1,
\end{equation}
where $\xi_y$, $y\in\Z^3$, are
i.i.d.\ bounded, mean zero, random variables,
in slight generalization of our example.

For the discretized wave equation the wave vector space is the unit torus
$\T^3$. If $\omega$ denotes the dispersion relation for
(\ref{eq:cubicNewt}), the Boltzmann transport equation becomes 
\begin{align}\label{eq:Btransporteq}
 & \partial_t W(x,k,t) + \frac{\nabla \omega(k)}{2\pi}\cdot
 \nabla_x W(x,k,t) \nonumber \\
 &\quad = 2\pi \E[\xi_0^2]
\int \! \rmd k'  \omega(k')^2 \delta(\omega(k)-\omega(k'))
\left( W(x,k',t) -  W(x,k,t) \right) .
\end{align}
We will establish that the disorder averaged
Wigner function on the kinetic
scale, space and time of order $\vep^{-1}$, 
is governed  by (\ref{eq:Btransporteq}).
In fact, we will
allow for more general elastic couplings between the crystal
atoms than given in (\ref{eq:cubicNewt}).  Our precise assumptions on
$\omega$ will be discussed in Section \ref{sec:assumptions}.

In passing, let us remark that, to compute the thermal conductivity of real
crystals, scattering from isotope disorder contributes only 
as one part.  At least
equally important are weak non-linearities in the elastic couplings, see
\cite{spohn05} for an exhaustive discussion.  In addition, at low temperatures,
roughly below $100^\circ \text{K}$
for silicon, lattice vibrations have to be quantized.
However, for isotope disorder as in (\ref{eq:defmi}) quantization would
not make any difference, since the corresponding 
Heisenberg equations of motion are also linear. 

In a loosely related work, Bal, Komorowski, and Ryzhik \cite{bal03} study the
high frequency limit of (\ref{eq:randomwaveeq}) and (\ref{eq:smalldisord}),
under the assumption that the initial 
data vary on a space scale $\gamma^{-1}$ with $\gamma\ll \vep\ll 1$.
They prove that the Wigner function is well approximated by a transport
equation of the form (\ref{eq:genBtransporteq}).  Only the Boltzmann
collision operator is to be replaced by its small angle approximation.
Thus according to the limit equation
the wave vector $k$ diffuses on the sphere $|k|=\text{\it const.}$, whereas in 
(\ref{eq:genBtransporteq}) it would be a random jump process.  Their method
is disjoint from ours and would not be able to cover the limiting case
$\gamma=\vep$.  Bal {\it et al.\/}\
also prove self-averaging of the limit Wigner
function, while our result will concern only 
the disorder averaged Wigner function.
We expect however to have self-averaging of the Wigner function also in
our case, see \cite{chen04} for the corresponding result for the 
lattice random Schr\"{o}dinger equation (\ref{eq:randomSch}).

Wave propagation in a random medium has been studied also away from the
weak disorder regime.  As the main novelty, at strong disorder, and at any
disorder in space dimension $1$, propagation is suppressed.  The wave
equation has localized eigenmodes.  We refer to the review article
\cite{klein04}.  The regime of extended eigenmodes is still unaccessible
mathematically. The kinetic limit can be viewed as yielding some, even
though rather modest, information on the delocalized eigenmodes, compare
with \cite{chen03}.

In the following section we provide a more precise definition of the model,
describe in detail our assumptions on the dispersion relation
$\omega$ and on the initial conditions, and state the main result.

\subsection*{Acknowledgements} 

Our interest in wave propagation in random media where triggered by
discussions of H.S.\ with H.-T.\ Yau during a common stay 
at the Institute for Advanced Study, Princeton  in the spring
2003.  We are grateful to
L\'{a}szl\'{o} Erd\H{o}s and Thomas Chen for their constant support and
encouragement. We also thank A.\ Kupiainen, A.\ Mielke, G.\ Panati, and S.\
Teufel for instructive discussions.

J.L. acknowledges support
from the Deutsche Forschungsgemeinschaft (DFG) project SP~181/19-1 and 
from the Academy of Finland in 2003.
This work has also been supported by 
the European Commission through its 6th Framework Programme ``Structuring
the European Research Area'' and the contract Nr. RITA-CT-2004-505493 for
the provision of Transnational Access implemented as Specific Support
Action.

\section{Main result}
\label{sec:main}

\subsection{Discrete wave equation}
\label{sec:model}

We will study the kinetic limit of the discrete wave equation 
\begin{align}\label{eq:defDyn}
 \frac{\rmd}{\rmd t} q_y(t) & = v_y(t), \nonumber \\
(1+\sqrt{\vep}\, \xi_y)^{-2} 
\frac{\rmd}{\rmd t} v_y(t) & = -\sum_{y'\in\Z^3} \alpha(y-y') q_{y'}(t)
\end{align}
with $y\in \Z^3$ and $q_y(t),v_y(t)\in \R$.  As a shorthand we set
$q(t)=(q_y(t),y\in\Z^3)$, $v(t)=(v_y(t),y\in\Z^3)$.
The mass of the atom at site $y$ is $(1+\sqrt{\vep}\, \xi_y)^{-2}$,
where $\xi=(\xi_y,y\in\Z^3)$ is a family of independent, identically
distributed random variables. Their common distribution is independent of
$\vep$, has zero mean and is supported on the interval 
$[-\bar{\xi},\bar{\xi}]$.  Expectation with respect to $\xi$ is denoted by
$\E$. We assume $\vep<\vep_0=\bar{\xi}^{-2}$ throughout. Hence
$1+\sqrt{\vep}\,\xi_y>0$ with probability one.

The coefficients $\alpha(y)$ are the elastic couplings between atoms, 
and we require them to have the following properties. 
\begin{jlist}[(E\thejlisti)]
\item\label{it:EC0} $\alpha(y)\ne 0$ for some $y\ne 0$.
\item\label{it:EC1} $\alpha(-y)=\alpha(y)$ for all $y$.
\item\label{it:EC2} There are constants $C_1,C_2>0$ such that for all $y$
\begin{align}\label{eq:expdec}
|\alpha(y)|\le C_1 \rme^{-C_2 |y|}.
\end{align}
\item\label{it:EC3} Let $\FT{\alpha}$ be the Fourier transform of $\alpha$,
which we define by
\begin{align}\label{eq:defFT}
 \FT{\alpha}(k) = \sum_{y\in \Z^3} \rme^{-\ci 2 \pi k \cdot y} \alpha(y).
\end{align}
Then $\FT{\alpha}:\T^3\to \R$, where $\T^3$ denotes the $3$-torus with unit
side length.  Mechanical stability demands $\FT{\alpha}\ge 0$. We require
here the somewhat stronger condition 
\begin{align}\label{eq:pinning}
\FT{\alpha}(k)> 0, \quad \text{for all }k\in \T^3.
\end{align}
\end{jlist}
If $\vep=0$, Eqs.~(\ref{eq:defDyn}) admit plane wave solutions with wave
vector $k\in\T^3$ and frequency
\begin{align}\label{eq:defom}
\omega(k)=\sqrt{\FT{\alpha}(k)}.
\end{align}
The function $\omega:\T^3\to \R$ is the {\em dispersion relation\/}.
Under our assumptions for $\alpha$, $\omega$ is real-analytic,
$\omega(-k)=\omega(k)$, $0<\ommin=\min_k\omega(k)$, and
$\ommax=\max_k\omega(k)<\infty$.

We solve the differential equations
(\ref{eq:defDyn}) as a Cauchy problem with initial data $q(0),v(0)$.
The time-evolution (\ref{eq:defDyn}) conserves the energy
\begin{align}\label{eq:defHam}
 E(q,v) = \frac{1}{2}  \Bigl( \sum_{y\in \Z^3} (1+\sqrt{\vep}\, \xi_y)^{-2} 
 v^2_y + \sum_{y,y'\in \Z^3} \alpha(y-y') q_y q_{y'} \Bigr).
\end{align}
The initial data are assumed to have finite energy, 
$E(q(0),v(0))<\infty$.  Since $\ommin >0$,
this implies that $q(0),v(0)\in\ell_2(\Z^3,\R)$.  
For any realization of $\xi$, the generator of the time-evolution
(\ref{eq:defDyn}) is a bounded operator on $\ell_2(\Z^3,\R^2)$. Therefore,
the Cauchy problem has a
unique, norm-continuous  solution which remains in $\ell_2(\Z^3,\R^2)$
for all $t\in \R$.

The energy depends on the realization of $\xi$, and it will be more 
convenient to switch to new variables such that the flat $\ell_2$-norm is
conserved.  For this purpose, let tilde denote the 
inverse Fourier transform, for which we adopt the convention
\begin{equation}
  \label{eq:definvFT}
  \IFT{f}_y = \int_{\T^3}\!\rmd k \, \rme^{\ci 2\pi y\cdot k}
  f(k),
\end{equation}
and let $\Omega$ denote the bounded operator on 
$\ell_2(\Z^3,\C)$ defined via
\begin{align}\label{eq:defOm}
(\Omega \phi)_y = \sum_{y'\in\Z^3} \IFT{\omega}_{y-y'} \phi_{y'}.
\end{align}
Since $q(t),v(t)\in\ell_2(\Z^3,\R)$, we can introduce the vector
$\psi(t)\in \ell_2(\Z^3,\C^2)$ through
\begin{align}\label{eq:defPsi}
  \psi(t)_{\sigma,y} =   \frac{1}{2} \left( (\Omega q(t))_y + 
  \ci \sigma (1+\sqrt{\vep}\, \xi_y)^{-1} v(t)_y \right),
\end{align}
where $\sigma = \pm 1$ and $y\in\Z^3$.  From now on, let
us denote $\ell_2=\ell_2(\Z^3,\C)$, and
$\hilb = \ell_2(\Z^3,\C^2)=\ell_2\oplus \ell_2$.

If we regard $\xi$ as a multiplication operator on $\ell_2$, i.e.,
if we define $(\xi\psi)_y = \xi_y \psi_y$, then
$\psi(t)$ satisfies the differential equation
\begin{align}\label{eq:hilbevol}
  \frac{\rmd}{\rmd t} \psi(t) = -\ci H_\vep \psi(t),
 \quad\text{with}\quad
   H_\vep  = H_0 + \sqrt{\vep} V,
\end{align}
where
\begin{align}\label{eq:defH0}
 H_0 = \begin{pmatrix} \Omega & 0\\ 
  0 & -\Omega \end{pmatrix},\quad
 V = \frac{1}{2} \begin{pmatrix} 
   \Omega\xi + \xi \Omega & 
   -\Omega\xi + \xi \Omega \\ 
   \Omega\xi - \xi \Omega & 
   -\Omega\xi - \xi \Omega \end{pmatrix} . 
\end{align}
Because $H_\vep$ is a self-adjoint operator on 
$\hilb$, the solution to
(\ref{eq:hilbevol}) generates a unitary group on $\hilb$.
Unitarity is equivalent to energy conservation, since for all $t$
\begin{align}\label{eq:normisE}
  \norm{\psi(t)}^2 = E(q(t),p(t)).
\end{align}
If $\psi(t)$ is one of the ``physical'' states obtained by
(\ref{eq:defPsi}), then it
satisfies $\psi(t)_{-,y}^*=\psi(t)_{+,y}$ for all $y$ and $t$ due to
$q(t),v(t)\in \R$.
We will discuss in Section \ref{sec:classical} how information about
$\psi(t)$ is transferred to $q(t)$, $v(t)$.

\subsection{Lattice Wigner function, initial conditions, and dispersion
  relation} 
\label{sec:assumptions}

The disorder has strength $\sqrt{\vep}$.  Since $\E[\xi_0]=0$,
effects of order $\sqrt{\vep}$ vanish in the mean, and 
a wave packet has a mean free path of the order of $\vep^{-1}$ lattice
spacings.  In the 
kinetic limit the speed of propagation of the waves is independent of 
$\vep$, indicating that the first time-scale, at which the
randomness becomes relevant, is also of the order of $\vep^{-1}$.
If $\xi=0$, this scaling corresponds to the semiclassical limit in which
the Wigner function satisfies the transport equation
\begin{align}
 & \partial_t W(x,k,t) + \frac{\nabla \omega(k)}{2\pi}\cdot
 \nabla_x W(x,k,t) = 0, \quad x\in\R^3,\ k\in \T^3 .
\end{align}
We refer to \cite{mielke05} for an exhaustive discussion.  From this 
perspective, the Wigner function is the natural object for studying the
kinetic limit.

Given a scale $\vep>0$, 
we define the Wigner function $W^\vep$ of any state
$\psi\in \hilb$ as the distributional Fourier transform
\begin{align}\label{eq:Wdef}
  W_{\sigma'\sigma}^\vep(x,k)   = 
  \int_{\R^3}\! \!\rmd p\, \rme^{\ci 2\pi x\cdot p}
  \FT{\psi}_{\sigma'}\Bigl(k-\frac{1}{2}\vep p\Bigr)^*
  \FT{\psi}_{\sigma}\Bigl(k+\frac{1}{2}\vep p\Bigr),
\end{align}
where $\sigma',\sigma \in \set{\pm 1}$,
$x\in\R^3$, $k\in \T^3$. This is also called the 
Wigner transform of $\psi$ and we denote it by $W^\vep[\psi]$.
The Fourier transform of $\psi$, denoted by
$\FT{\psi}$, is defined as in (\ref{eq:defFT}) and periodically
extended to a function on the whole of $\R^3$.
$W$ is an Hermitian $\M_2$-valued (complex $2{\times}2$-matrix) 
distribution. When integrated against a matrix-valued test-function 
$J\in \cals(\R^3\times\T^3,\M_2)$, (\ref{eq:Wdef}) becomes
\begin{align}\label{eq:defWpsi}
& \mean{J,W^\vep} =  
  \int_{\R^3}\! \!\rmd p 
  \int_{\T^3}\! \!\rmd k\,
  \FT{\psi}\Bigl(k-\frac{1}{2}\vep p\Bigr) \cdot
  \FT{J}(p,k)^*
  \FT{\psi}\Bigl(k+\frac{1}{2}\vep p\Bigr),
\end{align}
where $\FT{J}$ denotes the Fourier transform of $J$ in the first variable,
\begin{align}\label{eq:defFT1J}
 \FT{J}(p,k) = 
  \int_{\R^3} \!\rmd x\, \rme^{-\ci 2\pi p\cdot x}   J(x,k).
\end{align}
The notation $A^*$ denotes Hermitian conjugation, and the dot is used for
a finite-dimensional scalar product: $a\cdot b = \sum_i a_i^* b_i$.   
We have included a complex conjugation of 
the test-function in the definition in order to have the 
same sign convention for the Fourier transform of both test functions and
distributions. 

Let us choose now some initial conditions for (\ref{eq:hilbevol}).
In general, it will be $\vep$-dependent and we denote it by $\psi^\vep$.
The solution to (\ref{eq:hilbevol}) is then
\begin{align}\label{eq:defpsisol}
\psi(t) = \rme^{-\ci t H_\vep} \psi^\vep .
\end{align}
In the following we will be studying a limit where $\vep\to 0^+$ via some
arbitrary sequence of values. Our assumptions on the initial conditions are 
\begin{assumption}[Initial conditions]\label{th:initass}
For every $\vep$, there is $\psi^\vep \in\hilb$, independent of $\xi$,
such that 
\begin{jlist}[{\rm (IC\thejlisti)}]
\item\label{it:I1} $\displaystyle\sup_\vep \norm{\psi^\vep} < \infty$.
\item\label{it:I2} $\displaystyle\lim_{R\to\infty}
  \limsup_{\vep\to 0} \sum_{|y|> R/\vep} 
  |\psi^\vep_{y}|^2 = 0$.
\item\label{it:I3} There exists a positive bounded Borel measure $\mu_0$ on
$\R^3\times \T^3$ such that 
\begin{align}\label{eq:IClim}
\lim_{\vep\to 0}\mean{J,W^\vep_{++}[\psi^\vep]} =
 \int_{\R^3\times \T^3} \!\!\mu_0(\rmd x \,\rmd k)\, J(x,k)^*.
\end{align}
for all $J\in \cals(\R^3\times \T^3)$.
\end{jlist}
\end{assumption}
These assumptions are rather weak.  In fact, as discussed in the 
Appendix \ref{sec:appWigner}, if we assume 
(IC\ref{it:I1}), then 
the existence of the limit in (\ref{eq:IClim}) for all $J$  implies
already the existence of the measure $\mu_0$.
The condition (IC\ref{it:I2}) means that the sequence 
$|\psi^\vep_{y}|^2/\norm{\psi^\vep}^2$ of probability measures on $\Z^3$ is
tight on the kinetic scale $\vep^{-1}$.

Our second set of assumptions deals with the dispersion relation $\omega$. 
For this we need to introduce the notations
\begin{align}\label{eq:defsabs}
  \sabs{x} = \sqrt{1+x^2} \qand
\norm{f}_{N,\infty} = \sup_{|\alpha|\le N} \norm{D^\alpha\! f}_\infty ,
\end{align}
where $N=0,1,\ldots$ and $\alpha$ denotes a multi-index.
\begin{assumption}[Dispersion relation]\label{th:disprelass}
Let $\omega:\T^3\to\R$ satisfy all of the following:
\begin{jlist}[(DR\thejlisti)]
\item\label{it:DC1} $\omega$ is smooth and $\omega(-k)=\omega(k)$.
\item\label{it:DC2} $\ommin>0$ and 
$\ommax < \infty$ with $\ommin = \min_{k} \omega(k)$ and
$\ommax = \max_{k} \omega(k)$.
\item ({\em dispersivity})\label{it:suffdisp}\ \ 
There are constants
$d_1\in\N$ and $\comega>0$ such that 
for all $t\in \R$ and $f\in C^\infty(\T^3)$,
\begin{align}\label{eq:suffdisp}
 \left| \int_{\T^3} \rmd k\, f(k) \rme^{-\ci t \omega(k)}\right|
 \le  \frac{\comega}{\sabs{t}^{3/2}}\norm{f}_{d_1,\infty} .
\end{align}
\item\label{it:crossing} ({\em crossings are suppressed})\ \
There are constants $c_2>0$, $0<\gamma\le 1$ and $d_2\in\N$
such that for all $u\in \T^3$, $0<\beta\le 1$, $\alpha\in \R^3$,
and $\sigma\in\set{\pm 1}^3$,
\begin{align}\label{eq:crossingest}
&  \int_{(\T^3)^2} \rmd k_1\rmd k_2\, 
\frac{1}{|\alpha_1-\sigma_1 \omega(k_1)+\ci\beta|
|\alpha_2-\sigma_2 \omega(k_2)+\ci\beta|} 
\nonumber \\ & \qquad\times
\frac{1}{|\alpha_3-\sigma_3 \omega(k_1-k_2+u)+\ci\beta|}
 \le c_2 \beta^{\gamma-1} \sabs{\ln \beta}^{d_2} .
\end{align}
\end{jlist}
\end{assumption}

If $\omega$ has only isolated, non-degenerate critical points,
i.e., if $\omega$ is a Morse function, then the 
bound (\ref{eq:suffdisp}) with $d_1=4$ follows
by standard stationary phase methods.
The crossing condition is more difficult to verify.  We will discuss these
issues in detail in Sec.~\ref{sec:dispersion}, where examples satisfying
(DR\ref{it:suffdisp}) and (DR\ref{it:crossing}) are also provided.

\subsection{Main Theorem}
\label{sec:mainth}

The Boltzmann equation (\ref{eq:Btransporteq}) 
is the forward equation of a Markov jump process
$(x(t),k(t))$, $t\ge 0$.  $k(t)$, $t\ge 0$ is governed by the collision rate 
\begin{align}\label{eq:defnuk}
\nu_k(\rmd k') = \rmd k' \delta(\omega(k)-\omega(k')) 2\pi 
 \E[\xi_0^2]\omega(k')^2,
\quad k\in\T^3 ,
\end{align}
with a total collision rate
\begin{align}\label{eq:deftotcs}
  \totalcs(k) = \nu_k(\T^3).
\end{align}
As proved in Appendix \ref{sec:appBoltzmann}, since $\omega$ is continuous
and (DR\ref{it:suffdisp}) is satisfied, the map
\begin{align}\label{eq:cscont}
 k\mapsto \int\! \nu_k(\rmd k') g(k')
\end{align}
is continuous for every $g\in C(\T^3)$. In particular,
$\csmax = \sup_k \sigma(k) < \infty$.  Now given $k=k(0)$, 
one has $k(t)=k$ for $0\le t \le \tau$ with $\tau$ an exponentially
distributed random variable of mean $\totalcs(k)^{-1}$. At time $\tau$,
$k$ jumps to $\rmd k'$ with probability $\nu_k(\rmd k')/\sigma(k)$, etc.  
To define the joint
process $(x(t),k(t))$, $t\ge 0$, one sets
\begin{align}\label{eq:freebevol}
  \frac{\rmd}{\rmd t} x(t) = \frac{1}{2\pi} \nabla\omega(k(t)) .
\end{align}
We assume the process to start in the measure $\mu_0$ from 
(IC\ref{it:I3}).    Because of continuity in (\ref{eq:cscont}),
the process $(x(t),k(t))$, $t\ge 0$, is Feller.  Hence there is a
well-defined joint distribution at time $t$, which we denote by 
$\mu_t(\rmd x \,\rmd k)$.

We are now ready to state our main result.
\begin{theorem}\label{th:main}
Let the Assumptions  \ref{th:initass} and
\ref{th:disprelass} hold and let $\psi(t)$ denote the random vector 
determined by (\ref{eq:defpsisol}).
Then for all $t\ge 0$,  
$J\in \cals(\R^3\times \T^3)$, one has the  limit
\begin{equation}
  \label{eq:Wlimit}
  \lim_{\vep\to 0} \E[\mean{J,W_{++}^\vep[\psi(t/\vep)]}] =
 \int_{\R^3\times \T^3} \!\!\mu_t(\rmd x \,\rmd k)\, J(x,k)^*.
\end{equation}
\end{theorem}

As a complete theorem, one would have expected a limit for the Wigner
matrix, not just for the ($++$)-component, as stated above.
From the evolution equation (\ref{eq:hilbevol}) it follows immediately that
Theorem \ref{th:main} also holds for $W_{--}$.  One only has to assume
(IC\ref{it:I3}) for $W_{--}^\vep$, and replace everywhere
$\omega$ by $-\omega$.  As the rate kernel remains unchanged, 
this amounts to changing the sign in
(\ref{eq:freebevol}). For the deterministic initial data of Section
\ref{sec:model} the off-diagonal components $W^\vep_{+-}$ and $W^\vep_{-+}$
are fastly oscillating.  In general, they do not
have a pointwise limit, but vanish upon time-averaging, i.e., for any 
$t\ge 0$, $T>0$ and $\sigma=\pm 1$, one has
\begin{align}
 \lim_{\vep\to 0} \frac{1}{T}\int_{0}^T\!\! \rmd \tau\, 
 \E\!\left[W^\vep_{\sigma,-\sigma}[\psi( (t+\tau)/\vep)]\right]
 = 0.
\end{align}

Physically, one would like to avoid the assumption $\ommin>0$, since
elastic forces depend only on the relative distances between atoms, and thus
$\FT{\alpha}(0)=0$.  If $\FT{\alpha}(0)=0$, generically 
$\omega(k)\approx |k|$ for small $k$.  In addition, by (\ref{eq:defH0}),
the two bands of $H_0$ touch at $k=0$.  On a technical level, the
non-smooth crossing of the bands
adds another layer of difficulty which we wanted to
avoid here.

To give a brief outline: In the following section we exploit general 
properties about weak limits of lattice Wigner transforms to reduce the proof
of the main theorem into a Proposition stating that their Fourier transforms
converge to the characteristic functions of $(\mu_t)$. 
These properties concerning the Wigner
transform are valid under more general assumptions than those of the main
theorem, and we have separated their derivation to Appendix 
\ref{sec:appWigner}.  The core of the paper is the graphical expansion of
Section \ref{sec:graphs} where the proof of the above Proposition is done by
dividing it into several layers 
with ever more detailed Lemmas acting as links between
the different layers.  In particular, we have separated the analysis of the
non-vanishing parts of the graph expansion, so called simple graphs, to
Section  \ref{sec:simple}. 

The graph expansion follows the outline laid down in the works cited
earlier.  The new ingredients are the matrix structure
and the momentum dependence of the interaction.
We also develop here an alternative version for 
the so-called partial time-integration needed in the estimation of the
error terms.  The present version, described in Sec.~\ref{sec:duhamel},
facilitates the analysis of the error terms, allowing the use of same
estimates for both partially time-integrated and fully expanded 
graphs.  We also consider here more general dispersion relations and 
initial conditions than before, although it needs to be stressed that in
the case of 
the dispersion relation, the improvement is mainly a matter of more 
careful bookkeeping.  

The estimates, which
allow the division of the graphs into leading and subleading ones, rely on
the decay estimates (DR\ref{it:suffdisp}) and (DR\ref{it:crossing}).  In
section \ref{sec:dispersion} we discuss proving (DR\ref{it:crossing}) for a
given dispersion relation in more detail.  In particular, we show there
that the taking of the square root, which is necessary for obtaining the
dispersion relation from the elastic couplings, in general retains the
validity of the crossing estimate.  Finally, in the last
section we return to the original lattice dynamics (\ref{eq:defDyn}),
explain how (IC\ref{it:I1}) -- (IC\ref{it:I3}) relate to the initial
positions and velocities, and, in particular, discuss the propagation of the
energy density. 

\section{Proof of the Main Theorem}
\label{sec:proofmain}

In all of the results in this and the following two sections, unless stated
otherwise, we make the assumptions of Theorem
\ref{th:main}.  In addition, we assume that $\E[\xi_0^2]=1$.  This is not
a restriction, as it can always be achieved by rescaling $\xi$ by 
$\E[\xi_0^2]^{-\frac{1}{2}}$ and $\vep$ by $\E[\xi_0^2]$.  We study a given
sequence $(\vep_k)$, $k=1,2,\ldots$, such that $0<\vep_k<\vep_0$ and 
$\lim_k \vep_k=0$.   For notational 
simplicity, we will always denote the limits of the type
$\lim_{k\to\infty} f(\vep_k)$ by $\lim_{\vep\to 0} f(\vep)$.

We will study the limits of the mappings $\owvept$ 
defined by $\mean{J,\owvept }= 
\E[\mean{J,W_{++}^\vep[\psi(t/\vep)]}]$ where
$\psi(t) = \rme^{-\ci H_\vep t} \psi^\vep$. For any $\vep$ and
$t\ge 0$,
the mapping $\xi \mapsto \psi(t/\vep)$ lifts the probability measure for 
$\xi$ to a probability measure $\nu^\vep_t$ 
on the Hilbert space $\hilb$. For instance, $\nu^\vep_0$ 
is a Dirac measure concentrated at $\psi^\vep$.
Each of the measures $\nu^\vep_t$ is 
a weak Borel measure. In particular, let us prove next that every
$\psi(t/\vep)_{\sigma,y}$ is measurable. 
For any $R>0$, define
$\VR$ as the potential obtained by
neglecting far lying perturbations $\xi$, i.e., let
\begin{equation}
  \label{eq:defVL}
  \VR =  \sum_{\norm{y}_\infty \le R} \xi_{y} \Vy  
\end{equation}
where $\Vy$, $y\in\Z^3$, has a Fourier transform given by
the integral kernel
\begin{equation}  \label{eq:VFT}
   \FT{V}^{(y)}_{\sigma'\sigma}(k',k) = \rme^{-\ci 2\pi y\cdot(k'-k)}
 v_{\sigma'\sigma}(k',k) ,
\end{equation}
and $v\in L^2(\T^3\times\T^3,\M_2)$ is defined for 
$\sigma',\sigma\in \set{\pm1}$ and $k',k\in \T^3$ by
\begin{equation}
  \label{eq:vdef}
 v_{\sigma'\sigma}(k',k) = \frac{\sigma'\sigma}{2} 
  \left( \sigma' \omega(k') + \sigma \omega(k) \right).
\end{equation}
Then $\VR \to V$ strongly 
(i.e., for all $\psi\in\hilb$, $\norm{\VR\psi - V\psi}\to 0$)
when $R\to\infty$, and, as $\sup_R \norm{\VR}<\infty$, 
the same is true for any product of $\VR$:s and
bounded $R$-independent operators.  Therefore,
\begin{align}\label{eq:measurability}
\psi(t/\vep)_{\sigma,y} = \lim_{R\to\infty} \sum_{N=0}^R 
\frac{(-\ci t/\vep)^N}{N!} ( (H_0+\sqrt{\vep} \VR)^N\psi^\vep )_{\sigma,y} .
\end{align}
As the summand  is a complex function depending
only on finitely many of $(\xi_y)$, it is measurable.
For all $|\xi|\le \bar{\xi}$, $\psi(t/\vep)_{\sigma,y}$ is a convergent
limit of a sequence of such functions, 
which implies that also $\psi(t/\vep)_{\sigma,y}$ is measurable. 
In addition, by the unitarity of $\rme^{-\ci H_\vep t}$,
\begin{align}
  \norm{\psi(t/\vep)}^2 =   \norm{\psi^\vep}^2
\end{align}
which is uniformly bounded by (IC\ref{it:I1}).

By Theorem \ref{th:cwvep},
$\mean{J,W_{\nu_t^\vep}^\vep} = \int\!\nu_t^\vep(\rmd \psi)
\mean{J,W^\vep[\psi]}$ defines a
distribution in $\cals'(\R^3\times \T^3,\M^2)$
which we call the Wigner transform of the measure
$\nu_t^\vep$. These distributions behave
very similarly to probability measures on 
$\R^3\times \T^3$, and we have collected
their main properties in Appendix \ref{sec:appWigner}.
In particular, we can conclude that
$\owvept\in \cals'(\R^3\times\C^3)$, 
as for all $J\in \cals(\R^3\times\C^3)$,
\begin{align}
\mean{J,\owvept}=\mean{J,(W_{\nu_t^\vep}^\vep)_{++}}=
\int\!\nu_t^\vep(\rmd \psi)
\mean{J,W^\vep[\psi_+]}.
\end{align}
By Proposition \ref{th:Fnuprop},
the Fourier transform of $\owvept$ is determined by the functions
\begin{align}\label{eq:FvepFourier2}
&  F^\vep_t(p,n) =
  \E_{\nu_t^\vep}\!\!\left[ 
    \int_{\T^3}\! \!\rmd k\, \rme^{\ci 2 \pi n\cdot k}
  \FT{\psi}_+\!\Bigl(k-\frac{1}{2}\vep p\Bigr)^*
  \FT{\psi}_+\!\Bigl(k+\frac{1}{2}\vep p\Bigr)\right] 
\end{align}
where $p\in \R^3$ and $n\in \Z^3$. 
The assumptions (IC\ref{it:I1}) -- (IC\ref{it:I3}) allow then
applying Theorem 
\ref{th:weakimpliesborel} to conclude that $F^\vep_0$
converges pointwise to the Fourier transform of $\mu_0$ which, by
Theorem \ref{th:FimpliesWweak}, implies
\begin{lemma}\label{th:winitlim}
For all $p\in \R^3$ and $f\in C(\T^3)$,
\begin{align}
\lim_{\vep\to 0} \int_{\T^3} \!\!\rmd k f(k)\,
  \FT{\psi}^\vep_+\!\Bigl(k-\frac{1}{2}\vep p\Bigr)^*
  \FT{\psi}^\vep_+\!\Bigl(k+\frac{1}{2}\vep p\Bigr)
 =  \int_{\R^3\times \T^3}\!\! \!\!\mu_0(\rmd x \,\rmd k)\, 
\rme^{-\ci 2 \pi p\cdot x} f(k) .
\end{align}
\end{lemma}

Using time-dependent perturbation
expansion, we will prove in Section \ref{sec:graphs} that
\begin{proposition}\label{th:mainlim}
For all $\tmacro>0$, $\pmacro\in \R^3$ and $\nmacro\in \Z^3$
\begin{align}\label{eq:Fveptomut}
& \lim_{\vep\to 0}  F^\vep_{\tmacro}(\pmacro,\nmacro)
 =  \int_{\R^d\times \T^d}\!\! \!\!\mu_{\tmacro}(\rmd x \,\rmd k)\, 
\rme^{-\ci 2 \pi (\pmacro\cdot x- \nmacro\cdot k)}.
\end{align}
\end{proposition}
Then we can apply Theorem \ref{th:FimpliesWweak}, and conclude that for
any $t>0$, the sequence $(\owvept)_\vep$ converges
in the weak-$*$ topology to a bounded positive Borel measure whose
characteristic function coincides with the limit of $F^\vep_t$.
However, then by (\ref{eq:Fveptomut}) this measure is in fact equal to 
$\mu_t$.   This is sufficient to prove Theorem \ref{th:main}, since 
(\ref{eq:Wlimit}) is valid at $t=0$ by assumption (IC\ref{it:I3}).

\section{Graph expansion (proof of Proposition \ref{th:mainlim})}
\label{sec:graphs}

In this section we assume that all assumptions of Proposition
\ref{th:mainlim} are valid.  In particular,
$\pmacro\in \R^3$, $\nmacro\in \Z^3$ and $\tmacro>0$ 
will denote the fixed macroscopic parameters.
We first derive, using time-dependent perturbation theory, 
a way of splitting the time-evolved states into two parts,
\begin{align}
  \rme^{-\ci t H_\vep }\psi^\vep = \psimain(t) + \psierr(t).
\end{align}
The splitting is done in such the way that each part is component-wise
measurable, as before, and 
\begin{align}
& \lim_{\vep\to 0}\E\!\left[\norm{\psierr(\tmacro/\vep)}^2\right] = 0.
\end{align}
Then we will only need to inspect the limit of the main part.

\subsection{Duhamel expansion with soft partial time-integration}
\label{sec:duhamel}

We begin by deriving the above splitting.
Since both $H_0$ and $V$ are bounded operators for any realization of the
randomness, the Duhamel formula  
states that, for any $t\in \R$ we have as vector valued
integrals in $\banach(\hilb)$, 
\begin{gather}\label{eq:baseduh}
   \rme^{-\ci t H_\vep } = \rme^{-\ci t H_0} + 
  \int_0^{t}\!\! \rmd s\, \rme^{-\ci (t-s) H_\vep } (- \ci \sqrt{\vep} V)
  \rme^{-\ci s H_0} .
\end{gather}
This could be iterated to yield the full Dyson series which, however, would
become ill-behaved in the kinetic limit.  
Instead, we will expand the series only partially,
up to $N_0$ ``collisions''.  For the remainder we use a different
method, essentially a version of the ``partial time integration''
introduced in \cite{erdyau99} with a ``soft cut-off'' 
which allows easier analysis
of the error terms.  The results will be expressed 
in terms of the following (random) 
functions: 
\begin{definition}\label{th:defFGA}
For any $\vep$, and any $\kappa\ge 0$ and $s\in \R$ let
\begin{align}\label{eq:defW}
 W_{s} & = (- \ci \sqrt{\vep} V) \rme^{-\ci s H_0} 
 \qand W_{s,\kappa} = (-\ci \sqrt{\vep} V) 
 \rme^{-\ci s (H_0-\ci \kappa)},
\end{align}
and define
for any $\kappa\ge 0$, $t>0$, and $N,N',N_0\in\N$ with $N_0\ge 1$,
as vector valued integrals in $\banach(\hilb)$,
\begin{align}
F_{N}(t;\vep) & = \int_{\R_+^{N+1}}\!\rmd s \,
  \delta\Bigl(t-\sum_{\ell=1}^{N+1} s_\ell\Bigr) 
  \rme^{-\ci s_{N+1} H_0} W_{s_{N}}\cdots W_{s_1}, \label{eq:defFN} \\
 G_{N',N}(t;\vep,\kappa) & =   \int_{\R_+^{N+N'+1}}\!\!\!\!\!\!\!\!\!\!\! 
\rmd s \,
  \delta\Bigl(t-\!\!\!\!\sum_{\ell=1}^{N+N'+1}\!\!\!\! s_\ell\Bigr)
  \rme^{-\ci s_{N+N'+1} (H_0-\ci\kappa)}\!
  \prod_{j=N+1}^{N+N'}\!\!\!\! W_{s_{j},\kappa}
  \prod_{j=1}^{N} W_{s_{j}},   \label{eq:defGN} \\
A_{N',N_0}(t;\vep,\kappa) & =  \int_{\R_+^{N_0+N'}}\! \rmd s \,
  \delta\Bigl(t-\sum_{\ell=1}^{N_0+N'} s_\ell\Bigr)
  \prod_{j=N_0+1}^{N_0+N'} W_{s_{j},\kappa}
  \prod_{j=1}^{N_0} W_{s_{j}} \label{eq:defAN}.
\end{align}
Let us also define
\begin{align}\label{eq:defzeroval}
F_{N}(0;\vep)=\delta_{N0}\1,\quad
G_{N',N}(0;\vep,\kappa)=\delta_{N+N',0}\1,\quad\text{and}\quad 
A_{N',N_0}(0;\vep,\kappa) = 0.
\end{align}
\end{definition}
In these definitions, the notation $\rmd s\, \delta(t-\sum_{\ell=1}^N s_\ell)$,
with $t>0$ and $N\in \N_+$, 
refers to a bounded positive Borel measure on $\R_+^{N}$ defined naturally by
the $\delta$-function by integrating out one of the coordinates $s_\ell$.  
Explicitly, for any $f\in C(\R^{N})$ we have, for $N=1$,
$\int_{0}^\infty\! \rmd s\,  \delta(t-s) f(s) = f(t)$, and for $N\ge 1$,
\begin{align}
&  \int_{\R_+^{N}}\!\!\rmd s\,  \delta\Bigl(t-\sum_{\ell=1}^N s_\ell\Bigr)
f(s)
 =  \int_{\R_+^{N-1}}\!\!\!\!\!\!\!\rmd s  \,
\1\!\Bigl(\sum_{\ell=1}^{N-1} s_\ell\le t\Bigr) 
 f\Bigl(s_1,\ldots,s_{N-1},t-\sum_{\ell=1}^{N-1} s_\ell\Bigr).
\end{align}
The function $\1$ in the integrand restricts the integration region to 
the standard simplex in $\R^{N-1}$ scaled by the factor $t$.
This is a compact set and therefore, as long as the integrand is 
a continuous mapping from the simplex to a \frechet\  space, it
can be used to define vector valued integrals in the sense of
\cite{Rudin:FA}, Theorem 3.27.  The measure is invariant under permutations
of $(s_\ell)$ -- which proves that we could have integrated out
any of the coordinates, not only the last one  -- and it is bounded by
\begin{align}\label{eq:tNbound}
  \int_{\R_+^{N}}\!\rmd s \,\delta\Bigl(t-\sum_{\ell=1}^N s_\ell\Bigr)
 =  \frac{t^{N-1}}{(N-1)!}.
\end{align}

The proof that the integrands in the Definition \ref{th:defFGA} are
continuous, as 
well as a number of useful relations between the functions, are given in
the following:
\begin{lemma}\label{th:FGAcontin}
$W_s$ and $W_{s,\kappa}$ are continuous in $\banach(\hilb)$ in the variable
$s$, as well as are all of the functions defined in
(\ref{eq:defFN}) -- (\ref{eq:defzeroval}) in $t$.  They are also related by
the following equalities for all $N_0\ge 1$ and $N',N\ge 0$:
\begin{align}
F_N(t;\vep) & = G_{0,N}(t;\vep,0), \label{eq:FGrel}\\
F_{N_0}(t;\vep) & = 
\int_0^{t}\!\! \rmd r\, \rme^{-\ci (t-r) H_0} A_{0,N_0}(r;\vep,\kappa), 
\label{eq:FArel}\\
G_{N',N_0}(t ;\vep,\kappa) & =
\int_0^{t }\!\! \rmd r\,  \rme^{-\kappa (t -r)}
 \rme^{-\ci (t -r) H_0} A_{N',N_0}(r;\vep,\kappa),\label{eq:GArel} \\
A_{0,N_0+1}(t;\vep,\kappa) & =
\int_0^{t}\!\! \rmd r\,
 W_{t-r} A_{0,N_0}(r;\vep,\kappa), \label{eq:AA0rel} \\
A_{N'+1,N_0}(t;\vep,\kappa) & =
\int_0^{t}\!\! \rmd r\,
 W_{t-r,\kappa} A_{N',N_0}(r;\vep,\kappa). \label{eq:AArel}
\end{align}
\end{lemma}
\begin{proof}
As $H_0$ is bounded, 
$s\mapsto \rme^{-\ci s H_0}$ is norm-continuous for all $s\in \R$, and
so are then $W_s$ and $W_{s,\kappa}$. 
This proves that the integrands in the definitions 
(\ref{eq:defFN}) -- (\ref{eq:defAN})
are continuous functions for all real $s$, and thus all of  the vector valued
integrals are well-defined in  $\banach(\hilb)$.

We next need to prove the continuity of the functions $F(t)$, $G(t)$ and 
$A(t)$.
Since the proof is essentially identical in all three cases, we shall do it
only for $F$.  First, for $N=0$, we have
$F_0(t) = \rme^{-\ci t H_0}$ which is norm-continuous for
$t>0$, and $\lim_{t\to 0^+}F_0(t) =\1$, which proves that $F_0$ is
continuous also at $t=0$. When $N>0$, we have explicitly for all $t>0$
\begin{align}\label{eq:defFNexpl}
F_N(t) & =  \int_{s\in\R_+^{N}}\! \rmd s \, 
\1\Bigl(\sum_{\ell=1}^N s_\ell \le t\Bigr)
\rme^{-\ci \left(t-\sum_{\ell=1}^{N} s_\ell\right)H_0}
W_{s_{N}}\cdots W_{s_1}.
\end{align}
Since 
$\norm{W_s}, \norm{W_{s,\kappa}}\le \sqrt{\vep}\norm{V}$, 
we have by (\ref{eq:tNbound}),
$\norm{F_N(t)}\le (\sqrt{\vep}\norm{V} t)^{N}/N!$.  Therefore, 
$\lim_{t\to 0^+}F_N(t) =0$ which proves that $F_N$ is continuous at $0$.
On the other hand, for $t,h>0$
\begin{align}
& \norm{F_N(t+h)-F_N(t)}  \le \left(\sqrt{\vep}\norm{V}\right)^{N}
\Bigl[
 \int_{s\in\R_+^{N}}\! \rmd s \, 
\1\Bigl(\sum_{\ell=1}^N s_\ell \le t\Bigr) \norm{\rme^{-\ci h H_0}-\1}
\nonumber \\ & \qquad  +
 \int_{s\in\R_+^{N}}\! \rmd s \, 
\1\Bigl(t<\sum_{\ell=1}^N s_\ell \le t+h\Bigr) \Bigr].
\end{align}
The bound goes to $0$ when $h\to 0$ by dominated convergence, and we have
proven that $F_N$ is norm-continuous for all $N\in\N$.  

The integrands on the right hand side of 
equations (\ref{eq:FArel}) -- (\ref{eq:AArel}) are, therefore, continuous, and
each of the integrals is a vector valued integral in $\banach(\hilb)$.
Equation (\ref{eq:FGrel}) is obvious from the definitions, and if we can 
prove (\ref{eq:GArel}), then (\ref{eq:FArel}) follows from it (note that 
$A_{0,N_0}(r;\vep,\kappa)$ actually does not depend on $\kappa$).
To prove (\ref{eq:GArel}), apply an arbitrary functional 
$\Lambda\in\banach(\hilb)^*$ to the 
integral on the right hand side, and use the definition of $A$ to evaluate
$\Lambda[ \rme^{-\ci (t -r) H_0} A_{N',N_0}(r;\vep,\kappa)]$. 
Then Fubini's theorem allows
rearranging the integrals so that a change of variables $s_{N'+N_0+1}=t-r$
yields $\Lambda[G_{N',N_0}(t ;\vep,\kappa)]$.  
The proofs of equations (\ref{eq:AA0rel}) and (\ref{eq:AArel}) are very
similar and we skip them here. 
\end{proof}
\begin{theorem}\label{th:stoppedduh} 
Let $N_0\ge 1$, $N'_0\ge  0$, and $\kappa>0$ be given.   
Then for any $t>0$ and for any realization of $\xi$, 
we have as vector valued integrals in $\banach(\hilb)$
\begin{align}\label{eq:stoppedduh}
\rme^{-\ci t H_\vep } & = \sum_{N=0}^{N_0-1} F_{N}(t;\vep) +
\sum_{N'=0}^{N'_0-1} \kappa \int_0^{\infty}\!\!\rmd r\, \rme^{-\kappa (r-\ulr)}
\rme^{-\ci (t-\ulr) H_\vep } G_{N',N_0}(\ulr;\vep,\kappa)
\nonumber  \\ & \quad +
 \int_0^{t}\!\! \rmd r\,  \rme^{-\ci (t-r) H_\vep } 
  A_{N'_0,N_0}(r;\vep,\kappa)
\end{align}
where $\ulr = \min(t,r)$.
\end{theorem}
\begin{proof}
Let us suppress the dependence on $\vep$ from the
notation in this proof.  The first of the above integrals
is defined as $T\to\infty$ limit of
\begin{align}
& \int_0^{T}\!\rmd r\, \rme^{-\kappa (r-\ulr)}
\rme^{-\ci (t-\ulr) H} G_{N',N_0}(\ulr;\kappa)
\nonumber \\ & \quad  
= \int_0^{t}\!\rmd r\,
\rme^{-\ci (t-r) H} G_{N',N_0}(r;\kappa) +
 \int_t^{T}\!\rmd r\, \rme^{-\kappa (r-t)} G_{N',N_0}(t;\kappa),
\end{align}
which is well-defined as, by Lemma \ref{th:FGAcontin},
$G_{N',N_0}(r;\kappa)$ is continuous in $r$.  
By the same Lemma, also $A$
in the second integrand is continuous showing that the 
vector valued integral is well-defined.

If $N_0'=0$, Eq.~(\ref{eq:stoppedduh}) follows from (\ref{eq:baseduh})
by a straightforward induction in $N_0$ using (\ref{eq:FArel}) and
(\ref{eq:AA0rel}). Let us thus fix $N_0\ge 1$, and perform a second induction
in $N'_0\ge 0$.   Now for any $r'\ge 0$,
\begin{align}
  1 =  \kappa \int_{r'}^{\infty}\!\rmd r\, \rme^{-\kappa (r-r')},
\end{align}
which shows that
\begin{align}
& \int_0^{t}\!\! \rmd r'\,  \rme^{-\ci (t-r') H} 
  A_{N'_0,N_0}(r';\kappa) 
\nonumber \\ & \quad =  
\int_0^{t}\!\! \rmd r'\,
\kappa \int_{r'}^{\infty}\!\rmd r\, \rme^{-\kappa (r-\ulr)}
  \rme^{-\kappa (\ulr-r')}  \rme^{-\ci (t-\ulr) H}
 \rme^{-\ci (\ulr-r') H} A_{N'_0,N_0}(r';\kappa) 
\nonumber \\ & \quad =  
\kappa \int_{0}^{\infty}\!\rmd r\, \rme^{-\kappa (r-\ulr)}
 \rme^{-\ci (t-\ulr) H} 
\int_0^{\ulr}\!\! \rmd r'\,  \rme^{-\kappa (\ulr-r')}
 \rme^{-\ci (\ulr-r') H_0} A_{N'_0,N_0}(r';\kappa) 
\nonumber \\ & \qquad +
\int_0^{t}\!\! \rmd r'\,
\kappa \int_{r'}^{\infty}\!\rmd r\, 
\int_0^{\ulr-r'}\!\! \rmd s\,   \rme^{-\kappa (r-r')} 
 \rme^{-\ci (t-r'-s) H} W_s A_{N'_0,N_0}(r';\kappa) 
\end{align}
where we applied the Duhamel formula to the term
$\rme^{-\ci (\ulr-r') H}$, and all the manipulations
can the justified as before, by applying an arbitrary functional and 
then using Fubini's theorem.  By (\ref{eq:GArel}), 
the first term yields the new term to the sum over $N'$
in (\ref{eq:stoppedduh}).  In the second term we first change integration
variables from $s$ to $s'=s+r'$, and then use the identity
\begin{align}
 \1(r'\le s'\le \ulr) = \1(r'\le s') \1(s'\le t) \1(r\ge s')
\end{align}
and Fubini's theorem yielding the following form
for the second term:
\begin{align}
& \int_0^{t}\!\! \rmd s'\, \rme^{-\ci (t-s') H}
\int_0^{s'}\!\! \rmd r'\,
\kappa \int_{s'}^{\infty}\!\rmd r\,  \rme^{-\kappa (r-r')} 
 W_{s'-r'} A_{N'_0,N_0}(r';\kappa) 
\nonumber \\ & \quad =  
\int_0^{t}\!\! \rmd s'\, \rme^{-\ci (t-s') H}
\int_0^{s'}\!\! \rmd r'\, \rme^{-\kappa (s'-r')} 
 W_{s'-r'} A_{N'_0,N_0}(r';\kappa) 
\nonumber \\ & \quad =  
\int_0^{t}\!\! \rmd s'\, \rme^{-\ci (t-s') H}
A_{N'_0+1,N_0}(s';\kappa)
\end{align}
where we have used (\ref{eq:AArel}).  This completes the induction step
in $N_0'$. 
\end{proof}

Now we are ready to define how we the splitting is done.
\begin{definition}\label{th:defkappaetc}
Let $\gamma$ be a constant for which the dispersion relation $\omega$
satisfies the crossing assumption (IC\ref{it:crossing}), 
and let
\begin{align}\label{eq:defgammap}
\gamma' = \min\Bigl(\frac{1}{2},\gamma\Bigr),\quad
a_0  = \frac{\gamma'}{40}\quad\text{and}\quad
b_0 = 40 \Bigl( 1 + \frac{2}{\gamma'}\Bigr).
\end{align}
For any $\vep$ let us then define
\begin{align}\label{eq:chooseN0}
  N_0(\vep) 
  = \max\Bigl(1,\Bigl\lfloor\,\frac{a_0 \, |\ln \vep|}{\ln \sabs{\ln \vep}}\,
  \Bigr\rfloor\Bigr),
\quad N'_0(\vep) = 8 N_0(\vep) \quad\text{and}\quad \kappa(\vep) = 
\vep \sabs{\ln \vep}^{b_0} ,
\end{align}
where $\lfloor x\rfloor$ denotes the integer part of $x\ge 0$,
and let
\begin{align}
\psimain(t;\vep) =  \sum_{N=0}^{N_0(\vep)-1} F_{N}(t;\vep)\psi^\vep
 \quad\text{and}\quad
\psierr(t;\vep) = 
\rme^{-\ci t H_\vep }\psi^\vep - \psimain(t;\vep).
\end{align}
\end{definition}
For this choice of parameters, in the limit  $\vep\to 0$
we have $N_0\to \infty$, $\kappa\to 0$, and
\begin{align}\label{eq:N0limits}
  c^N  N! \sabs{\ln \vep}^{N+d} \vep^{\gamma'} \to 0, 
\quad\text{and}\quad
 \vep^{-2}  \left(\frac{\vep}{\kappa}\right)^{N_0}
  c^N  N! \sabs{\ln \vep}^{N+d} \to 0,
\quad
\end{align} 
where $N=r N_0(\vep)$, with $0\le r < 20$ and $c,d\ge 0$ being
arbitrary constants.
\begin{definition}
For $p\in\R^3$ and $f\in C(\T^3,\M_2)$ let
$B_{p,f}$ denote the operator defined for all $\phi,\psi\in\hilb$ by
\begin{align}
\braket{\phi}{B_{p,f}\psi} = 
  \int_{\T^3}\! \!\rmd k\, 
  \FT{\phi}(k-p/2)\cdot
  f(k) \FT{\psi}(k+p/2).
\end{align}
\end{definition}
Clearly, then
\begin{align}\label{eq:Bpfnorm}
  \left| \braket{\phi}{B_{q,f}\psi} \right| \le 
  \norm{\phi}\,  \norm{\psi}\, \norm{f}_{\infty}
\end{align}
thus $B_{q,f}\in\banach(\hilb)$ and $\norm{B_{q,f}}\le  \norm{f}_{\infty}$.
Let us also point out that $B_{0,\1}=\1$, and 
\begin{align}\label{eq:FvepFourier3}
&  F^\vep_{\tmacro}(\pmacro,\nmacro) =
  \E_{\nu_{\tmacro}^\vep}\!\!\left[ \braket{\psi}{
      B_{\vep\pmacro,e_{\nmacro}\! P_{++}}\psi}\right]
  = \E\!\left[ \braket{\rme^{-\ci \tmacro H_\vep/\vep }\psi^\vep}{
      B_{\vep\pmacro,e_{\nmacro}\! P_{++}}
      \rme^{-\ci \tmacro H_\vep/\vep }\psi^\vep}\right]
\end{align}
where $e_n(k)=\rme^{\ci 2 \pi n\cdot k}$ and $P_{++}$ denotes the
projection onto the $+$-subspace.

Suppose that
\begin{align}\label{eq:errtermbound}
& \lim_{\vep\to 0}\E\!\left[\norm{\psierr(\tmacro/\vep;\vep)}^2\right] = 0.
\end{align}
Then we only need to consider the terms coming from $\psimain$,
i.e., to inspect the limit of
\begin{align}\label{eq:FvepFourier4}
&  \Fmain(\pmacro,\nmacro,\tmacro) =
  \E\!\left[ \braket{\psimain(\tmacro/\vep;\vep)}{
      B_{\vep\pmacro,e_{\nmacro}\! P_{++}}\psimain(\tmacro/\vep;\vep)}\right].
\end{align}
To see this, first note that by (\ref{eq:FvepFourier3}) and 
$\norm{B_{\vep\pmacro,e_{\nmacro}\! P_{++}}}\le 1$,
\begin{align} \label{eq:deltaFbound} 
& | F^\vep_{\tmacro}(\pmacro,\nmacro)-\Fmain(\pmacro,\nmacro,\tmacro)|
\nonumber \\ & \quad 
\le 2 \E[\norm{\psierr(\tmacro/\vep;\vep)}^2]^{\frac{1}{2}}
  \E[\norm{\psimain(\tmacro/\vep;\vep)}^2]^{\frac{1}{2}}
 + \E[\norm{\psierr(\tmacro/\vep;\vep)}^2] .
\end{align}
On the other hand, by unitarity and the assumption (IC\ref{it:I1}), then
\begin{align}
  \sup_{\vep}\E[ \norm{\psimain(\tmacro/\vep;\vep)}^2] \le  
  2 (\sup_\vep \norm{\psi^\vep}^2 + \sup_\vep
 \E[ \norm{\psierr(\tmacro/\vep;\vep)}^2]) <\infty,
\end{align}
and the bound in (\ref{eq:deltaFbound}) goes to zero as
$\vep \to 0$. 

To prove (\ref{eq:errtermbound}), we apply Theorem \ref{th:stoppedduh}
with $\kappa=\kappa(\vep)$ and $N'_0=N'_0(\vep)$.  By the Schwarz
inequality, then
\begin{align}
& \E\!\left[\norm{\psierr(t;\vep)}^2\right] \le
2\Bigl( N'_0 \sum_{N'=0}^{N'_0-1} \kappa^2 \,
 \E\Bigl[\Bigl(\int_0^{\infty}\!\!\rmd r\, \rme^{-\kappa (r-\ulr)}
 \norm{G_{N',N_0}(\ulr;\vep,\kappa)\psi^\vep}\Bigr)^2\Bigr]
\nonumber  \\ & \quad 
 +  \E\Bigl[ \Bigl(
 \int_0^{t}\!\! \rmd r\,
 \norm{A_{N'_0,N_0}(r;\vep,\kappa)\psi^\vep}\Bigr)^2 \Bigr]  \Bigr) .
\end{align}
Now we can use Schwarz again in the form
\begin{align}
& \Bigl(\int_0^{\infty}\!\!\rmd r\, \rme^{-\kappa (r-\ulr)}
 \norm{G_{N',N_0}(\ulr;\vep,\kappa)\psi^\vep}\Bigr)^2
\nonumber \\ &\quad 
 \le 
\int_0^{\infty}\!\!\rmd r'\, \rme^{-\kappa (r'-\min(t,r'))}
\int_0^{\infty}\!\!\rmd r\, \rme^{-\kappa (r-\ulr)}
 \norm{G_{N',N_0}(\ulr;\vep,\kappa)\psi^\vep}^2,
\end{align}
and similarly for the term containing $A$, and we then obtain the bound
\begin{align}\label{eq:errbnd2}
& \E\!\left[\norm{\psierr(\tmacro/\vep;\vep)}^2\right] \le
 2 \tmacro^2 \vep^{-2}
\sup_{0\le r\le \tmacro/\vep} \E\!\left[
\norm{A_{N'_0,N_0}(r;\vep,\kappa)\psi^\vep}^2\right]
 \nonumber \\ &  \qquad +
2 (\tmacro \kappa/\vep + 1)^2 (N'_0)^2
\sup_{\substack{0\le N'\le N_0'-1\\ 0\le r\le \tmacro/\vep}}
\E\!\left[\norm{G_{N'\!,N_0}(r;\vep,\kappa)\psi^\vep}^2\right].
\end{align}

Let $\Emax = \sup_\vep \norm{\psi^\vep}^2$ which is finite by (IC\ref{it:I1}).
In the following sections we shall prove that
\begin{proposition}\label{th:GNbound}
There are constants $c$ and $c'$ and $\vep_1$, which depend only on $\omega$
and $\bar{\xi}$, such that, if $0<\vep\le \vep_1$ 
and $\tmacro>0$, then for all $0\le t\le \tmacro/\vep$
and $0\le N'<N'_0$
\begin{align}\label{eq:GNbound}
& \E\!\left[\norm{G_{N'\!,N_0}(t;\vep,\kappa)\psi^\vep}^2\right]
 \le  c'  \Emax
  (c T)^{\!\frac{\bar{N}}{2}}
\nonumber \\ & \qquad \times
\Bigl[
  \bar{N}!
  \Bigsabs{\ln \frac{T}{\vep}}^{\bar{N}+\max(2,d_2)}
  \bar{N}^{\max(1,d_1)} 
  \frac{T}{\sabs{\bar{t}/\vep}^{\gamma'}} 
+ \frac{1}{\lfloor N_0/2\rfloor!}\Bigr].
\end{align}
where $N_0$, $N'_0$ and $\kappa$ are as in Definition 
\ref{th:defkappaetc}, and we have denoted
$T=\sabs{\tmacro}$ and $\bar{N}=2(N_0+N'_0)= 18 N_0$.
$d_1,d_2$ are the constants in (DR\ref{it:suffdisp}) and 
(DR\ref{it:crossing}).
\end{proposition}
\begin{proposition}\label{th:ANbound} 
There are constants $c$ and $c'$ and $\vep_1$, which depend only on $\omega$
and $\bar{\xi}$, such that, if $0<\vep\le \vep_1$ 
and $\tmacro>0$, then for all $0\le t\le \tmacro/\vep$
\begin{align}\label{eq:ANbound}
\E\bigl[
\norm{A_{N'_0,N_0}(t;\vep,\kappa)\psi^\vep}^2\bigr]
& \le  c' \Emax \, (c T)^{\frac{\bar{N}}{2}}  
\bar{N}!\, \Bigsabs{\ln \frac{T}{\vep}}^{\! \bar{N}} 
 \Bigl[ \vep^3 + \Bigl(\frac{\vep}{\kappa}\Bigr)^{N_0} \Bigr]
\end{align}
where $N_0$, $N'_0$ and $\kappa$ are as in Definition 
\ref{th:defkappaetc},
$T=\sabs{\tmacro}$ and $\bar{N}=18 N_0$.
\end{proposition}
Using these bounds in (\ref{eq:errbnd2})
and then applying (\ref{eq:N0limits}) shows that indeed
then (\ref{eq:errtermbound}) holds: for the term containing 
$\lfloor N_0/2\rfloor!$ this can be seen using, for instance,
the property that
$|\ln \vep|\le (N_0+1)^2$ for all sufficiently small $\vep$.
Therefore, to complete the proof of Theorem \ref{th:mainlim},
we only need to prove 
the Propositions \ref{th:GNbound} and \ref{th:ANbound} 
and that 
\begin{align}\label{eq:Fveptomut2}
& \lim_{\vep\to 0}  \Fmain(\pmacro,\nmacro,\tmacro)
 =  \int_{\R^d\times \T^d}\!\! \!\!\mu_{\tmacro}(\rmd x \,\rmd k)\, 
\rme^{-\ci 2 \pi (\pmacro\cdot x- \nmacro\cdot k)}.
\end{align}

\subsection{Graph representation}\label{sec:graphrep}

To prove the remaining Propositions, we use a representation of the
expectation values as a sum over a finite number of
graphs each contributing a term
whose magnitude can be estimated.  We first present two Lemmas, the first of
which is used compute the expectation values, and the second is a standard
tool in time-dependent perturbation theory for manipulation of oscillatory
integrals. 
\begin{lemma}[Representation of expectation values]\label{th:mainrep}
Let $N',N\ge 0$ and $\vep>0$ be given, and let $s\in\R^{N+1}$ and
$s'\in\R^{N'+1}$.  Let also $\psi\in\hilb$ be some non-random vector.
Then for all $p\in\R^3$ and $f\in C^{\infty}(\T^3,\M_2)$,
\begin{align}\label{eq:psiampL5}
& \E\left[\braket{\rme^{-\ci s'_{N'+1} H_0} W_{s'_{N'}} \cdots
 W_{s'_{1}} \psi}{ B_{p,f} \rme^{-\ci s_{N+1} H_0} W_{s_{N}} \cdots
 W_{s_{1}} \psi}\right]
 \nonumber \\ & 
 =  (-\ci)^{N-N'} \vep^{\frac{N'+N}{2}}
 \!\!\!\!\!\!\sum_{S\in\pi(I_{N,N'})}
  \prod_{A\in S} C_{|A|}
 \!\!\!\!\!\!\sum_{\sigma \in \set{\pm 1}^{N+N'+2}}
  \int_{\T^3} \rmd \eta_0
  \,\FT{\psi}_{\sigma_1}\!(\eta_0)\, 
  \FT{\psi}_{\sigma'_{1}}\!(\eta_0-p)^*
\nonumber \\ & \quad \times
  \int_{\T^{3(N+N'+1)}}\! \rmd\eta\,
  \delta(\eta_{N+1}+p)
  \prod_{A\in S} \delta\Bigl(\sum_{\ell\in A} \eta_\ell\Bigr)
f_{\sigma_{N+2},\sigma_{N+1}}\!\Bigl(k_{N+1}-\frac{1}{2} p\Bigr)
 \nonumber \\ & \quad \times
\prod_{\ell=1}^{N}
 v_{\sigma_{\ell+1}\sigma_\ell}(k_{\ell+1},k_\ell)
\prod_{\ell=1}^{N'}
 v_{\sigma'_{\ell+1}\sigma'_\ell}(k'_{\ell+1},k'_\ell)
\prod_{\ell=1}^{N+1}
\rme^{-\ci s_\ell \sigma_\ell \omega(k_\ell)}
\prod_{\ell=1}^{N'+1}
\rme^{\ci s'_\ell \sigma'_\ell \omega(k'_\ell)}
\end{align}
where $I_{N,N'}=\set{1,\ldots,N}\cup\set{N+2,\cdots,N+1+N'}$, 
and $\pi(I)$ denotes 
the set of all partitions of the finite set $I$.  In addition,
$k_\ell$ and and $k'_\ell$ are functions of $\eta$: for all
$\ell=1,\ldots,N+N'+2$, we define
\begin{align}\label{eq:defkell}
 k_\ell(\eta) = \sum_{n=0}^{\ell-1} \eta_n
\end{align}
and for all $\ell=1,\ldots,N'+1$, we let
$k'_\ell(\eta)=k_{N+N'+3-\ell}(\eta)$
and $\sigma'_\ell = \sigma_{N+N'+3-\ell}$.
\end{lemma}
$\pi(I)$ is defined explicitly in Appendix \ref{sec:appComb}, in Definition
\ref{th:defPiI}.
The delta-functions here are a convenient notation for denoting
restrictions of the integration into subspaces.  Like the earlier
time-integration delta-functions, they can be resolved by integrating
formally out one of the variables: for each $A\in S$ 
we choose
$n\in A$, remove the integral over $\eta_n$ and set
$\eta_n = -\sum_{n'\in A:n'\ne n} \eta_{n'}$.  In particular,
always $k_1=\eta_0$ and $k'_1=\eta_0-p$.  
\begin{proof}
Both sides of the equality (\ref{eq:psiampL5}) are continuous in $\psi$.
Therefore, it is enough to prove the Lemma for $\psi$ which
have a compact support.  Assume such a vector $\psi$.  
Using (\ref{eq:defVL}) -- (\ref{eq:vdef}), we define for any $R>0$
\begin{align}
 W^R_{s} & = (- \ci \sqrt{\vep} \VR ) \rme^{-\ci s H_0} 
 \qand
 \Wy_{s} = (- \ci \sqrt{\vep} \Vy ) \rme^{-\ci s H_0} .
\end{align}
As already mentioned in Sec.~\ref{sec:proofmain}, any finite product of 
$\VR$:s and arbitrary $R$-inde\-pendent
bounded operators converge strongly when $R\to\infty$ to the expression with 
$\VR$ replaced by $V$.  In addition, since $\norm{\VR}\le \norm{V}\le
2 \ommax  \bar{\xi}$, 
we can now apply dominated convergence
to prove that 
\begin{align}\label{eq:Rcutoff}
& \E\!\left[\braket{\rme^{-\ci s'_{N'+1} H_0} W_{s'_{N'}} \cdots
 W_{s'_{1}} \psi}{ B_{p,f} \rme^{-\ci s_{N+1} H_0} W_{s_{N}} \cdots
 W_{s_{1}} \psi}\right]
 \nonumber \\ & \quad 
 =  \lim_{R\to \infty} 
\E\!\left[\braket{\rme^{-\ci s'_{N'+1} H_0} W^R_{s'_{N'}} \cdots
 W^R_{s'_{1}} \psi}{ B_{p,f} \rme^{-\ci s_{N+1} H_0} W^R_{s_{N}} \cdots
 W^R_{s_{1}} \psi}\right] .
\end{align}
For a fixed $R>0$, let us define
$\Lambda_R=\defset{y\in\Z^3}{\norm{y}_\infty\le R}$, and use
(\ref{eq:defVL}) to the term on the right yielding
\begin{align}
&   \sum_{y'\in \Lambda_R^{N'}\!\!,  y\in\Lambda_R^{N} }
  \E\Bigl[\prod_{\ell'=1}^{N'} \xi_{y'_{\ell'}}
  \prod_{\ell=1}^N \xi_{y_\ell}\Bigr]
 \\ & \qquad \times
\braket{\rme^{-\ci s'_{N'+1} H_0} W^{(y'_{N'})}_{s_{N'}} \cdots
 W^{(y'_{1})}_{s'_{1}}
 \psi}{ B_{p,f}
\rme^{-\ci s_{N+1} H_0} W^{(y_{N})}_{s_{N}} \cdots
 W^{(y_{1})}_{s_{1}} \psi} .  \nonumber
\end{align}
Then we can denote $y_{N+2+N'-\ell'}=y'_{\ell'}$,
define the new index set $I=I_{N,N'}$ and apply
the moments-to-cumulants formula, Lemma \ref{th:momtocum}, to find that
this is equal to
\begin{align}\label{eq:remscalp}
 &  \sum_{S\in\pi(I)} \sum_{x\in \Lambda_R^{S}}
\prod_{A\in S} C_{|A|} 
\sum_{y\in (\Lambda_R)^{I_{N,N'}} }
\prod_{A\in S} \prod_{\ell\in A} \delta_{y_\ell,x_A}
 \\ & \quad \times
\braket{\rme^{-\ci s'_{N'+1} H_0} W^{(y_{N+2})}_{s_{N'}} \cdots
 W^{(y_{N+1+N'})}_{s'_{1}} \psi}{ B_{p,f}
\rme^{-\ci s_{N+1} H_0} W^{(y_{N})}_{s_{N}} \cdots
 W^{(y_{1})}_{s_{1}} \psi} .  \nonumber
\end{align}
Evaluation of the remaining scalar product in Fourier space yields
the following integral representation for it: 
\begin{align}
&  (-\ci)^{N-N'} \vep^{\frac{N'+N}{2}}
 \sum_{\sigma \in \set{\pm 1}^{N+N'+2}}
  \int_{\T^3} \!\rmd h \, f_{\sigma_{N+2},\sigma_{N+1}}\!(h)
  \int_{(\T^3)^{N}} \!\rmd k \int_{(\T^3)^{N'}} \!\rmd k'
\nonumber \\ & \quad \times
\FT{\psi}_{\sigma_1}\!(k_1)\, \FT{\psi}_{\sigma'_{1}}\!(k'_1)^*
  \prod_{\ell=1}^N \left[
    \rme^{-\ci 2 \pi y_\ell\cdot (k_{\ell+1}-k_\ell)}
    \right] 
  \prod_{\ell=1}^{N'} \left[
    \rme^{\ci 2 \pi y_{N+2+N'-\ell}\cdot (k'_{\ell+1}-k'_\ell)}
    \right] 
 \nonumber \\ & \quad \times
\prod_{\ell=1}^{N}
 v_{\sigma_{\ell+1}\sigma_\ell}(k_{\ell+1},k_\ell)
\prod_{\ell=1}^{N'}
 v_{\sigma'_{\ell+1}\sigma'_\ell}(k'_{\ell+1},k'_\ell)
\prod_{\ell=1}^{N+1}
\rme^{-\ci s_\ell \sigma_\ell \omega(k_\ell)}
\prod_{\ell=1}^{N'+1}
\rme^{\ci s'_\ell \sigma'_\ell \omega(k'_\ell)}
\end{align}
where we have defined $k_{N+1}= h + p/2$, and 
$k'_{N'+1}= h - p/2$, and $\sigma'_\ell = \sigma_{N+3+N'-\ell}$ for
all $\ell=1,\ldots,N'+1$.  We then change integration variables, 
first $h=  k_{N+1}-p/2$, and then from $k$ to
\begin{align}
  \eta_{\ell} = \begin{cases}
    k_1, & \text{ for }\ell=0,\\
    k_{\ell+1} - k_{\ell}, &\text{ for }\ell=1,\ldots,N,\\
    k'_{N+2+N'-\ell} - k'_{N+3+N'-\ell}, &\text{ for }
    \ell=N+2,\ldots,N+1+N',\\
  \end{cases}
\end{align}
and we also define $\eta_{N+1} = k'_{N'+1}-k_{N+1}= -p$.
The inverse of this transformation is given by (\ref{eq:defkell}), 
and thus the change of variables 
has a Jacobian equal to one.
In the new variables we have
\begin{align}
  \prod_{\ell=1}^N \left[
    \rme^{-\ci 2 \pi y_\ell\cdot (k_{\ell+1}-k_\ell)}
    \right] 
  \prod_{\ell=1}^{N'} \left[
    \rme^{\ci 2 \pi y_{N+2+N'-\ell}\cdot (k'_{\ell+1}-k'_\ell)}
    \right] 
    =   \prod_{n\in I}
    \rme^{-\ci 2 \pi y_n\cdot \eta_n}.
\end{align}
As for each $\ell\in I$ there is a unique $A\in S$ such that $\ell \in A$,
we can perform the sum over $y$ in (\ref{eq:remscalp}).  Thus 
(\ref{eq:remscalp}) is equal to 
\begin{align}
 & (-\ci)^{N-N'} \vep^{\frac{N'+N}{2}}
  \sum_{S\in\pi(I)} \prod_{A\in S} C_{|A|} 
 \sum_{\sigma \in \set{\pm 1}^{N+N'+2}} \sum_{x\in \Lambda_R^{S}} 
  \int_{\T^3} \rmd \eta_0 \int_{(\T^3)^I} \!\rmd \eta\,
  \FT{\psi}_{\sigma_1}\!(\eta_0)
\nonumber  \\ & \quad \times
  \FT{\psi}_{\sigma'_{1}}\!\Bigl(
  \eta_0-p+\sum_{n\in I}\eta_n\Bigr)^*
 f_{\sigma_{N+2},\sigma_{N+1}}\!(k_{N+1}-p/2)
  \prod_{A\in S}  \rme^{-\ci 2 \pi x_A\cdot \sum_{n\in A}\eta_n}
 \nonumber \\ & \quad \times
\prod_{\ell=1}^{N}
 v_{\sigma_{\ell+1}\sigma_\ell}(k_{\ell+1},k_\ell)
\prod_{\ell=1}^{N'}
 v_{\sigma'_{\ell+1}\sigma'_\ell}(k'_{\ell+1},k'_\ell)
\prod_{\ell=1}^{N+1}
\rme^{-\ci s_\ell \sigma_\ell \omega(k_\ell)}
\prod_{\ell=1}^{N'+1}
\rme^{\ci s'_\ell \sigma'_\ell \omega(k'_\ell)} .
\end{align}
Then we can do one more change of variables by choosing for each $A$ a
representative $n_A\in A$, and changing the integration variable
$\eta_{n_A}$ to $q_A = \sum_{n\in A}\eta_n$ (with unit Jacobian).
Then we are left with sums of the form
\begin{align}
   \sum_{x\in \Lambda_R^{S}} \int_{(\T^3)^S}\! \rmd q\,
   \rme^{-\ci 2 \pi \sum_{A\in S} x_A\cdot q_A} F(q), 
\end{align}
where $F$ denotes the result from first integrating out all the remaining 
$\eta$-integrals.  Since $\psi$ has compact support, $\FT{\psi}$ is smooth,
and so is the rest of the $\eta$-integrand, by assumption (DR\ref{it:DC1}).
Therefore, by compactness of the integration region, 
$F$ is a smooth function of $q$, and thus
its Fourier transform is pointwise invertible, implying
\begin{align}
  \lim_{R\to\infty}
   \sum_{x\in \Lambda_R^{S}} \int_{(\T^3)^S}\! \rmd q\,
   \rme^{-\ci 2 \pi \sum_{A\in S} x_A\cdot q_A} F(q) = F(0).
\end{align}
Then it is a matter of inspection to check that indeed
\begin{align}
& \lim_{R\to\infty}
\E\left[\braket{\rme^{-\ci s'_{N'+1} H_0} W^R_{s'_{N'}} \cdots
 W^R_{s'_{1}} \psi}{ B_{p,f} \rme^{-\ci s_{N+1} H_0} W^R_{s_{N}} \cdots
 W^R_{s_{1}} \psi}\right]
\end{align}
is equal to the right hand side of (\ref{eq:psiampL5}).
\end{proof}
\begin{lemma}\label{th:Knprop}
For any $N\ge 1$, define $K_N:(0,\infty)\times \C^{N} \to \C$
by
\begin{gather} \label{eq:defKn}
K_N(t,w) =   \int_{\R_+^{N}}\! \rmd s \,
  \delta\Bigl(t-\sum_{\ell=1}^{N} s_\ell\Bigr) \prod_{\ell=1}^N
 \rme^{-\ci s_\ell w_\ell} .
\end{gather}
Then all of the following hold:
\newcounter{tempnumi}
\begin{enumerate} 
\item\label{it:Knrecur} $\displaystyle K_{N+1}(t,w) = \int_0^t \!\rmd r\,
\rme^{-\ci (t-r)w_{N+1} } K_{N}(r,w_1,\ldots,w_{N})$.
\item\label{it:Knbound} 
  $\displaystyle\left|K_N(t,w)\right| \le
  \frac{t^{N-1}}{(N-1)!} \rme^{R N t}$, where 
  $R=\max(0,\im w_1,\ldots,\im w_N)$.
\item\label{it:KncountourI} Let $D\subset \C$ be compact and let 
  $\Gamma$ be a closed path which
  goes once anticlockwise around $D$ without intersecting it.  Then for all
  $w\in D^{N}$ and $t>0$,
\begin{gather} \label{eq:Knrep1}
K_N(t,w) = - \oint_\Gamma \frac{\rmd z}{2\pi} \rme^{-\ci t z}
  \prod_{\ell=1}^{N} \frac{\ci}{z-w_\ell} .
\end{gather}
\end{enumerate}
\end{lemma} 
\begin{proof} 
Now $K_1(t,w) = \rme^{-\ci t w}$, for which the properties
\ref{it:Knbound} and \ref{it:KncountourI} hold
trivially. When $N\ge 2$, the definition of $K_N$ is explicitly
\begin{align}
K_N(t,w) = \int_{\R_+^{N-1}}\! \rmd s \, 
\1\!\Bigl(\sum_{\ell=1}^{N-1} s_\ell\le t\Bigr)
\rme^{-\ci \sum_{\ell=1}^{N-1} s_\ell w_\ell}
 \rme^{-\ci (t-\sum_{\ell=1}^{N-1} s_\ell) w_{N}},
\end{align}
from which \ref{it:Knrecur} can be proven by induction. 
The property in \ref{it:Knbound} 
follows then by induction from \ref{it:Knrecur}. 
So does also \ref{it:KncountourI}, after one notices that if $N\ge 2$ and
$t=0$, the right 
hand side of (\ref{eq:Knrep1}) is equal to zero since Cauchy's theorem
allows taking the path $\Gamma$ to infinity.  
\end{proof}

\begin{figure}
  \begin{center}
    \myfigure{width=0.9\textwidth}{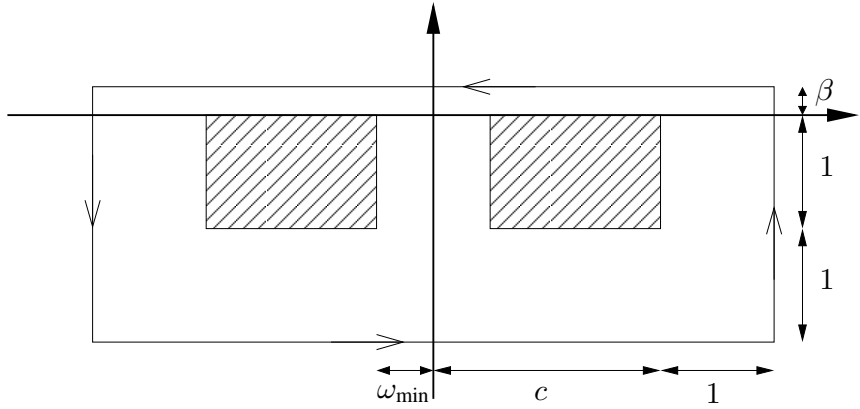} 
    \caption{The integration path $\gpath(c)$. If $c=\ommax$,
      the shaded area contains all values of the type
       $\pm\omega(k)-\ci \kappa$ for all $0\le \kappa\le 1$.}
    \label{fig:intpath}
  \end{center}
\end{figure}
To apply the above Lemma we will choose the integration path $\Gamma$ as
follows: 
For any $c>0$ and $0<\beta\le 1$, 
let $\gpath(c)$ denote the integration contour which follows the path
shown in Fig.~\ref{fig:intpath}.  Let $\gpath=\gpath(\ommax)$,
and we will choose $\Gamma=\gpath$ for some $\beta$.
By construction, (\ref{eq:Knrep1}) then
holds for all $w_\ell$ of the form
$\pm\omega(k_\ell)-\ci \kappa_\ell$ with $k_\ell\in \T^3$ and 
$0\le\kappa_\ell\le 1$.

\begin{lemma}\label{th:EGG} 
Let $t>0$, $\kappa\ge 0$, and $N',N_1,N_2\in\N$ be given. Then for all 
$p\in\R^3$, and $f\in C^{\infty}(\T^3,\M_2)$,
\begin{align}\label{eq:EGG}
  & \E\!\left[\braket{G_{N',N_2}(t;\vep,\kappa)\psi^\vep}{
      B_{p,f} G_{N',N_1}(t;\vep,\kappa)\psi^\vep}\right] 
 \nonumber \\ & \quad
 =  \sum_{S\in\pi(I)}  \prod_{A\in S} C_{|A|} \, 
 \Kpart(t,S;(N',N_2),(N',N_1),\vep,\kappa,p,f)
\end{align}
where $I=I_{N'+N_1,N'+N_2}$, and
for any partition $S\in \pi(I)$ we have defined 
\begin{align}\label{eq:defKpart}
& \Kpart (t,S;(N',N_2),(N',N_1),\vep,\kappa,p,f)
\nonumber \\ & \quad
= (-\ci)^{\bar{N}_1-\bar{N}_2} \vep^{\bar{N}/2}
 \int_{\T^3} \rmd \eta_0
 \oint_\gpath \frac{\rmd z}{2\pi} 
 \oint_\gpath \frac{\rmd z'}{2\pi} \,
 \rme^{-\ci t (z+z')}
\nonumber \\ & \qquad \times
  \int_{(\T^{3})^{\bar{N}+1}}\! \rmd\eta\,
  \delta(\eta_{\bar{N}_1+1}+p)
  \prod_{A\in S} \delta\Bigl(\sum_{\ell\in A} \eta_\ell\Bigr)
 \, \FT{\psi}^\vep(\eta_0-p) \cdot
\nonumber \\ & \quad
\Bigl[
  \prod_{\ell=\bar{N}_1+3+N'}^{\bar{N}+2} \Bigl(
 \frac{\ci}{z'+\ho(k_\ell)}  v(k_{\ell},k_{\ell-1}) \Bigr)
  \prod_{\ell=\bar{N}_1+3}^{\bar{N}_1+2+N'} \Bigl(
 \frac{\ci}{z'+\ci \kappa+\ho(k_\ell)} 
 v(k_{\ell},k_{\ell-1}) \Bigr)
 \nonumber \\ & \quad \times
 \frac{\ci}{z'+\ci \kappa+\ho(k_{\bar{N}_1+2})} 
f\Bigl(k_{\bar{N}_1+1}-\frac{1}{2} p\Bigr)
 \frac{\ci}{z+\ci \kappa-\ho(k_{\bar{N}_1+1})} 
 \nonumber \\ & \quad \times
  \prod_{\ell=N_1+1}^{N_1+N'} \Bigl(
 v(k_{\ell+1},k_{\ell})
 \frac{\ci}{z+\ci \kappa-\ho(k_\ell)} 
 \Bigr)
  \prod_{\ell=1}^{N_1} \Bigl(
 v(k_{\ell+1},k_{\ell})
 \frac{\ci}{z-\ho(k_\ell)} 
 \Bigr) \FT{\psi}^\vep(\eta_0) \Bigr]
\end{align}
with $\bar{N}_i=N_i+N'$ for $i=1,2$, and $\bar{N}=\bar{N}_1+\bar{N}_2$.
$\ho$ and $v$ are matrix-valued functions, $v$ is defined by 
(\ref{eq:vdef}) and
$\ho(k)_{\sigma'\sigma} = \delta_{\sigma'\sigma} \sigma \omega(k)$.
$0<\beta\le 1$ is arbitrary, and
$\gpath$ denotes the corresponding integration path.
\end{lemma}
\begin{proof}
First we use the definition of the two $G$-operators, 
(\ref{eq:defGN}), to express them as integrals over time-variables which
we denote by $s$ and $s'$.  Since these are vector valued integrals,
the scalar product can be taken inside the time-integrations.
Then Fubini's theorem allows swapping the order of the time-integrations and
the expectation value. For this we need to have measurability
with respect to the product measure, which
can be proven by showing, as in (\ref{eq:Rcutoff}), that
the integrand is a limit of a sequence of measurable functions.
We use Lemma \ref{th:mainrep} to express the remaining expectation
value as an integral over the $\eta$-variables, and apply
Fubini's theorem to exchange the order of the $s$- and $s'$-integrations 
and the $\eta$-integration.  By Lemma \ref{th:Knprop}:\ref{it:KncountourI}
we can express the $s$ and $s'$ -integrals as integrals over $z$ and $z'$,
and then summing over the $\sigma$-variables we arrive at the integrand in
(\ref{eq:defKpart}).  The only remaining step is to
reorder the $\eta$- and $z$-, $z'$-integrals as given in
(\ref{eq:defKpart}) which is allowed by
Fubini's theorem.
\end{proof}
\begin{corollary}\label{th:EFF} 
\begin{align}\label{eq:EFF}
 & \Fmain(\pmacro,\nmacro,\tmacro) 
 \\  \nonumber & \quad
 =  \sum_{N_1,N_2=0}^{N_0(\vep)-1}
  \sum_{S\in\pi(I_{N_1,N_2})}  \prod_{A\in S} C_{|A|} \, 
 \Kpart(\tmacro/\vep,S;(0,N_2),(0,N_1),\vep,0,\vep \pmacro,e_{\nmacro}P_{++}).
\end{align}
\end{corollary}
\begin{proof} By definition (\ref{eq:FvepFourier4}),
\begin{align}
  \Fmain(\pmacro,\nmacro,\tmacro) =
  \sum_{N_1,N_2=0}^{N_0(\vep)-1}
  & \E\!\left[\braket{F_{N_2}(\tmacro/\vep;\vep)\psi^\vep}{
      B_{\vep p,e_{\nmacro}P_{++}} F_{N_1}(\tmacro/\vep;\vep)\psi^\vep}\right] 
\end{align}
which yields (\ref{eq:EFF}) by using first (\ref{eq:FGrel}) and then
Lemma \ref{th:EGG}.
\end{proof}
\begin{lemma}\label{th:EAA}
Let $t,\kappa>0$, and $N'_0,N_0\in\N$, with $N_0\ge 1$
be given. Then 
\begin{align}\label{eq:EAA}
  & \E\!\left[\norm{A_{N'_0,N_0}(t;\vep,\kappa)\psi^\vep}^2\right] 
 =  \sum_{S\in\pi(I)} \prod_{A\in S} C_{|A|} \,
  \Kpamp (t,S;N'_0,N_0,\vep,\kappa)
\end{align}
where $I=I_{N,N}$ with $N=N'_0+N_0$, and
for any partition $S\in \pi(I)$,
\begin{align}\label{eq:defKpamp}
&   \Kpamp (t,S;N'_0,N_0,\vep,\kappa)
= \vep^{N} \int_{\T^3} \rmd \eta_0
 \oint_\gpath \frac{\rmd z}{2\pi} 
 \oint_\gpath \frac{\rmd z'}{2\pi} 
 \rme^{-\ci t (z+z')}
\nonumber \\ & \quad \times
  \int_{(\T^{3})^{2 N+1}}\! \rmd\eta\,
  \delta(\eta_{N+1})
  \prod_{A\in S} \delta\Bigl(\sum_{\ell\in A} \eta_\ell\Bigr)
 \, \FT{\psi}^\vep(\eta_0) \cdot
\nonumber \\ & \quad
\Bigl[
  \prod_{\ell=N+3+N'_0}^{2 N+2} \Bigl(
 \frac{\ci}{z'+\ho(k_\ell)}  v(k_{\ell},k_{\ell-1}) \Bigr)
  \prod_{\ell=N+3}^{N+2+N'_0} \Bigl(
 \frac{\ci}{z'+\ci \kappa+\ho(k_\ell)} 
 v(k_{\ell},k_{\ell-1}) \Bigr)
 \nonumber \\ & \quad \times
  \prod_{\ell=N_0+1}^{N_0+N'_0} \Bigl(
 v(k_{\ell+1},k_{\ell})
 \frac{\ci}{z+\ci \kappa-\ho(k_\ell)} 
 \Bigr)
  \prod_{\ell=1}^{N_0} \Bigl(
 v(k_{\ell+1},k_{\ell})
 \frac{\ci}{z-\ho(k_\ell)} 
 \Bigr) \FT{\psi}^\vep(\eta_0) \Bigr]
\end{align}
where the matrix-valued functions $\ho$ and $v$, and the path
$\gpath$ are defined as in Lemma \ref{th:EGG} and $0<\beta\le 1$ is arbitrary.
\end{lemma}
\begin{proof} 
By following the same steps as in the proof of Lemma \ref{th:EGG}. 
\end{proof}

\begin{figure}
  \begin{center}
    {
      \myfigure{width=0.9\textwidth}{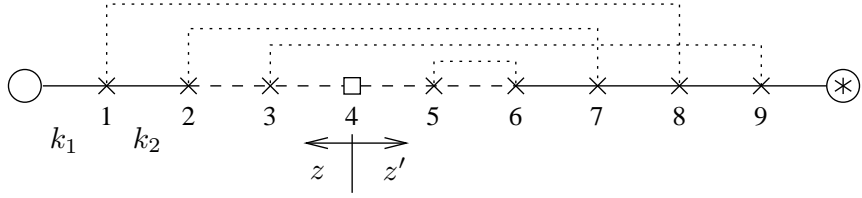} }
    \caption{An example of a graph for $\Kpart$ 
      corresponding to $N_1=1$, $N_2=3$, $N'=2$, and to a partition
      $S$ with $|S|=4$.  We have also indicated how we chose to label
      the interaction
      vertices, and a few momenta belonging to the propagator lines.  
      See the text for a description of the precise meaning
      of the different components of the graph.}
    \label{fig:graphex}
  \end{center}
\end{figure}
$\Kpart$ is called the {\em amplitude\/} of the partition, or graph, $S$.  
It will be helpful to think of the amplitudes in terms of planar graphs,
where the structure of the graph encodes the inter-dependence of the momenta
$k_\ell$, as imposed by the product of delta-functions 
$\prod_{A\in S} \delta(\sum_{\ell\in A} \eta_\ell)$. The graph
is constructed by starting from the left with a circle, denoting the rightmost
$\FT{\psi}^\vep$, and then representing the different factors in the matrix
product (\ref{eq:defKpart}), in the order they are acting, so that a solid line
represents a term $\ci/(z-H)$, a cross a term $v$,
a dashed line a term $\ci/(z+\ci\kappa-H)$, until we reach the
{\em observable\/} $f$, 
which will be denoted by a square.  After this the same
procedure is repeated, except now a solid line denotes $\ci/(z'+H)$
and a dashed line, $\ci/(z'+\ci\kappa+H)$.  The line terminates at a
circled asterisk which corresponds to $(\FT{\psi}^\vep)^*$.  
Each of the fractionals
is called a {\em propagator,\/} and a cross is called an 
{\em interaction vertex.\/}  Finally, all interaction vertices
belonging to the same cluster in the partition $S$ are joined by a dotted
line. 
Fig.\ \ref{fig:graphex} gives an illustration of such a graph.

The graph for $\Kpamp$ is constructed similarly.  As
$\Kpamp$ is missing the propagators attached to the
observable, it is called the amplitude of an {\em amputated\/} graph.
We divide the graphs into the following categories:
\begin{definition}
Let $N',N\ge 0$ be given, and let
$S\in\pi(I_{N,N'})$.  We call the partition $S$ {\em irrelevant\/} if it
contains a singlet, i.e., if there is $A\in S$ such that $|A|=1$.
Otherwise, the partition $S$ is called {\em relevant,\/} and then it is
\begin{description}
  \setlength{\itemsep}{0pt} 
\item[higher order,] if there is $A\in S$ such that $|A|>2$.
\item[crossing,] if it is a pairing which contains two pairs
 crossing each other, i.e., there are
$\set{i_1,i_2}$, $\set{j_1,j_2}\in S$ such that $i_1<j_1<i_2<j_2$.
\item[nested,] if it is neither of the above, but there is a pairing
$\set{i_1,i_2}\in S$ which is completely on one side of the observable, but 
which is not a nearest neighbour pairing, i.e.,
such that $i_1+1<i_2$ and either $i_1\ge N+2$ or $i_2\le N$.
\item[simple,] otherwise.
\end{description}
\end{definition}

It will turn out that only the simple partitions 
related to the main term contribute to the kinetic scaling
limit.  The proof that all other partitions can be neglected will rely on
the following Lemmas whose proofs are the most involved part of the
analysis and will be given in Sections \ref{sec:pflemmas} and \ref{sec:simple}.
As before, we let here $\Emax = \sup_\vep \norm{\psi^\vep}^2$.
\begin{lemma}[Basic A-estimate]\label{th:basicAest}
There are constants $c$ and $c'$, which depend only on $\omega$,
such that for any 
$t>0$, $\kappa, \vep\in (0,\frac{1}{2}]$, $N_0,N_0'\in\N$ with $N_0\ge 1$,
and a relevant $S\in\pi(I_{N,N})$,
\begin{align}\label{eq:Abound}
& \left| \Kpamp (t,S;N'_0,N_0,\vep,\kappa) \right| \le 
 c' \Emax \, \vep^{N-|S|} \Bigl(\frac{\vep}{\kappa}\Bigr)^{n_S}
  \left(c \sabs{\vep t}\right)^{N}
  \Bigsabs{\ln \frac{\sabs{\vep t}}{\vep}}^{2N} 
\end{align}
where $N=N'_0+N_0$ and
\begin{align}\label{eq:defNS}
  n_S =  \left| \defset{\max A}{A\in S} \cap 
    \set{N_0,\ldots,N_0+1+2 N'_0}\right|. 
\end{align}
\end{lemma}
This estimate suffices to prove the bound for the amputated expectation value.
\begin{proofof}{Proposition \ref{th:ANbound}}
Let
$\kappa=\kappa(\vep)$ and $N'_0 = 8 N_0$ as in Definition \ref{th:defkappaetc},
and denote $N=N_0+N'_0 = 9 N_0$.  Let also
$a=2\bar{\xi}(3\bar{\xi}^2+1)$ as in Lemma \ref{th:highosum}, 
and assume that $\vep_1 \le \min(1/a^2,1/2)$ 
is chosen so that $\kappa(\vep)\le 1/2$ 
for all $\vep\le \vep_1$.  We then consider an arbitrary 
$\vep\le \vep_1$.  

We can then apply  Lemma \ref{th:basicAest}, together 
with $\sabs{\vep t}\le \sabs{\tmacro}=T$, arriving at
\begin{align}
& \left| \Kpamp (t,S;N'_0,N_0,\vep,\kappa) \right| \le 
 c' \Emax \, \vep^{N-|S|} \Bigl(\frac{\vep}{\kappa}\Bigr)^{n_S}
  (c T)^{N}
  \Bigsabs{\ln \frac{T}{\vep}}^{2N} .
\end{align}
We still need to estimate the sum over the partitions $S$.  
First we sum over all partitions containing a cluster of size 
at least $7$.  In this case, we estimate $\vep/\kappa\le 1$ and, 
using $\sqrt{\vep}\le 1/a$, apply Lemma \ref{th:highosum} which proves
\begin{align}
& \sum_{\substack{S\in\pi(I_{N,N}),\\ \exists A\in S : |A|>6}}
 \prod_{A\in S}  \left|C_{|A|} \right| \,
\left| \Kpamp (t,S;N'_0,N_0,\vep,\kappa) \right| 
\nonumber \\  & \qquad
\le  c' \Emax (c T)^{N}  \Bigsabs{\ln \frac{T}{\vep}}^{2N} 
 (2 N)! \, a^6 \vep^3 .
\end{align}
Let then $S$ be such that for all 
$A\in S$, $|A|\le 6$.  Then $|S|\ge (2 N)/6= 3 N_0$, and thus 
$n_S\ge |S| - 2 N_0\ge N_0$.  Therefore,  
$(\vep/\kappa)^{n_S}\le (\vep/\kappa)^{N_0}$.  To estimate the remaining sum
over the partitions, we can neglect the restriction on the size of 
the clusters. We use Lemma \ref{th:highosum} to bound the sum over 
higher-order partitions, and compute the estimate for 
pairings explicitly.  Since the number of possible pairings is 
$(2N)!/(2^N N!) \le (2N)!$ we get
\begin{align}
& \sum_{S\in\pi(I_{N,N})}
 \prod_{A\in S}  \left|C_{|A|} \right| \, \vep^{N-|S|}
\le  (2 N)! ( 1 + \vep a^2 ) \le  2 (2 N)!.
\end{align}
We have thus proven that 
\begin{align}
&  \sum_{S\in\pi(I_{N,N})}
 \prod_{A\in S}  \left|C_{|A|} \right| \,
\left| \Kpamp (t,S;N'_0,N_0,\vep,\kappa) \right| 
\nonumber \\  & \qquad
\le  c' \Emax (c T)^{N}  \Bigsabs{\ln \frac{T}{\vep}}^{2N} 
 (2 N)! \left( a^6 \vep^3 + 2 \Bigl(\frac{\vep}{\kappa}\Bigr)^{N_0} \right)
\end{align}
from which  (\ref{eq:ANbound}) follows by Lemma \ref{th:EAA}
after redefinition of the constant $c'$.  The bound is trivially
valid for $t=0$ since $\norm{A_{N'_0,N_0}(t)}=0$.
\end{proofof}

For the other estimates, we no longer need the additional decay provided by
the terms containing $\kappa$.  We do however need to make sure that its
presence does not spoil any of the estimates for $G$.
In the following Lemmas, whose proofs will be postponed until
Section \ref{sec:pflemmas},
$N',N_1,N_2\ge 0$ and $N_0\ge 1$,
and we use the notations $\bar{N}_1=N'+N_1$, $\bar{N}_2=N'+N_2$, 
$\bar{N}=\bar{N}_1+\bar{N}_2$.  Let also $I=I_{\bar{N}_1,\bar{N}_2}$.  
\begin{lemma}[Basic estimate]\label{th:basicest}
There are constants $c$ and $c'$, which depend only on $\omega$,
such that for any $t>0$, $\kappa, \vep\in (0,\frac{1}{2}]$,
and every relevant $S\in\pi(I)$,
\begin{align}\label{eq:basicbound}
& \left|  \Kpart(t,S;(N',N_2),(N',N_1),\vep,\kappa,p,f) \right| 
 \nonumber \\ & \quad
\le  c' \norm{f}_\infty \Emax
  \left(c\vep \sabs{t}\right)^{\!\frac{\bar{N}}{2}}
  \Bigsabs{\ln \frac{\sabs{\vep t}}{\vep}}^{\bar{N}+2}
  \vep^{\1(\bar{N}>0)(\bar{N}-2 |S|)/2} .
\end{align}
\end{lemma}
\begin{lemma}[Crossing partition]\label{th:crossingest}
Let the assumptions of Lemma \ref{th:basicest} be satisfied.  
There is a constant $c''$, depending only on $\omega$, such that if 
$S\in\pi(I)$ is crossing, then the
bound on the right hand side of (\ref{eq:basicbound}) is valid also if
it is multiplied by
\begin{align}\label{eq:crossingextra}
c'' \sabs{t}^{-\gamma} 
  \Bigsabs{\ln \frac{\sabs{\vep t}}{\vep}}^{\max(0,d_2-2)},
\end{align}
where $\gamma$ and $d_2$ are as in the assumption (DR\ref{it:crossing}).
\end{lemma}
\begin{lemma}[Nested partition]\label{th:nestedest}
Let the assumptions of Lemma \ref{th:basicest} be satisfied.  
There are constants $c''_1$, $c''_2$, depending only on $\omega$, such that if 
$S\in\pi(I)$ is nested, then the
bound on the right hand side of (\ref{eq:basicbound}) is valid also if
it is multiplied by
\begin{align}\label{eq:nestedextra}
 c''_1 (c''_2)^{\frac{\bar{N}}{2}} \bar{N}^{d_1} \!\sabs{t}^{-\frac{1}{2}}.
\end{align}
\end{lemma}
These immediately yield the following estimate for the contribution from
non-simple partitions:
\begin{corollary}\label{th:nonsimple}
There are constants $c$ and $c'$ and $\vep'$, which depend only on $\omega$
and $\bar{\xi}$, such that, if $0<\vep\le \vep'$, $0\le \kappa\le 1/2$,
$t>0$, and $\bar{N}>0$, then
\begin{align}\label{eq:nonsimplebound}
  &  \sum_{\substack{S\in\pi(I),\\ S \text{ not simple}}}
 \prod_{A\in S} |C_{|A|}| \, 
 |\Kpart(t,S;(N',N_2),(N',N_1),\vep,\kappa,p,f)| 
\nonumber \\ & \quad
\le  c' \norm{f}_\infty \Emax
  \left(c\vep \sabs{t}\right)^{\!\frac{\bar{N}}{2}}
  \bar{N}!
  \Bigsabs{\ln \frac{\sabs{\vep t}}{\vep}}^{\bar{N}+\max(2,d_2)}
  \bar{N}^{d_1} 
  \Bigl(\frac{\sabs{\vep t}}{\sabs{t}}\Bigr)^{\gamma'}
\end{align}
where $\gamma'$ is defined in (\ref{eq:defgammap}), and
$d_1,d_2$ are the constants in (DR\ref{it:suffdisp}) and 
(DR\ref{it:crossing}).
\end{corollary}
\begin{proof}
Let $a=2\bar{\xi}(3\bar{\xi}^2+1)$ as in Lemma \ref{th:highosum}, 
and let $\vep'=\min(1/a^2,1/2)$.  Then for any $0<\vep\le \vep'$,
we get from Lemmas \ref{th:basicest} and \ref{th:highosum},
\begin{align}
& \sum_{\substack{S\in\pi(I),\\ \exists A\in S : |A|>2}}
 \prod_{A\in S}  \left|C_{|A|} \right| \,
\left| \Kpart(t,S;(N',N_2),(N',N_1),\vep,\kappa,p,f) \right| 
\nonumber \\  & \qquad
\le  c' \norm{f}_\infty \Emax
  \left(c\vep\sabs{t}\right)^{\!\frac{\bar{N}}{2}}
  \Bigsabs{\ln \frac{\sabs{t}}{\vep}}^{\bar{N}+2}
    \bar{N}! \sabs{a}^2 \sqrt{\vep} .
\end{align}
There are at most $\bar{N}!$ pairings, which can be combined with
Lemmas \ref{th:crossingest} and \ref{th:nestedest} to prove
(\ref{eq:nonsimplebound}) after redefinition of constants. 
\end{proof}
\begin{figure}
  \begin{center}
    {
      \myfigure{width=0.9\textwidth}{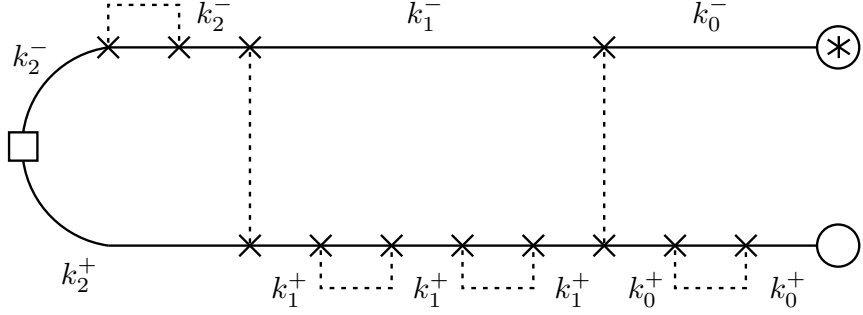} }
    \caption{An example of a simple graph $S_m(n,n')$ 
      with $\kappa=0$, $m=2$, and $n=(1,2,0)$,
      $n'=(0,0,1)$.  We have also depicted the dependence of the momenta 
      outside  the gates, see Sec.\ \ref{sec:gensimple} for details.}
    \label{fig:gensimple}
  \end{center}
\end{figure}
We need one more estimate before we can complete the proof of Proposition
\ref{th:GNbound}.
\begin{lemma}[Simple partition]\label{th:simpleest}
For any $m\in \N$ and $n,n'\in \N^{m+1}$,
let $S_m(n',n)$ denote the partition which consists a ladder of
$m$ ``rungs'' and where the components of $n$ and $n'$ define
the number of  ``gates''
between the rungs, see Fig.~\ref{fig:gensimple}.

$S\in\pi(I_{\bar{N}_1,\bar{N}_2})$ is simple if and only if 
there are $m$ and $n,n'$, with 
$m+2\sum_{j=1}^{m+1} n_j = \bar{N}_1$ and
$m+2\sum_{j=1}^{m+1} n'_j = \bar{N}_2$, such that $S=S_m(n,n')$.  In
addition, there are constants $c$, $c'$, and $c_i$, $i=0,1,2,3$, 
depending only on $\omega$,
such that for any $t>0$, $\kappa, \vep\in (0,\frac{1}{2}]$,
\begin{align}\label{eq:simplebound}
& \left|  \Kpart(t, S_m(n,n');(N',N_2),(N',N_1),\vep,\kappa,p,f) \right| 
 \nonumber \\ & \quad 
\le  \frac{ \Emax \norm{f}_\infty (c \vep t)^{\bar{N}/2}}{
 [((\bar{N}_1+m)/2)!((\bar{N}_2+m)/2)!]^{1/2}} 
 + c' \norm{f}_\infty    \Emax
(c_0\vep \sabs{t})^{\bar{N}/2} 
 \nonumber \\ & \quad
\times
\Bigsabs{\ln \frac{\sabs{\vep t}}{\vep}}^{3}
(\bar{N}+1) \Bigl(1 + c_1 \frac{|p|}{\sqrt{\vep}}
  + c_2  \frac{\kappa}{\vep} 
 + c_3 \sqrt{1+\kappa/\vep}\Bigr) 
 \frac{ \sabs{\vep t}}{\sabs{t}^{1/2}}.
\end{align}
\end{lemma}

\begin{proofof}{Proposition \ref{th:GNbound}}
Let $\kappa=\kappa(\vep)$ and $N'_0 = 8 N_0$ as in Definition 
\ref{th:defkappaetc}.  Let also
$a=2\bar{\xi}(3\bar{\xi}^2+1)$ as in Lemma \ref{th:highosum}, 
and assume that $\vep_1 \le \min(1/a^2,1/2)$ 
is chosen so that $\kappa(\vep)\le 1/2$ 
for all $\vep\le \vep_1$.  We then consider an arbitrary 
$0<\vep\le \vep_1$, $0\le N'<N'_0$ and $0<t\le \bar{t}/\vep$.

By Lemma \ref{th:EGG},
$\E\!\left[\norm{G_{N'\!,N_0}(t;\vep,\kappa)\psi^\vep}^2\right]$ is 
bounded by
\begin{align}
&
 \sum_{S\in\pi(I)}  \prod_{A\in S} \left|C_{|A|}\right| \, 
 | \Kpart(t,S;(N',N_0),(N',N_0),\vep,\kappa(\vep),0,\1)|
\end{align}
where $I=I_{N_1,N_1}$ with
$N_1=N'+N_0$. Let also $\bar{N}=2(N'+N_0)$ when $1<\bar{N} \le 18 N_0$.
$\vep$ is small enough for applying
Corollary \ref{th:nonsimple} and Lemma \ref{th:simpleest},
yielding an upper bound
\begin{align}\label{eq:GNbound2}
&
  c' \Emax
  \left(c\vep \sabs{t}\right)^{\!\frac{\bar{N}}{2}}
  \bar{N}!
  \Bigsabs{\ln \frac{\sabs{\vep t}}{\vep}}^{\bar{N}+\max(2,d_2)}
  \bar{N}^{d_1} 
  \Bigl(\frac{\sabs{\vep t}}{\sabs{t}}\Bigr)^{\gamma'}
\nonumber \\ & \quad
 + c'' \Emax
(c_0'\vep \sabs{t})^{\!\frac{\bar{N}}{2}} (\bar{N}+1)!
\Bigsabs{\ln \frac{\sabs{\vep t}}{\vep}}^{3} \sabs{\ln\vep}^{b_0}
 \frac{ \sabs{\vep t}}{\sabs{t}^{1/2}}
\nonumber \\ & \quad
 +   \Emax \sum_{S_m(n,n')\in\pi(I)} 
 \frac{ (c_0 \vep t)^{\bar{N}/2}}{((N_1+m)/2)!} 
\end{align}
where we have estimated the number of simple pairings by $\bar{N}!$ and used 
the explicit form of $\kappa(\vep)$.  Now, if $0<\vep_1\le \vep'$ 
is such that $N_0(\vep)\ge b_0+1$ for all $0<\vep\le \vep_1$, then for
these $\vep$, the sum of the first two terms is bounded by
\begin{align}
&   c''' \Emax 
  \left(c''_0\vep \sabs{t}\right)^{\!\frac{\bar{N}}{2}}
  \bar{N}!
  \Bigsabs{\ln \frac{\sabs{\vep t}}{\vep}}^{\bar{N}+\max(2,d_2)}
  \bar{N}^{\max(1,d_1)} 
  \frac{\sabs{\vep t}}{\sabs{t}^{\gamma'}} 
\nonumber \\ & \quad
 \le  c''' \Emax 
  \left(c''_0\vep \sabs{\bar{t}/\vep}\right)^{\!\frac{\bar{N}}{2}}
  \bar{N}!
  \Bigsabs{\ln \frac{\sabs{\bar{t}}}{\vep}}^{\bar{N}+\max(2,d_2)}
  \bar{N}^{\max(1,d_1)} 
  \frac{\sabs{\bar{t}}}{\sabs{\bar{t}/\vep}^{\gamma'}} 
\end{align}
where we have used $\bar{N}>1\ge 2\gamma'$ to justify the estimate 
$\sabs{t}^{\bar{N}/2-\gamma'} \le \sabs{\bar{t}/\vep}^{\bar{N}/2-\gamma'}$.  
Here applying first 
$\vep \sabs{\bar{t}/\vep} \le \sabs{\bar{t}}= T$ and then 
$\bar{N}\le 18 N_0$ yields the first term in (\ref{eq:GNbound}).

We estimate the last term in (\ref{eq:GNbound2}) using 
\begin{align}
 & \sum_{S_m(n,n')\in\pi(I)} 
 \frac{2^{-\bar{N}/2}}{((N_1+m)/2)!} 
\le  \sum_{m=0}^{N_1} \sum_{n\in \N^{m+1}} 
 \sum_{n'\in \N^{m+1}}   \frac{1}{((N_1+m)/2)!} 
\nonumber \\ & \qquad
\1\Bigl(N_1=m+2\sum_{j=1}^{m+1} n_j\Bigr)
 \1\Bigl(N_1=m+2\sum_{j=1}^{m+1} n'_j\Bigr) 
 \frac{1}{2^m} \prod_{j=0}^m \Bigl(\frac{1}{2^{n_j}}\frac{1}{2^{n'_j}}\Bigr)
\nonumber \\ & \quad
\le  \frac{1}{\lfloor N_1/2\rfloor!} \sum_{m=0}^{\infty} 2^{m+2}
 \frac{1}{\lfloor m/2\rfloor!} 
\end{align}
where the sum over $m$ is finite.  Applying $N_1\ge N_0$ and 
$2 c_0 \vep t \le \sabs{2 c_0} T$, and readjusting the constants
finishes then the proof of the Proposition.
\end{proofof}

\subsection{Consequences of dispersivity}
\label{sec:mainDRestimates}

In the derivation of the above Lemmas, 
we will heavily rely on the following estimates, which follow from the
assumed sufficiently strong dispersivity of $\omega$.
\begin{lemma} \label{th:morseprop}
Let $\omega \in L^{\infty}(\T^3)$ be such that it satisfies
the assumption (DR\ref{it:suffdisp}) with a constant $\comega$, 
and assume that $\ommax = \sup_k |\omega(k)|<\infty$.
Then, for any $\kappa\ge 0$ and $\beta>0$ such that $\kappa+\beta\le 1$,
all of the following propositions hold for 
$\gpath=\gpath(\ommax)$ and $n\in\N$:
\begin{enumerate}
\item\label{it:m0}
 $\displaystyle \sup_{\sigma=\pm 1,k\in\T^3,z\in\gpath}
  \frac{1}{|z+\ci\kappa-\sigma \omega(k)|} 
  \le \frac{1}{\beta+\kappa}$, 
\item\label{it:mz1}
 $\displaystyle \sup_{\sigma=\pm 1,k\in\T^3}
  \oint_{\gpath} \frac{|\rmd z|}{2\pi} 
  \frac{1}{|z+\ci\kappa-\sigma \omega(k)|} 
  \le |\gpath| \sabs{\ln (\beta+\kappa)}$. 
\item\label{it:mk1} 
  $\displaystyle\sup_{\sigma=\pm 1,z\in\gpath} \int_{\T^3} \rmd k\, 
    \frac{1}{|z+\ci\kappa-\sigma \omega(k)|} \le 
 12 \comega \sabs{\ln (\beta+\kappa)}$. 
\item\label{it:mk2}  
  For all $n\ge 2$, $\displaystyle \sup_{\sigma=\pm 1,z\in\gpath}
  \int_{\T^3} \rmd k\, \frac{1}{|z+\ci\kappa-\sigma \omega(k)|^n}
  \le \frac{3 \comega}{(\beta+\kappa)^{n-1}}$.
\item\label{it:mpure} For any smooth function $f$, 
\begin{align}
 \sup_{\sigma=\pm 1,z\in\gpath}
 \left| \int_{\T^3} \rmd k\, 
  \frac{f(k)}{z+\ci\kappa-\sigma \omega(k)} \right|
  \le 3 \comega \norm{f}_{d_1,\infty},
\end{align}
and for all $n\ge 2$, and $n_1,n_2\ge 0$ such that $n_1+n_2=n$,
\begin{align}
 & \sup_{\sigma=\pm 1,z\in\gpath}
 \left|  \int_{\T^3} \rmd k\, 
  \frac{f(k)}{(z-\sigma \omega(k))^{n_1}
  (z+\ci\kappa-\sigma \omega(k))^{n_2}} \right|
 \le \frac{3 \comega \norm{f}_{d_1,\infty} }{\beta^{n-3/2}}.
\end{align}
\end{enumerate}
\end{lemma}
These results are similar to those used in \cite{chen03,erdos02,erdyau04}, 
as are the ideas behind the proofs.  However, we present 
here a more 
straightforward way of doing the analysis.  
The main additional ingredient we need is the following Lemma:
\begin{lemma}\label{th:absest}
For any $r\in\R$ and $0<\beta\le 1$,
\begin{align}\label{eq:absvalrep}
 \frac{1}{\sqrt{r^2+\beta^2}} = \int_{-\infty}^\infty\!\! \rmd s \,
 \rme^{\ci s r} \int_0^\infty\! \frac{\rmd x}{\pi} \frac{1}{\sqrt{1+x^2}} 
 \rme^{-\beta |s| \sqrt{1+x^2}}
\end{align}
where for any $s\in\R$, $s\ne 0$,
\begin{align}\label{eq:absvalbound}
& \int_0^\infty\! \frac{\rmd x}{\pi} \frac{1}{\sqrt{1+x^2}} 
 \rme^{-\beta |s| \sqrt{1+x^2}} \le
 \sabs{\ln\beta} \rme^{-\beta |s|} + \1(|s|\le 1)\ln |s|^{-1} 
\end{align}
and the function on the right hand side belongs to $L^1(\rmd s)$.
\end{lemma}
\begin{proof}
Suppose we have proven the bound in (\ref{eq:absvalbound}).  Since 
$\int_{-1}^1 \rmd s \ln|s|^{-1}=2$, the bound on the right hand side 
belongs to $L^1(\rmd s)$, and we can apply Fubini's theorem in 
(\ref{eq:absvalrep}) to swap
the $s$ and $x$ integrals.  Then 
\begin{align}
  \int_{-\infty}^\infty\!\! \rmd s \,
 \rme^{\ci s r-\beta  \sqrt{1+x^2} |s|} = 
 \frac{2 \beta  \sqrt{1+x^2}}{\beta^2(1+x^2)+r^2}
\end{align}
from which (\ref{eq:absvalrep}) follows immediately.

We thus only need to prove the bound (\ref{eq:absvalbound}).  Let first
$0<|s|\le 1$.  Then $\beta |s|\le 1$, and
\begin{align}
& \int_0^\infty\!\frac{\rmd x}{\sqrt{1+x^2}} 
 \rme^{-\beta |s| \sqrt{1+x^2}}
\le
 \int_0^{(\beta|s|)^{-1}}\!\!\!\frac{\rmd x}{\sqrt{1+x^2}} 
 \rme^{-\beta |s|}
+ \int_{(\beta|s|)^{-1}}^\infty\!\!\rmd x\, \beta|s|
 \rme^{-\beta |s| x} 
\nonumber \\ & \quad  =
{\rm arsinh}((\beta|s|)^{-1} ) \rme^{-\beta |s|} + \rme^{-1}
\le (2+\ln\beta^{-1}) \rme^{-\beta |s|} + \ln |s|^{-1}
\nonumber \\ & \quad 
\le 2\sqrt{2} \sabs{\ln\beta} \rme^{-\beta |s|} + \ln |s|^{-1} 
\le \pi (\sabs{\ln\beta} \rme^{-\beta |s|} + \ln |s|^{-1} )
\end{align}
where we used the fact that for all $x\ge 1$
\begin{align}
{\rm arsinh}\, x = \ln(x+\sqrt{x^2+1}) \le 1 + \ln x.
\end{align}
If $|s|\ge 1$, similarly
\begin{align}
& \int_0^\infty\!\!\frac{\rmd x}{\sqrt{1+x^2}} 
 \rme^{-\beta |s| \sqrt{1+x^2}} \le
 \int_0^{1/\beta}\!\!\frac{\rmd x}{\sqrt{1+x^2}}  \rme^{-\beta |s|}
+ \int_{1/\beta}^\infty\!\!\rmd x\, \beta
 \rme^{-\beta |s| x} 
\nonumber \\ & \quad =
{\rm arsinh}(1/\beta )\rme^{-\beta |s|} + \rme^{-|s|}
\le (2+\ln\beta^{-1})\rme^{-\beta |s|}
\le \pi \sabs{\ln\beta}\rme^{-\beta |s|}
\end{align}
which completes the proof of inequality (\ref{eq:absvalbound}).
\end{proof}
\begin{proofof}{Lemma \ref{th:morseprop}}
The integration path $\gpath$ consists of two pieces: the uppermost part
parameterized by
$[-1-\ommax ,\ommax +1]\ni \alpha \mapsto z=-\alpha+\ci \beta$, and the
remainder whose distance from the set 
$\defset{|\re z|\le \ommax}{-1\le \im z\le 0}$ is at least $1$, see Fig.\
\ref{fig:intpath}. 
Therefore, on the first part $|z+\ci\kappa-\sigma \omega(k)|$ 
is bounded from below by $\beta+\kappa\le 1$,
and on the second part by $1$.  This proves item \ref{it:m0}.

For item \ref{it:mz1}, we first separate 
a segment of length two in the
uppermost part of $\gpath$, corresponding to
$|\alpha-\sigma\omega(k)|\le 1$.
The integral over the remaining part is then bounded by
$(|\gpath|-2)/(2\pi)$.  The value of the integral over
the segment is equal to $1/\pi$ times
\begin{align}
& \int_{0}^1\! \rmd r \, \frac{1}{\sqrt{r^2+(\beta+\kappa)^2}} 
=  \Bigl/_{\!\!\!0}^1 \,
\ln(r +\sqrt{(\beta+\kappa)^2+r^2}) 
\le 1+ |\ln (\beta+\kappa)|.
\end{align}
Thus the total integral is bounded by
\begin{align}
  \frac{|\gpath|+ 2 |\ln (\beta+\kappa)|}{2\pi}
  \le  |\gpath| \frac{1+ |\ln (\beta+\kappa)|}{2\pi}
  \le |\gpath| \sabs{\ln (\beta+\kappa)}. 
\end{align}

In all of the estimates in items \ref{it:mk1}--\ref{it:mpure} it is
sufficient to assume that $z$ belongs to the uppermost part of the integration
path, i.e.,  $z=\alpha+\ci \beta$ for some $\alpha\in\R$, since otherwise
we trivially have bounds by $1$ for items \ref{it:mk1} and \ref{it:mk2},
and by $\norm{f}_\infty$ for item \ref{it:mpure}.  Let us also 
denote $\beta'=\beta+\kappa$.
Then in item \ref{it:mk1}, the bound follows from applying Lemma 
\ref{th:absest}, the assumption (DR\ref{it:suffdisp}),
and the relations
\begin{align}\label{eq:int0bounds}
\int_0^1 \!\rmd s\, \ln |s|^{-1} =1 \qand
\int_0^\infty \!\rmd s \, \frac{1}{\sabs{s}^{3/2}} \le 
1+\int_1^{\infty}\!\rmd s\, s^{-3/2}=3.
\end{align}
For item \ref{it:mk2}, we first use the trivial bound in item \ref{it:mz1}
to see that the integrand is bounded by 
$(\beta')^{2-n}/|\alpha-\sigma \omega(k)+\ci\beta'|^2$.  Then
the estimate follows from applying the equality
\begin{align}\label{eq:deltaappr}
 \frac{1}{r^2+(\beta')^2} =  \frac{1}{2\beta'}
 \int_{-\infty}^\infty\!\! \rmd s \,
 \rme^{\ci s r-\beta' |s|} 
\end{align}
valid for all $r\in\R$, and then estimating the result using the assumption
(DR\ref{it:suffdisp}).

Finally, for item \ref{it:mpure}, we use the fact that in all of the terms
the imaginary part is strictly positive, proving
\begin{align}
 & \int_{\T^3} \rmd k\, 
  \frac{f(k)}{(\alpha+\ci\beta -\sigma \omega(k))^{n_1}
  (\alpha+\ci(\beta+\kappa)-\sigma \omega(k))^{n_2}}
\nonumber \\ & \quad
 = (-\ci)^n \int_{\R_+^n}\!\rmd s\, 
 \rme^{-\sum_{\ell=1}^{n_2} s_{\ell} \kappa}
 \rme^{-(\beta-\ci \alpha) \sum_{\ell=1}^{n} s_{\ell} } 
\int_{\T^3} \rmd k\, f(k)
 \rme^{-\ci \sigma \omega(k) \sum_{\ell=1}^{n} s_{\ell} } .
\end{align}
By (DR\ref{it:suffdisp}), this is bounded by
\begin{align}
 & \int_{\R_+^n}\!\rmd s\, 
 \rme^{-\beta \sum_{\ell} s_{\ell}} 
 \frac{\comega}{\Bigsabs{ \sum_{\ell} s_{\ell}}^{3/2}}
  \norm{f}_{d_1,\infty} 
 =   \frac{\comega \norm{f}_{d_1,\infty} }{(n-1)!} 
\int_0^\infty\!\rmd s
\frac{s^{n-1}}{\sabs{s}^{3/2}} \rme^{-\beta s}.
\end{align}
If $n=1$, the integral over $s$ is bounded by $3$.  
Otherwise, it is bounded by
\begin{align}
& 1 + \int_1^\infty\!\rmd s \, s^{n-1-3/2} \rme^{-\beta s}
\le 1 + \beta^{3/2-n} 
\left(1+\int_0^\infty\!\rmd r \, r^{n-2} \rme^{-r}\right) 
\nonumber \\ & 
\le 3 \beta^{3/2-n} (n-1)!
\end{align}
where we have used $\beta\le 1$.  
This finishes the proof of item \ref{it:mpure}.
\end{proofof}
\begin{corollary} \label{th:Mmorseprop}
Under the same assumptions as in Lemma \ref{th:morseprop},
for both $\sigma=\pm 1$,
\begin{enumerate}
\item\label{it:Nm0}
 $\displaystyle \sup_{k\in\T^3,z\in\gpath}
  \Bigl\Vert\frac{1}{z+\ci\kappa-\sigma \ho(k)} \Bigr\Vert
  \le \frac{1}{\beta+\kappa}$, 
\item\label{it:Nmz1}
 $\displaystyle \sup_{k\in\T^3}
  \oint_{\gpath} \frac{|\rmd z|}{2\pi} 
  \Bigl\Vert\frac{1}{z+\ci\kappa-\sigma \ho(k)} \Bigr\Vert
  \le 2 |\gpath| \sabs{\ln \beta}$. 
\item\label{it:Nmk1} 
  $\displaystyle\sup_{z\in\gpath} \int_{\T^3} \rmd k\, 
  \Bigl\Vert\frac{1}{z+\ci\kappa-\sigma \ho(k)} \Bigr\Vert
 \le 
 24 \comega \sabs{\ln \beta}$. 
\end{enumerate}
\end{corollary}
\begin{proof}
Item \ref{it:Nm0} is clear.  The other two follow by
first estimating the integrands by
\begin{align}\label{eq:normup}
  \Bigl\Vert\frac{1}{z+\ci\kappa -\sigma \ho(k)} \Bigr\Vert
  \le \sum_{\sigma'=\pm 1}
  \Bigl|\frac{1}{z+\ci\kappa-\sigma' \omega(k)} \Bigr|
\end{align}
and then applying the Lemma.
\end{proof}

\subsection{Derivation of the basic bounds}
\label{sec:pflemmas}

Let us now fix the way of resolving the momentum
delta-functions.  Given a partition $S\in \pi(I_{N,N'})$, 
we define $M(S)=\defset{\max A}{A\in S}$, and for any index
$\ell\in I_{N,N'}$, 
let $A(\ell)$ denote the unique cluster in $S$ which contains $\ell$.
For each $A\in S$ we integrate out $\eta_\ell$ with $\ell=\max A$.
Then every $\eta_\ell$ with $\ell\in I_{N,N'}\setm M(S)$
is {\em free\/}, i.e., it is
integrated over the whole of $\T^3$ independently of the values of the 
other integration variables, and for $\ell\in M(S)$ we have
\begin{align}
  \eta_\ell = - \sum_{n\in A(\ell):n<\ell} \eta_{n}.
\end{align}
Given a propagator index $\ell\in \set{1,\ldots,N+N'+2}$, 
we call a cluster $A\in S$ {\em broken at $\ell$\/} if
$\min A<\ell \le \max A$, and we call an index 
$n \in I_{N,N'}$ {\em free at $\ell$\/} if $n<\ell$ and $\max A(n)\ge \ell$.  The first terminology is
explained by Figure \ref{fig:graphex}, and the second comes from the fact
that the function
$k_\ell(\eta)$ depends only on those free integration variables 
$\eta_n$ which are free at $\ell$.
Explicitly, by (\ref{eq:defkell}) we have for all $\ell$
\begin{align}\label{eq:standardkell}
 k_\ell(\eta) = \eta_0 + \1(\ell>N+1) \eta_{N+1} +
 \sum_{\substack{n\in I_{N,N'}: n<\ell\\ \max A(n)\ge \ell} } \eta_n.
\end{align}

The following Lemma will allow estimating most of the $\eta$-integrals:
\begin{lemma}\label{th:iterstep}
Let $N',N\in \N$ be given, and assume
$S\in\pi(I_{N,N'})$.  Define
$M'=M(S)\cup\set{N+1}$, and for each
$\ell=2,\ldots,N+N'+1$ let
$f_\ell\in L^1(\T^3)\cap L^\infty(\T^3)$ with $f_\ell\ge 0$.
Then, if $n\in\set{1,\ldots,N+N'}$ is such that $|I_{n}\setm M'|>1$,
we have for $n'=\max (I_n \setm M')-1$
\begin{align}\label{eq:fiterest}
&\int_{(\T^3)^{I_n\setm M'}}\!\!\!\!\!\rmd \eta\, \prod_{\ell=2}^{n+1} 
f_\ell(k_\ell(\eta))
\le \norm{f_{n'+2}}_1 \prod_{\ell =n'+3}^{n+1} \norm{f_{\ell}}_\infty
\int_{(\T^3)^{I_{n'}\setm M'}}\!\!\!\! \rmd \eta\, \prod_{\ell=2}^{n'+1} 
f_\ell(k_\ell(\eta)) .
\end{align}
\end{lemma}
\begin{proof}
Since $|I_{n}\setm M'|>1$, $\max (I_n \setm M')\ge 2$ and thus
$1\le n'< n$.  Clearly, then also $I_{n'}\setm M'$ is a non-empty, proper
subset of $I_{n}\setm M'$. If $n'< n-1$, 
we first use  $f_\ell\le \norm{f_\ell}_\infty$
and positivity of $f_\ell$, to estimate the product of the terms with 
$\ell=n'+3,\ldots,n+1$.  Since $I_{n}\setm M' =
\set{n'+1} \cup (I_{n'}\setm M')$,
we find that the left hand side of 
(\ref{eq:fiterest}) is less than or equal to 
\begin{align}
& \prod_{\ell =n'+3}^{n+1} \norm{f_{\ell}}_\infty
\int_{(\T^3)^{I_{n'}\setm M'}}\! \rmd \eta\, \Bigl[
\int_{\T^3}\! \rmd \eta_{n'+1}\, \prod_{\ell=2}^{n'+2} 
f_\ell(k_\ell(\eta)) \Bigr].
\end{align}
However, if $\ell\le n'+1$, then $k_\ell$ does not depend on $\eta_{n'+1}$,
and therefore
\begin{align}
&
\int_{\T^3}\! \rmd \eta_{n'+1}\, \prod_{\ell=2}^{n'+2} f_\ell(k_\ell(\eta)) =
 \prod_{\ell=2}^{n'+1}  f_\ell(k_\ell(\eta))
 \int_{\T^3}\! \rmd \eta_{n'+1}\, f_{n'+2}(k_{n'+2}(\eta)).
\end{align}
As $k_{n'+2}(\eta) = \eta_{n'+1} +( \text{term independent of }\eta_{n'+1})$,
the remaining integral is equal to $\norm{f_{n'+2}}_1$.
\end{proof}
The point of including $N+1$ to $M'$ is that then
$I_{N,N'}\setm M(S)=I_{N+N'}\setm M'$ for all $N,N'\ge 0$. 
Estimating the missing case
with $|I_{n}\setm M'|=1$ similarly and using induction proves
\begin{corollary}\label{th:genbasicb}
If $I_{N,N'}\setm M(S) \ne \emptyset$,
\begin{align}
&\int_{(\T^3)^{I_{N,N'}\setm M(S)}}\! \rmd \eta\, \prod_{\ell=2}^{N+N'+1} 
f_\ell(k_\ell(\eta))
\le \prod_{n\in I_{N+N'}\setm M'} \norm{f_{n+1}}_1 
 \!\!\prod_{n\in M'\cap I_{N+N'}}\!\! \!\!\norm{f_{n+1}}_\infty .
\end{align}
\end{corollary}
This is sufficient to prove the basic estimates.
\begin{proofof}{Lemma \ref{th:basicAest}} 
Applying the above resolution of delta-functions
to (\ref{eq:defKpamp}), and then
taking absolute values inside the remaining integrals shows that 
\begin{align}\label{eq:defKpamp2}
&  | \Kpamp (t,S;N'_0,N_0,\vep,\kappa)|
\le (2\ommax)^{2 N}\vep^{N} 
\int_{\T^3} \rmd \eta_0 \norm{\FT{\psi}^\vep(\eta_0)}^2
\nonumber \\ & \times
 \oint_\gpath \frac{|\rmd z|}{2\pi} 
 \oint_\gpath \frac{|\rmd z'|}{2\pi} 
 \rme^{t \im (z+z')}
  \Bigl\Vert \frac{1}{z-\ho(\eta_0)} \Bigr\Vert \,
  \Bigl\Vert \frac{1}{z'+\ho(\eta_0)} \Bigr\Vert
\nonumber \\ & \times
\int_{(\T^3)^{I_{N,N}\setm M(S)}}\! \rmd \eta\, 
 \prod_{\ell=2}^{N} 
  \Bigl\Vert\frac{1}{z+\ci \kappa_\ell -\ho(k_\ell)} \Bigr\Vert
 \prod_{\ell=N+3}^{2 N+1} 
  \Bigl\Vert\frac{1}{z'+\ci \kappa_\ell +\ho(k_\ell)} \Bigr\Vert
\end{align}
where we used the bound $\norm{v(k',k)}\le 2\ommax $, and defined
$\kappa_\ell = \kappa$ for $N_0+1\le \ell\le N_0+2+2 N'_0$ 
and zero elsewhere.  

We shall now choose
\begin{align}\label{eq:defbeta}
 \beta=\frac{1}{2\sabs{t/2}},
\end{align}
when $\beta+\kappa \le 1$.
Since $N>0$ and there are no singlets in $S$, we have
$1\in I_{N,N}\setm M(S)$.  Therefore, by Corollaries
\ref{th:Mmorseprop} and \ref{th:genbasicb}, the last line of
(\ref{eq:defKpamp2}) is bounded by
\begin{align}
(24 \comega \sabs{\ln \beta})^{2 N-|S|-1} 
\frac{1}{(\beta+\kappa)^{n_S}} \frac{1}{\beta^{|S|-n_S}}.
\end{align}
where $n_S$ is defined by (\ref{eq:defNS}). To arrive at 
this bound, first note that $f_{N+1},f_{N+2}=1$ and, as by $N>0$ we have
$N+1\in M'\cap I_{2N}$, one of the
$L^\infty$-estimates is $\norm{f_{N+2}}_{\infty}=1$.
For $f_{N+1}$ we have used the property that all bounds coming from Corollary
\ref{th:Mmorseprop} are greater than one.  The remaining integrals over
$z$ and $z'$ are then estimated using
$t\im(z+z')\le 2 t \beta\le 2$ and Corollary \ref{th:Mmorseprop}.

Since for all $c\ge 1$ and $x\in \R$, 
$\sabs{c x} \le c \sabs{x}$, now 
$\vep/\beta \le 2\vep \sabs{t\vep/(2\vep)} \le \sabs{t\vep}$,
and thus also $0<-\ln \beta \le \ln (\sabs{t\vep}\!/\vep)$. 
Using these bounds and $\comega\ge 1$ proves (\ref{eq:Abound}) for 
$c=\sabs{48 \ommax \comega}^2$ and $c'= \rme^2 2^4 (2 \ommax+5)^2$. 
\end{proofof}
For the rest of this section, we make the assumptions in Lemma
\ref{th:basicest}.  In particular, we assume that $N_1$, $N_2$ and $N'$ 
are given as in the Lemma, and we let $\bar{N}_1=N_1+N'$, 
$\bar{N}_2=N_2+N'$, and $N=(\bar{N}_1  + \bar{N}_2)/2$.  Let us also define 
$\kappa_\ell = \kappa$ for $N_1+1\le \ell\le N_1+2+2 N'$ 
and zero elsewhere.
\begin{proofof}{Lemma \ref{th:basicest}} 
The proof is almost identical to the one above, except we can ignore the
sharper bounds coming from $\kappa>0$.  We start from
\begin{align}\label{eq:Kpartb1}
& \left|  \Kpart(t,S;(N',N_2),(N',N_1),\vep,\kappa,p,f) \right|
\nonumber \\ & \quad 
\le \norm{f}_\infty (2\ommax)^{2 N} \vep^{N} \int_{\T^3} \rmd \eta_0 
\norm{\FT{\psi}^\vep(\eta_0)}\norm{\FT{\psi}^\vep(\eta_0-p)} 
\nonumber \\ & \quad \times
 \oint_\gpath \frac{|\rmd z|}{2\pi} 
 \oint_\gpath \frac{|\rmd z'|}{2\pi} 
 \rme^{2}
  \Bigl\Vert \frac{1}{z+\ci \kappa_1-\ho(\eta_0)} \Bigr\Vert \,
  \Bigl\Vert \frac{1}{z'+\ci \kappa_{2 N+2}+\ho(\eta_0-p)} \Bigr\Vert
\nonumber \\ & \quad
\int_{(\T^3)^{I_{\bar{N}_1,\bar{N}_2}\!\setm M(S)}}\! \rmd \eta\, 
 \prod_{\ell=2}^{\bar{N}_1+1} 
  \Bigl\Vert\frac{1}{z+\ci \kappa_\ell -\ho(k_\ell)} \Bigr\Vert
 \prod_{\ell=\bar{N}_1+2}^{2 N+1} 
  \Bigl\Vert\frac{1}{z'+\ci \kappa_\ell +\ho(k_\ell)} \Bigr\Vert .
\end{align}
If $N=0$, then $N'=0=N_1=N_2$ and the last line in the
above formula is equal to one, and estimating the first two lines
by Corollary \ref{th:Mmorseprop} yields the estimate in
(\ref{eq:basicbound}) with $c'= \rme^2 2^4 (2 \ommax+5)^2$. 

Let then $N>0$.  As $S$ is relevant and thus contains no singlets, we have
$I_{\bar{N}_1,\bar{N}_2}\!\setm M(S)\ne \emptyset$ and $|S|\le N$.  
We can thus apply Corollaries \ref{th:Mmorseprop} and \ref{th:genbasicb} 
and show that the last line in (\ref{eq:Kpartb1}) is bounded by
\begin{align}\label{eq:b1b2}
(24 \comega \sabs{\ln \beta})^{|I_{2 N}\setm M'|} 
\beta^{-|M'\cap I_{2 N}|}.
\end{align}
Now $|M'\cap I_{2 N}|\le |S|$ as $|M'|=|S|+1$ and
$2 N+1 \in M'\setm I_{2 N}$
(if $\bar{N}_2=0$, then $2 N +1  = \bar{N}_1+1$, and otherwise
$2 N +1 \in I_{\bar{N}_1,\bar{N}_2}$).
Using also $|I_{2 N}\setm M'|\le 2 N$, we thus find that (\ref{eq:b1b2}) is
bounded by $(24 \comega)^{2 N} \sabs{\ln \beta}^{2 N} \beta^{-|S|}$.
The remainder of the integral can be estimated as when $N=0$,
and the terms containing $\beta$ majorized as in the previous
proof. 
This proves (\ref{eq:Abound}) for the same $c'$ as above and 
$c=\sabs{48 \ommax \comega}^2$. 
\end{proofof}
\begin{proofof}{Lemma \ref{th:crossingest}} 
Let us denote here $I=I_{\bar{N}_1,\bar{N}_2}$, and recall the earlier
definitions of $M(S)$ and $M'$.  We begin
the estimation of the amplitude of
the crossing partition $S$ from (\ref{eq:Kpartb1}).
A sequence of
two pairings $(P,P')$, $P,P'\in S$, is called {\em crossing\/} 
if $\min P<\min P'<\max P<\max P'$.  For
convenience we have included the ordering of the pairings in the definition.
Furthermore, a crossing sequence is called {\em loose,\/} 
if for every $n\in I$ with $\min P'<n<\max P$ we have
$\min P'<\min A(n)<\max A(n)<\max P$. In other words, a loose
crossing sequence has no pairings connecting the
inside of the ``crossing interval'' to its outside.  For instance, in 
Fig.\ \ref{fig:graphex} the crossing sequence $(\set{2,7},\set{3,9})$
is loose, while $(\set{1,8},\set{3,9})$ is not.

We begin by proving that, if the sequence $(P_0, P'_0)$ is crossing,
then there is a loose crossing sequence $(P,P')$ such that 
$\min P'_0\le \min P'<\max P\le \max P_0$.  To do this, let us define the
functions $\overline{f},\underline{f}:S\to I\cup\set{0}$ so that 
\begin{align}
\overline{f}(P) & = 
\max \defset{n\in I}{n=\min A(n)<\max P<\max A(n)},\\
\underline{f}(P) & = \min \defset{n\in I}{\min A(n)<\min P<\max A(n)=n},
\end{align}
and $\overline{f}(P)=0$, respectively $\underline{f}(P)=0$, 
if there are no indices $n$ satisfying the corresponding condition.
Starting from $(P_0, P'_0)$, we define $(P_1, P'_1)$ by
$P'_1 = P'_0$ and $P_1 = A(\underline{f}(P'_0))$ -- 
this is well-defined since, $(P_0, P'_0)$ being
crossing, $\underline{f}(P'_0)\ne 0$. Then $(P_1, P'_1)$
is a crossing sequence for which 
$\min P'_0= \min P'_1<\max P_1\le \max P_0$.  Next we construct
$(P_2, P'_2)$ by the formula 
$P_2 = P_1$ and $P'_2 = A(\overline{f}(P_1))$, when it will be
a crossing sequence with 
$\min P_2=\min P_1 < \min P'_1\le \min P'_2 < \max P_2= \max P_1$.

Iterating these two steps, we obtain a sequence of crossing pairings 
$(P_n, P'_n)$, $n=0,1,2,\ldots$, which satisfy 
$\min P'_{n} \le \min P'_{n+1} < \max P_n$ and 
$\min P'_{n} < \max P_{n+1} \le \max P_{n}$ for all $n$.
Therefore, $(\min P'_{n})$ is an increasing sequence of integers
which is bounded from above by $\max P_0$, and $(\max P_{n})$
is a decreasing sequence bounded from below by $\min P'_0$.
Thus there is $n_0$ such that for all $n \ge n_0$ both sequences remain
constant, which implies that, if we denote $P=P_{n_0}$ and $P'=P'_{n_0}$,
then for all $n\ge n_0$, $(P_n, P'_n)=(P,P')$.  Then it is straightforward
to check that $(P,P')$ is a loose crossing sequence with 
$\min P'_0\le \min P'<\max P\le \max P_0$.

Let thus $(P,P')$ be a loose crossing sequence and denote $i_1=\min P$,
$i_2=\max P$, $j_1=\min P'$ and $j_2=\max P'$.  Then 
$i_1<j_1<i_2<j_2\le\bar{N}+1$, and by looseness and $\eta_{\bar{N}_1+1}=-p$,
\begin{align}
\sum_{n=j_1+1}^{i_2-1} \eta_n 
= - \1(j_1<\bar{N}_1+1<i_2) p
\end{align}
which is a constant we denote by $-p'$.
On the other hand, $\eta_{i_1},\eta_{j_1}$ are free variables, and
$\eta_{i_2}=-\eta_{i_1}$.  Therefore,
\begin{align}\label{eq:kminP}
k_{i_2+1}=k_{j_1+1} -\eta_{i_1} - p' .
\end{align}

Then we iterate Lemma \ref{th:iterstep} until $n'=j_1$,
with the exception that we do not take the $L^\infty$-norm of 
$f_{i_2+1}$ -- this is possible as by (\ref{eq:kminP})
$f_{i_2+1}$ depends only on free variables with index $n\le j_1$.
We change variables from $\eta_{j_1}$ to $k=k_{j_1+1}$ which yields the
integral $\int_{\T^3} \rmd k\, f_{j_1+1}(k) f_{i_2+1}(k-p'-\eta_{i_1})$,
depending only on the free variable $\eta_{i_1}$.
Then we iterate Lemma \ref{th:iterstep} further until $n'=i_1$,
change integration variables from $\eta_{i_1}$ to 
$k'= k_{i_1+1} = \eta_{i_1} + k_{i_1}$, and take supremum over $k_{i_1}$.
This yields a factor which depends only on $z$ and $z'$, namely
\begin{align}\label{eq:crint2}
& \sup_{u\in\T^3} \int_{(\T^3)^2}\!\! \rmd k\,\rmd k'\, 
f_{j_1+1}(k) f_{i_2+1}(k-k'+u) f_{i_1+1}(k') .
\end{align}
The remaining integral can then be estimated as before, by iterating Lemma
\ref{th:iterstep}.  The resulting upper bound will be the same as for the
corresponding 
basic estimate, except we have replaced one $L^\infty$- and two $L^1$-norms
by (\ref{eq:crint2}).  This yields an upper bound 
for the last line of (\ref{eq:Kpartb1})
which is equal to (\ref{eq:b1b2}) times
\begin{align}\label{eq:crint3}
& (24 \comega \sabs{\ln \beta})^{-2} \beta
\sup_{u\in\T^3} \int_{(\T^3)^2}\!\! \rmd k\,\rmd k'\, 
f_{j_1+1}(k) f_{i_2+1}(k-k'+u) f_{i_1+1}(k') .
\end{align}
Each of the $f_\ell$-terms is of the form 
$\norm{1/(w+\ci \kappa_\ell\pm H(k))}$ where $w$ is either of the integration
variables  $z$ or $z'$.  If $w$ does not belong to the uppermost part of
the path $\gpath$, then $f_\ell\le 1$ and we can use this bound to remove
the corresponding term from the integrand.  However, if any of the $f_\ell$
terms is missing, (\ref{eq:crint3}) is bounded by $\beta$.  
In the only remaining case, all of the $w$ 
are of the form $w=\alpha+\ci \beta$ for some $\alpha\in \R$.
Then, by (\ref{eq:normup}) and the assumption
(DR\ref{it:crossing}), (\ref{eq:crint3}) is bounded by 
\begin{align}
 (24 \comega)^{-2}  2^3 c_2 \beta^{\gamma} \sabs{\ln \beta}^{d_2-2} .
\end{align}
Since $\gamma\le 1$, we can conclude 
that there is a constant $c''$, depending only on $\omega$,
such that for all $z,z'\in \gpath$, (\ref{eq:crint3}) is bounded by
$c'' \beta^{\gamma} \sabs{\ln \beta}^{\max(0,d_2-2)}$.  This again
is bounded by (\ref{eq:crossingextra}) since $\beta\le \sabs{t}^{-1}$.
Then we can estimate the rest of the integral in (\ref{eq:Kpartb1}) as
before, and we 
have proven that $|\Kpart|$ can be bound by the right hand
side of (\ref{eq:basicbound}) times (\ref{eq:crossingextra}).
\end{proofof}

For the remaining Lemmas we need to analyze the integrals more carefully,
in particular, it will not be possible to take the norms inside all of the
integrals.  This is the case for a {\em gate,\/} or immediate
recollision, which corresponds to $A=\set{\ell, \ell+1}$ for some
index $\ell\in I_{N,N'}$.  Then $\eta_\ell$ is free only at $\ell+1$
implying that only $k_{\ell+1}$ depends
on it. In addition, the
momenta before and after the gate are forced to be equal, since now
$\sum_{n=0}^{\ell+1} \eta_n  = \sum_{n=0}^{\ell-1} \eta_n$. Therefore, after
we integrate over $\eta_\ell$, we can replace each gate by a certain matrix
factor. This factor is $g(k_{\ell};z+\ci\kappa_\ell)$ if $\ell<N$, 
and $-g(k_{\ell};-z'-\ci\kappa_\ell)$ if $\ell>N+1$.
Here $g$ is the following matrix-valued function:
\begin{definition}\label{th:defgate}
We define for all $k\in\T^3$, and $w\in\C\setm[-\ommax,\ommax]$,
\begin{align}
& g(k;w)  = \int_{\T^3}\!\! \rmd k' v(k,k')
\frac{\ci}{w-\ho(k')} v(k',k) .
\end{align}
\end{definition}
Explicitly, the $\sigma_1 \sigma_2$-component of $g$ is then given by
\begin{align}\label{eq:gcomponents}
\sum_{\sigma'=\pm 1}
\int_{\T^3}\! \rmd k' \frac{\ci}{w-\sigma'\omega(k')}
\frac{\omega(k)+\sigma_1 \sigma'\omega(k')}{2}
\frac{\omega(k)+\sigma_2 \sigma'\omega(k')}{2} .
\end{align}

We need to study this function in fairly great detail, and for this we will
also need certain properties of the level set measures of $\omega$, 
derived in  Appendix \ref{sec:appBoltzmann}.
\begin{lemma}\label{th:gateint}
As a function of $k$, $g(k;w)$ is a second order polynomial in $\omega(k)$,
with coefficients uniformly bounded for all
$w\in \gpath+\ci \kappa$ and
$\beta,\kappa>0$, with $\kappa+\beta\le 1$.
In addition, 
the following limit converges for all $k\in \T^3$ and $\sigma=\pm 1$:
\begin{align}
  \Theta_\sigma(k) = \lim_{\beta\to 0^+} 
  g_{\sigma\sigma}(k;\sigma \omega(k)+\ci \beta).
\end{align}
The functions $\Theta_\sigma$ are H\"{o}lder continuous with 
exponent $\frac{1}{2}$, $\Theta_- = \Theta_+^*$, 
\begin{align}\label{eq:reTheta}
\re \Theta_+(k) =  \pi \omega(k)^2
\int_{\T^3} \!\rmd k' \, \delta(\omega(k')-\omega(k)) = 
\frac{1}{2} \nu_k(\T^3),
\end{align}
for all $k\in\T^3$, and there are constants
$c'_i$, $i=1,2,3$, such that for all $\beta,\kappa$ as above,
$k,k'\in \T^3$, $\sigma=\pm 1$, and $w\in\gpath+\ci \kappa$,
\begin{align}\label{eq:gtotheta}
 & \left|g_{\sigma\sigma}(k';w)-  \Theta_\sigma(k) \right|
  \le c'_1 |\omega(k')-\omega(k)| +
  \frac{c'_2}{\sqrt{\beta}} |w-\sigma\omega(k')|
  + c'_3 \sqrt{\beta+\kappa} .
\end{align}
\end{lemma}
\begin{proof}
By (\ref{eq:gcomponents}) and Lemma \ref{th:morseprop}:\ref{it:mpure},
$k\mapsto g_{\sigma_1\sigma_2}(k;w)$ 
is a second order polynomial in $\omega(k)$ with uniformly bounded 
coefficients.  In particular, also
\begin{align}\label{eq:defbarg}
& \bar{g} = \sup_{\beta,\kappa} \sup_{z\in \gpath, k\in \T^3}
\left\Vert g(k;z+\ci\kappa)\right\Vert <\infty.
\end{align}
To study the limit of small $\beta$, we apply to (\ref{eq:gcomponents})
the following equality, which is valid for all $\alpha\in\R$,
$\beta\ge\beta'>0$, 
\begin{align}
  \frac{\ci}{\alpha+\ci \beta'}
 -  \frac{\ci}{\alpha+\ci \beta}
 = \int_{\beta'}^{\beta} \rmd \lambda\,
  \frac{1}{(\alpha+\ci \lambda)^2}.
\end{align}
Then by Lemma \ref{th:morseprop}, 
there is $c''$ such that for $0<\beta'<\beta\le 1$,
$\sigma=\pm 1$, and $k\in\T^3$,
\begin{align}
 & \left|g_{\sigma\sigma}(k;\sigma\omega(k)+\ci \beta')-
   g_{\sigma\sigma}(k;\sigma\omega(k)+\ci \beta)\right|
 \le \frac{c''}{2} \int_{\beta'}^\beta\rmd \lambda\, \lambda^{-1/2} 
 \le c''\sqrt{\beta}.
\end{align}
This proves that the limits $\Theta_\sigma(k)$ exist for all $k$ and
$\sigma$, and that 
\begin{align}\label{eq:Thetaest}
 & \sup_{k,\sigma}\left|\Theta_\sigma(k)-
   g_{\sigma\sigma}(k;\sigma\omega(k)+\ci \beta)\right|
 \le c''\sqrt{\beta}.
\end{align}
As
$g_{--}(k;-\omega(k)+\ci \beta) =   g_{++}(k;\omega(k)+\ci \beta)^*$,
we have then $\Theta_- = \Theta_+^*$.

By the same Lemma, there is a constant $c'_1$ such that
\begin{align}
\sup_{w\in\gpath+\ci\kappa,\sigma=\pm 1}
\left|g_{\sigma\sigma}(k';w) - g_{\sigma\sigma}(k;w) \right|
 \le \left|\omega(k)-\omega(k')\right| c'_1.
\end{align}
Suppose for a moment that $0<\beta\le 1$ and
$\alpha\in\R$ satisfies $|\alpha|\le \ommax+1$.
Since for all $a,a'\in \R$,
\begin{align}
 &  \frac{1}{ a'+\ci \beta}
 -   \frac{1}{ a+\ci \beta}
 = \int_0^1\!\! \rmd\lambda \frac{a-a'}{(a+
  \lambda(a'-a)+\ci \beta)^2}
\end{align}
we can apply Lemma \ref{th:morseprop}:\ref{it:mpure} with $n=2$
and conclude that there is a constant $c''_2$, depending only on the function
$\omega$, such that
\begin{align}\label{eq:gvaryzbound}
 & \left| g_{\sigma\sigma}(k;\alpha+\ci \beta)- 
   g_{\sigma\sigma}(k;\sigma\omega(k)+\ci \beta)
\right|
 \le c''_2 |\alpha-\sigma\omega(k)| \beta^{-1/2}
\nonumber \\ & \quad 
 \le c''_2 |\alpha+\ci \beta-\sigma\omega(k)| \beta^{-1/2}+
 c''_2 \beta^{1/2}.
\end{align}
Let then $\kappa,\beta$ be as in the assumptions of the Lemma.
Then for $w\in\gpath+\ci \kappa$, either $w$ is of the already 
considered form, or $|w-\sigma\omega(k)|\ge 1$. But
in the latter case
$\left| g_{\sigma\sigma}(k;w)- 
g_{\sigma\sigma}(k;\sigma\omega(k)+\ci \beta)\right|
\le 2\bar{g} |w-\sigma\omega(k)| \beta^{-1/2}$, and the inequalities proven
so far imply  
the inequality (\ref{eq:gtotheta}).

For the continuity of $\Theta_\sigma$, let $h$ be such that $|h|<1$, and
define $\beta=|h|$.  Then 
\begin{align}  
 & \left|\Theta_\sigma(k+h)-\Theta_\sigma(k) \right|
 \le (c'_2+c'_3) \sqrt{\beta} + 
 \left|\Theta_\sigma(k+h)-g_{\sigma\sigma}(k;\sigma\omega(k)+\ci\beta) \right|
\nonumber  \\ & \quad
  \le 2 (c'_2+c'_3)\sqrt{\beta} + 
  (c'_1+\frac{c'_2}{\sqrt{\beta}}) |\omega(k+h)-\omega(k)| .
\end{align}
Since $\omega$ is smooth, there thus is a constant $c$ such that
\begin{align}
\left|\Theta_\sigma(k+h)-\Theta_\sigma(k) \right|\le c \beta^{1/2}=
c |h|^{1/2}
\end{align} 
which proves that the function is H\"{o}lder-continuous with an exponent
$1/2$. 

Finally, to prove (\ref{eq:reTheta}) note that
$\re \Theta_+(k)$ is the $\beta\to 0^+$ limit of 
\begin{align}
 \sum_{\sigma'=\pm 1} \int_{\T^3} \rmd k'\! \frac{\beta}{
 (\omega(k)-\sigma'\omega(k'))^2+\beta^2}
\left(\frac{\omega(k)+\sigma'\omega(k')}{2}\right)^2 .
\end{align}
As $|\omega(k)|\ge \ommin > 0$, 
the $\sigma'=-1$ term is $\order{\beta}$, and for the
$\sigma'=+1$ we use
\begin{align}
\frac{\omega(k)+\omega(k')}{2} = 
\omega(k) + \frac{\omega(k')-\omega(k)}{2} 
\end{align}
to expand the square.  The term having $(\omega(k')-\omega(k))^2$
is $\order{\beta}$, and the term with $\omega(k')-\omega(k)$ vanishes 
by dominated convergence, justifiable by the estimate (\ref{eq:bxunif}).
The only non-vanishing term is 
\begin{align}
  \omega(k)^2 \int_{\T^3}\!\rmd k'\,
  \frac{\beta}{(\omega(k)-\omega(k'))^2+\beta^2} 
\end{align}
which converges to the middle formula in (\ref{eq:reTheta}).  The last
equality follows then 
from Proposition \ref{th:crosssectprop}:\ref{it:movecontel}.
\end{proof}
\begin{proofof}{Lemma \ref{th:nestedest}} 
We call a pairing $P\in S$ {\em nesting\/} if $\min P > \bar{N}_1+1$
or $\max P<\bar{N}_1+1$, and there is $P'\in S$ 
such that $\min P<\min P'<\max P'<\max P$ -- the first condition is to
exclude nests going over $\bar{N}_1+1$ which will contribute to the main term. 
A nesting is called {\em minimal\/} if the nest contains only gates, 
i.e., $\min P<n<\max P$ implies $A(n)$ is a gate.  We claim that, if 
$P_0\in S$ is nesting, then there is a minimal nesting $P\in S$ such that
$\min P_0\le \min P<\max P\le \max P_0$.

To prove this, we start from $P_0$, and
iterate the following procedure for $j=0,1,\ldots$: 
If there is $n\in I$ such that $\min P_j < n<\max P_j$ and $A(n)$ is not a
gate, then we let $n'$ to be the smallest of such indices and  
define $P_{j+1}=A(n')$.  If there is no such $n$, we define
$P_{j+1}=P_j$.  As $S$ is not crossing, we must have
$\min P_j\le \min P_{j+1} < \max P_{j+1}\le \max P_{j}$ and, since $P_{j+1}$
is not a gate, there is $n$ such that $\min P_{j+1} < n< \max P_{j+1}$.
By the non-crossing assumption, 
$\min P_{j+1} < \min A(n)<\max A(n)< \max P_{j+1}$ and $P_{j+1}$ is a
nesting pairing such that $\min P_j\le \min P_{j+1}<\max P_0$.  
Since $(\min P_j)$ forms an increasing sequence of integers bounded from
above, it is constant from some $j_0$ onwards.  Then $P=P_{j_0}$ is a minimal
nesting pairing with the required properties.

Let us thus assume that $P$ is a minimal nesting pairing, and let
$i_1=\min P$, $i_2=\max P$.  Then $P$ nests $m=(i_2-i_1-1)/2$ gates
with $0<m<\bar{N}/2$.  As $i_1$ is free at $\ell$ only when 
$i_1< \ell \le i_2$,  
and the addition of the gates does not change the momentum,
we can first integrate over the gate momenta and then over $\eta_{i_1}$
before integrating any of the other free variables.  
This yields a matrix factor
\begin{align}
 G'_m & =\int_{\T^3}\!\rmd k \, v(k',k) 
 \prod_{j=1}^{m} \Bigl( 
 \frac{\tau \ci}{w + \ci \kappa_{i_1+2 j+1} -\tau \ho(k)} 
 g(k;\tau w + \tau \ci \kappa_{i_1+2 j})
 \Bigr) 
 \nonumber \\ & \quad \times
 \frac{\ci}{w + \ci \kappa_{i_1+1} -\tau \ho(k)} v(k,k')
\end{align}
where $k=k_{i_1+1}$ and $k'=k_{i_1}$, and $w=z$, $\tau=+1$,
if $i_2\le \bar{N}_1$ and $w=z'$, $\tau=-1$, otherwise.
In any case, $w\in \gpath$ and we will choose $\beta$ as in (\ref{eq:defbeta}).

We then expand out the components of $G'_m$ which yields
for $\sigma'_1,\sigma'_2\in \set{\pm 1}$,
\begin{align}
 & (G'_m)_{\sigma'_2\sigma'_1}  = \tau^m 
\sum_{\sigma\in \set{\pm 1}^{m +1}}
\int_{\T^3}\!\rmd k \, v_{\sigma'_2 \sigma_{m +1}}(k',k) 
   v_{\sigma_{1}\sigma'_1}(k,k') 
 \nonumber \\ & \quad \times
 \prod_{j=1}^{m} 
 g_{\sigma_{j+1}\sigma_j}(k;\tau w + \tau \ci \kappa_{i_1+2 j})
 \prod_{j=1}^{m+1}
 \frac{\ci}{w + \ci \kappa_{i_1+2 j-1} -\tau \sigma_j \omega(k)} .
\end{align}
Consider first a term in the sum where $\sigma_j=\sigma_1$ for all $j$.
Then there are $n_1$, $n_2$, $n'_1$, $n'_2\in \N$ such that $n_1+n_2 = m+1$
and $n'_1+n'_2=m$ and the summand is equal to
\begin{align}
&  \ci^{m+1}
\int_{\T^3}\!\rmd k \, 
\frac{v_{\sigma'_2 \sigma_{1}}(k',k) 
   v_{\sigma_{1}\sigma'_1}(k,k') 
 g_{\sigma_{1}\sigma_1}(k;\tau w)^{n'_1}
 g_{\sigma_{1}\sigma_1}(k;\tau w + \tau \ci \kappa)^{n'_2} }{
 (w -\tau \sigma_1 \omega(k))^{n_1}
 (w + \ci \kappa -\tau \sigma_1 \omega(k))^{n_2}  }.
\end{align}
By  Lemma \ref{th:morseprop}:\ref{it:mpure},
its absolute value has an upper bound
$3 \comega \norm{f}_{d_1,\infty}\beta^{1/2-m}$ where $f$ denotes the
numerator in the above integrand.  Employing the Leibniz rule and induction, 
it is possible to prove that for any $n,d_1\in \N_+$,
\begin{align}
\Bigl\Vert\prod_{j=1}^n f_j\Bigr\Vert_{d_1,\infty}\le n^{d_1} 
\prod_{j=1}^n \norm{f_j}_{d_1,\infty} .
\end{align}
Thus we can conclude from Lemma \ref{th:gateint}
that there is $c_1$, depending only on $\omega$, 
such that $\norm{f}_{d_1,\infty} \le (m+2)^{d_1} c_1^{m+2}$.
Therefore, we have proven that for the two terms, in which all components
of $\sigma$ are equal, the summand is bounded by 
\begin{align}
3 \comega (m+2)^{d_1} c_1^{m+2} \beta^{\frac{1}{2}-m}.
\end{align}

Consider then the remaining case, when there is $j_0> 1$ such that 
$\sigma_{j_0}\ne \sigma_1$. Then, by using $1/(ab) = (1/a-1/b)/(b-a)$ and
$\ommin >0$, we find that
\begin{align}\label{eq:oneoverab}
& \left|\frac{1}{w + \ci \kappa_{i_1+1} -\tau \sigma_1 \omega(k)} 
\frac{1}{w + \ci \kappa_{i_1+2 j_0-1} -\tau \sigma_{j_0} \omega(k)}
\right|  \nonumber \\ & \quad 
\le \frac{1}{2 \ommin} \Bigl(
\frac{1}{|w + \ci \kappa_{i_1+1} -\tau \sigma_1 \omega(k)|} +
\frac{1}{|w + \ci \kappa_{i_1+2 j_0-1} -\tau \sigma_{j_0} \omega(k)|}
\Bigr) .
\end{align}
But, since $|g|,|v|\le c_1$, we can use the trivial bound for the rest of
the terms, and then bound the remaining integral by Lemma 
\ref{th:morseprop}:\ref{it:mk1}. This shows that
the absolute value of the summand is bounded by
\begin{align}
& \frac{c_1^{m+2}}{2\ommin}   \beta^{1-m}
24 \comega \sabs{\ln \beta} .
\end{align}
Since there are less than $2^{m+1}$ such terms, we have proven that
\begin{align}
& \norm{G'_m} \le 2 \max_{\sigma',\sigma}{|(G'_m)_{\sigma'\sigma}|}
\le (2 c_1)^{m+2} 12 \comega \beta^{\frac{1}{2}-m} \Bigl(
  \frac{2}{\ommin} \beta^{\frac{1}{2}} \sabs{\ln \beta} + 
 (m+2)^{d_1} \Bigr).
\end{align}
Here $\beta^{\frac{1}{2}} \sabs{\ln \beta} \le 1$ since $0<\beta\le 1$,
and we find that there are constants $c'_1$, $c'_2$, which depend
only on $\omega$, such that 
\begin{align}\label{eq:xxb1}
& \norm{G'_m} \le c'_1 (c'_2)^{m+1} (m+2)^{d_1} \beta^{\frac{1}{2}-m} .
\end{align}

This upper bound is a constant, and we can take it out of all of 
the remaining integrals.
Estimating the remainder by Corollary \ref{th:genbasicb} yields a bound
which is the basic bound times 
\begin{align}\label{eq:xxb2}
(2\ommax)^{-2(m+1)} \beta^m (24 \comega\sabs{\ln \beta})^{-m-1} 
\end{align}
since there are $m+1$ missing $L^1$-norms,  $m$ missing $L^\infty$-norms
and $2(m+1)$ missing bounds for $v$.  To finish the proof of the Lemma, 
we multiply (\ref{eq:xxb1}) with (\ref{eq:xxb2}), and then 
use $2 m+2\le \bar{N}$ and $\beta\le 1/\sabs{t}$.
\end{proofof} 

\section{Simple partitions}\label{sec:simple}

\subsection{General bound (proof of Lemma \ref{th:simpleest})}
\label{sec:gensimple}

Clearly, every $S_m(n,n')$, which has $m$, $n$ and
$n'$ such that
$m+2\sum_{j=1}^{m+1} n_j = \bar{N}_1$ and
$m+2\sum_{j=1}^{m+1} n'_j = \bar{N}_2$, belongs to 
$\pi(I_{\bar{N}_1,\bar{N}_2})$ and is simple.  Let us next prove that also the
converse holds.  Suppose $S$ is simple.  
If $A\in S$ is such that $\max A<\bar{N}_1+1$, then 
$A$ must be a gate since otherwise it would form either a nest or crossing
for some $A(n)$ with $\min A<n<\max A$.  Similarly, if 
$\min A>\bar{N}_1+1$, then $A$ must also be a gate.  The remaining pairings
form a subset $S'=\defset{A\in S}{\min A<\bar{N}_1+1<\max A}$, 
and let $m=|S'|$.  If $S'$ is empty, $S=S_0(\bar{N}_1/2,\bar{N}_2/2)$.
Otherwise, let us order $A'\in S'$ into a sequence such that 
$\min A'_j<\min A'_{j+1}$ for all $j=1,\ldots,m$.  Then
$\max A'_j>\max A'_{j+1}$ for all $j$, since otherwise 
$(A'_j, A'_{j+1})$ is crossing.  Let also $A'_0=\set{0,\bar{N}+2}$,
$A'_{m+1}=\set{\bar{N}_1+1}$, and
define, for $i=0,\ldots,m$,  $n_i$ as the number of gates $A$
with $\min A'_i< \min A <\min A'_{i+1}$ and $n'_i$
as the number of gates $A$
with $\max A'_i> \min A >\max A'_{i+1}$.  Then $S=S_m(n,n')$,
with $m$, $n$, $n'$ satisfying the condition given in the Lemma.
We have begun indexing the components of $n$ and
$n'$ from $0$, not from $1$ -- this will become convenient later.

Consider then $\Kpart(t, S;(N',N_2),(N',N_1),\vep,\kappa,p,f)$ 
corresponding to such $S$.  
First we integrate out the gates, each yielding a factor 
$\pm g$, as before. The remaining free indices, if any, are
$\eta_{r_j}$ with $r_j = \min A'_j=\sum_{j'=0}^{j-1} (2 n_{j'}+1)$, for $j=1,\ldots,m$. 
We then make a change of variables to 
\begin{align}
  k''_j = \sum_{j'=0}^{j-1} \eta_{\ell_{j'}} - \frac{1}{2} p
\end{align}
where $j=1,\ldots,m+1$.  This implies that for all $\ell\le \bar{N}_1$, 
which are not inside a gate,  $k_\ell = k''_j + p/2$ for some $j$, 
and for all  $\ell> \bar{N}_1+1$, which are not inside a gate, 
$k_\ell = k''_j - p/2$  for some $j$.  For an explicit example, see Fig.\
\ref{fig:gensimple}. 

To write the result in a convenient form, let us define 
$r'_j = \max A'_j=\bar{N}+2-\sum_{j'=0}^{j-1} (2 n'_{j'}+1)$,
for $j=1,\ldots,m$, and $r_0=0$, $r'_{0}=\bar{N} +2$, and let then
$\kappa_{j,i}=\kappa_{r_j+1+i}$ and $\kappa'_{j,i}=\kappa_{r'_j-i}$
for any appropriate choice of indices $j,i$.
Dropping the double-primes, and using the short-hand notations
\begin{align}
  k^{\pm}_j = k_j \pm \frac{1}{2} p,
\end{align}
we obtain the following representation for $\Kpart$
\begin{align}\label{eq:notsosimplesum2}
&  \vep^{m}
 \int_{(\T^3)^{I'_m}} \!\!\rmd k\,
 \oint_\gpath \frac{\rmd z}{2\pi} 
 \oint_\gpath \frac{\rmd z'}{2\pi} \, \rme^{-\ci t (z+z')}
\FT{\psi}^\vep(k_0^-)\cdot 
\nonumber \\ & \quad
\Bigl\{\prod_{j=m}^0 \Bigl[ 
  \frac{\ci}{z'+\ci\kappa'_{j,0}+H(k_j^-)} 
\prod_{i=n'_j}^{1}
\Bigl(g(k_j^- ;-z'-\ci\kappa'_{j,2 i-1}) 
\frac{\ci \vep }{z'+\ci\kappa'_{j,2 i}+H(k_j^-)} \Bigr)
\nonumber \\ & \quad
\times v(k_j^-,k_{j+1}^-)^{\1(j\ne m)} \Bigr]  f(k_m)
\prod_{j=0}^m \Bigl[  v(k_{j+1}^+,k_j^+)^{\1(j\ne m)}
\nonumber \\ & \quad
 \prod_{i=1}^{n_j}
\Bigl(\frac{-\ci \vep }{z+\ci\kappa_{j,2 i}-H(k_j^+)}
g(k_j^+ ;z+\ci\kappa_{j,2 i-1}) \Bigr)
  \frac{\ci}{z+\ci\kappa_{j,0}-H(k_j^+)}
\Bigr] \FT{\psi}^\vep(k_0^+)  \Bigr\}.
\end{align}
where the index set $I'_m=\set{0,\ldots,m}$, we have defined $M^0=\1$ 
for all matrices $M$, and
we have used the equality 
\begin{align}
(-\ci)^{\bar{N}_2-\bar{N}_1} \vep^{(\bar{N}_1+\bar{N}_2)/2} = \vep^m \prod_{j=0}^{m}
\left((-\vep)^{n_j} (-\vep)^{n'_j} \right).
\end{align}

The following Lemma, whose proof we postpone for the moment,
is used also in the computation of the limit of the main term. 
\begin{lemma}\label{th:maindiff}
\begin{align}\label{eq:simplebound4}
& \Bigl|   \Kpart(t, S_m(n,n');(N',N_2),(N',N_1),\vep,\kappa,p,f) 
 \nonumber \\ & \qquad
- \Kpmain(t, S_m(n,n'),N',\vep,\kappa,p,f) 
 \Bigr| 
 \le c' \norm{f}_\infty    \Emax
(c \vep \sabs{t})^{\!\frac{\bar{N}}{2}}
 \nonumber \\ & \quad
\times
\Bigsabs{\ln \frac{\sabs{\vep t}}{\vep}}^{3}
(\bar{N}+1) \Bigl(1 + c_1 \frac{|p|}{\sqrt{\vep}}
  + c_2  \frac{\kappa}{\vep} 
 + c_3 \sqrt{1+\kappa/\vep}\Bigr) 
 \frac{ \sabs{\vep t}}{\sabs{t}^{1/2}}.
\end{align}
where 
\begin{align}\label{eq:defKmain}
& \Kpmain(t, S_m(n,n'),N',\vep,\kappa,p,f) =
 \sum_{\sigma',\sigma\in \set{\pm 1}}
  \vep^{m} \int_{(\T^3)^{I'_m}} \!\!\rmd k\,
 \FT{\psi}^\vep_{\sigma'}(k_0^-)^*  
 \FT{\psi}^\vep_{\sigma}(k_0^+)
\nonumber \\ & \quad \times
f_{\sigma'\sigma}(k_m)
(\sigma'\sigma)^{m} \prod_{j=1}^{m} \omega(k_{j})^2
(-\vep \Theta_\sigma(k_0))^{\sum_{j=0}^m n_j}
(-\vep \Theta_{-\sigma'}(k_0))^{\sum_{j=0}^m n'_j}
\nonumber \\ & \quad \times
 K_{(\bar{N}_2+m)/2+1}(t,w'_{\sigma'}(k^-))
 K_{(\bar{N}_1+m)/2+1}(t,w_{\sigma}(k^+)) 
\end{align}
with 
\begin{align}
(w_{\sigma}(k))_{j,i} = \sigma \omega(k_j)-\ci \kappa_{j,2 i}\qand
(w'_{\sigma}(k))_{j,i} = -\sigma \omega(k_j)-\ci \kappa'_{j,2 i}.
\end{align}
\end{lemma}
Then we can finish the proof by using the following upper bound
for $|\Kpmain|$ 
\begin{align}\label{eq:kpmainbnd}
&\norm{f}_\infty \ommax^{2m} \bar{\Theta}^{\bar{N}/2-m}
\vep^{\bar{N}/2}
\Bigl[ \int_{(\T^3)^{I'_m}} \!\!\rmd k\,
 \norm{\FT{\psi}^\vep(k_0)}^2 
|K_{N'_1}(t,w_{+}(k))|^2
 \nonumber \\ & \qquad \times
 \int_{(\T^3)^{I'_m}} \!\!\rmd k\,
 \norm{\FT{\psi}^\vep(k_0)}^2 
| K_{N'_2}(t,w'_{+}(k))|^2 \Bigr]^{\frac{1}{2}}
\end{align}
where $N'_1=(\bar{N}_1+m)/2+1$, $N'_2=(\bar{N}_2+m)/2+1$,  and
$\bar{\Theta} = \sup_{k}|\Theta_+(k)|<\infty$.  To get the bound we 
have applied the
Schwarz inequality, then shifted all integration variables by $\pm p/2$
and, finally, used $|K_{N}(t,w_{-}(k))|=|K_{N}(t,w_{+}(k))|$. 

If $m=0$, we apply Lemma \ref{th:Knprop}:\ref{it:Knbound} to find that
the square root in (\ref{eq:kpmainbnd}) is bounded by
$\Emax t^{\bar{N}/2}((\bar{N}_1/2)! (\bar{N}_2/2)!)^{-1/2}$.  Therefore, 
for $m=0$,
\begin{align}\label{eq:Kpmb1}
& |\Kpmain(t, S_m(n,n'),N',\vep,\kappa,p,f)|\le
 \frac{ \Emax \norm{f}_\infty (c \vep t)^{\bar{N}/2}}{
 [((\bar{N}_1+m)/2)!((\bar{N}_2+m)/2)!]^{1/2}} 
\end{align}
with $c=\ommax^2 \bar{\Theta}$.  Let then $m>1$.   Denoting 
$N=N'_1$, we need to inspect
\begin{align}
& |K_{N}(t,w_{+}(k))|^2
= \int_{\R_+^{N}}\! \rmd s \,
  \delta\Bigl(t-\sum_{\ell=1}^{N} s_\ell\Bigr)
\int_{\R_+^{N}}\! \rmd s' \,
  \delta\Bigl(t-\sum_{\ell=1}^{N} s'_\ell\Bigr)
 \nonumber \\ & \qquad \times
 \prod_{\ell=1}^N \rme^{-\ci (s_\ell-s'_\ell) \omega(k_{j(\ell)})}
 \prod_{\ell=1}^N
 \rme^{-\kappa_{j(\ell),i(\ell)} (s_\ell+s'_\ell)}.
\end{align}
where $(j(\ell),i(\ell))$ define the natural index mapping from $I_N$ to
allowed $(j,i)$ such that $1\mapsto (0,0)$, $r_1+1\mapsto (1,0)$, etc.
Then we use Fubini's theorem to integrate out $k_j$ with $j\ge 1$, and
estimate the integral by (DR\ref{it:suffdisp}).  This shows that
\begin{align}\label{eq:KL2est}
 &\int_{(\T^3)^{I_m}} \!\!\rmd k\, |K_{N}(t,w_{+}(k))|^2
\le \comega^m \int_{\R_+^{N}}\! \rmd s \,
  \delta\Bigl(t-\sum_{\ell=1}^{N} s_\ell\Bigr)
\int_{\R_+^{N}}\! \rmd s' \,
  \delta\Bigl(t-\sum_{\ell=1}^{N} s'_\ell\Bigr)
 \nonumber \\ & \qquad
 \prod_{j=1}^m \Bigl\langle
 \sum_{i=0}^{n_j}(s_{\ell(j,i)}-s'_{\ell(j,i)})\Bigr\rangle^{-\frac{3}{2}}.
\end{align}
Let us next define
\begin{align}\label{eq:stoab}
  a_\ell = \vep\frac{s_\ell+s'_\ell}{2}\qand b_\ell = s_\ell - s'_\ell.
\end{align}
when $s_\ell=a_\ell/\vep +\frac{1}{2} b_\ell$, 
$s'_\ell = a_\ell/\vep  -\frac{1}{2} b_\ell$.
If we first resolve the delta-functions by integrating out $s_1$ and
$s'_1$, and then make the above change of variables, the Jacobian is 
$\vep^{-(N-1)}$, and we find that the right hand side of 
(\ref{eq:KL2est}) is equal to 
\begin{align}
& \vep^{-(N-1)} \comega^m \int_{\R_+^{N}}\!\!\! \rmd a\, 
\delta\Bigl(\vep t-\sum_{\ell=1}^{N} a_\ell\Bigr) 
\nonumber \\ & \qquad \times 
  \int_{\R^{I_N\setm\set{1}}}\!\!\! \rmd b \,
  \1\Bigl(\Bigl|\sum_{\ell=2}^{N}b_\ell \Bigr| \le 2\frac{a_1}{\vep} \Bigr)
  \prod_{\ell=2}^{N} \1\!\left(|b_\ell|\le 2 \frac{a_\ell}{\vep}\right)
 \prod_{j=1}^m \Bigl\langle
 \sum_{i=0}^{n_j} b_{\ell(j,i)}\Bigr\rangle^{-\frac{3}{2}}.
\end{align}
Here we use the trivial bound $\1(\cdot)\le 1$ to remove the 
characteristic functions containing $a_{r_j+1}$ for $j=0,1,\ldots,m$, and
estimate the integrals over $b_{r_j+1}$,
$j=1,\ldots,m$, by the bound (\ref{eq:int0bounds}).  
Then we can integrate the remaining $N-1-m$ 
integrals over $b_\ell$, use the bounds $a_\ell\le \vep t$, and then finally 
estimate the $a$-integral by Lemma \ref{th:Knprop}:\ref{it:Knbound}.
This shows that
\begin{align}
 &\int_{(\T^3)^{I_m}} \!\!\rmd k\, |K_{N}(t,w_{+}(k))|^2
\le \vep^{-(N-1)}\comega^m 6^m (4 t)^{N-1-m} \frac{ (\vep t)^{N-1}}{(N-1)!}
 \nonumber \\ & \quad 
\le \frac{(6\comega t)^{\bar{N}_1}}{((\bar{N}_1+m)/2)!}.
\end{align}
where we used $N-1=(\bar{N}_1+m)/2$.  Since the above argument works for any
partition $n$ and $\kappa_{i,j}$, we can now also conclude that
\begin{align}
 &\int_{(\T^3)^{I_m}} \!\!\rmd k\, |K_{N'_2}(t,w'_{+}(k))|^2
\le \frac{(6\comega t)^{\bar{N}_2}}{((\bar{N}_2+m)/2)!}.
\end{align}
Therefore (\ref{eq:Kpmb1}) is valid also in this case for
$c=6 \comega\ommax^2 \bar{\Theta}$  which is larger than the $c$ for the 
$m=0$ case. Combined with Lemma \ref{th:maindiff} 
we obtain (\ref{eq:simplebound}) and this finishes the proof of Lemma
\ref{th:simpleest}.

We still need to prove Lemma \ref{th:maindiff}. This
will be based on the following result which shows that
removing any of the denominators improves the estimate:
\begin{lemma}\label{th:missingpropag}
For any $\sigma_{j,i}, \sigma'_{j,i}= \pm 1$,
the following integral
\begin{align}\label{eq:absval}
 & \int_{(\T^3)^{I'_m}} \!\!\rmd k\,
 |\FT{\psi}^\vep_{\sigma'_{0,0}}\!(k_0^-)|
 |\FT{\psi}^\vep_{\sigma_{0,0}^{\phantom a}}\!(k_0^+)| 
 \oint_\gpath \frac{|\rmd z|}{2\pi} 
 \oint_\gpath \frac{|\rmd z'|}{2\pi}
 \nonumber \\ & \qquad \times
 \prod_{j=0}^{m} \Bigl(
   \prod_{i=0}^{n_j}  \frac{1}{|z+\ci \kappa_{j,2 i}-\sigma_{j,i} \omega(k^+_j)|}
   \prod_{i=0}^{n'_j}  
   \frac{1}{|z'+\ci \kappa'_{j,2 i}+\sigma'_{j,i} \omega(k^-_j)|}
 \Bigr).
\end{align}
with any $0<\beta\le 1-\kappa$ is bounded by
\begin{align}\label{eq:sbound0}
 \Emax |\gpath|^2  \sabs{\ln \beta}^2
 (3 \comega)^m \beta^{-m-\sum_j(n_j+n'_j)}.
\end{align}
If $m+\sum_j n_j > 0$ and
the integrand is multiplied 
by $|z+\ci \kappa_{j,2 i}-\sigma_{j,i} \omega(k^+_j)|$
for some pair of indices $j\in I'_m$, $i\in\set{0,\ldots,n_j}$, then
the integral has an upper bound which is given by  
(\ref{eq:sbound0}) times
\begin{align}\label{eq:sbound1}
4\beta \sabs{\ln\beta}.
\end{align}
The same is true whenever $m+\sum_j n'_j > 0$, and
the integrand is multiplied 
by $|z'+\ci \kappa'_{j,2 i}+\sigma'_{j,i} \omega(k^-_j)|$ for
some pair of indices $j\in I'_m$, $i\in\set{0,\ldots,n'_j}$.
\end{lemma}
\begin{proof}
Using Lemma \ref{th:morseprop}:\ref{it:m0},
we find an upper bound 
\begin{align}
 & \beta^{-\sum_j (n_j +n'_j)}
  \int_{(\T^3)^{I'_m}} \!\!\rmd k\,
 |\FT{\psi}^\vep_{\sigma_{0,0}^{\phantom a}}\!(k_0^+)| 
 |\FT{\psi}^\vep_{\sigma'_{0,0}}\!(k_0^-)|
 \oint_\gpath \frac{|\rmd z|}{2\pi} 
 \oint_\gpath \frac{|\rmd z'|}{2\pi}
 \nonumber \\ & \qquad \times
 \prod_{j=0}^{m} \Bigl( \frac{1}{|z+\ci \kappa_{j,0}-\sigma_{j,0} \omega(k^+_j)|}
   \frac{1}{|z'+\ci \kappa'_{j,0}+\sigma'_{j,0} \omega(k^-_j)|}
 \Bigr).
\end{align}
We estimate the $k_j$ integrals for $j=1,\ldots,m$ by
\begin{align}
 & \int_{\T^3} \rmd k_j  \frac{1}{|z+\ci \kappa_{j,0}-\sigma_{j,0} \omega(k^+_j)|}
   \frac{1}{|z'+\ci \kappa'_{j,0}+\sigma'_{j,0} \omega(k^-_j)|}
\le \frac{3 \comega}{\beta}
\end{align}
which follows from the Schwarz inequality and Lemma 
\ref{th:morseprop}:\ref{it:mk2}.
Then we estimate $z$ and $z'$-integrals by
Lemma \ref{th:morseprop}:\ref{it:mz1}, after which 
the remaining $k_0$-integral can be bound by
the Schwarz inequality.  This proves (\ref{eq:sbound0}). 

Assume then that $m+\sum_j n_j > 0$, for some index pair
$j,i$.  If $n_j>0$, the only change needed to be made to the
above steps is to retain one of the
remaining factors depending on $k_j$.
This will yield a bound which is better than
(\ref{eq:sbound0}) by a full factor of $\beta$.  
If $n_j=0$, we necessarily have $m>0$.  If $j=0$, let $j'=1$, otherwise let
$j'=j$.  We use the trivial estimate for all terms with $i >0$,
and estimate also the remaining factors independent of $k_0$ and $k_{j'}$
as before.  Then we can apply Lemma \ref{th:morseprop} to estimate the
remaining integrals in the following order:
first the $z$-integral, then the $k_{j'}$-integral, $z'$-integral, 
and finally $k_0$-integral.
This yields a bound which is 
(\ref{eq:sbound0}) times $4\beta \sabs{\ln\beta}\ge \beta$. 
The remaining case, where a $z'$-factor is cancelled instead of a $z$-factor
follows by identical reasoning.
\end{proof}
\begin{proofof}{Lemma \ref{th:maindiff}}
Let us begin by writing the $2\times 2$ matrix product in
(\ref{eq:notsosimplesum2}) in component form, and let 
$\sigma_{j,i }$ and $\sigma'_{j,i }$ denote the component attached to the
factor with $\kappa_{j,i}$, respectively $\kappa'_{j,i}$.  We also use
$\beta=(2\sabs{t/2})^{-1}$, as before.  For the absolute value of 
any term in the resulting 
sum over $\sigma'$ and $\sigma$ we then have an upper bound:
\begin{align}
 \norm{f}_\infty  \vep^{\frac{\bar{N}}{2}} \bar{g}^{\frac{\bar{N}}{2}-m} \ommax^{2 m}
  \rme^2 \times \text{(\ref{eq:absval})}
\end{align}
where $\bar{g}$ is the finite constant in (\ref{eq:defbarg}), for which also
$\sup_k |\Theta_+(k)|\le \bar{g}$.

Suppose that there is an index pair $(j,i)$ such that 
$\sigma_{j,i}=-\sigma_{0,0}$.  Then we take the absolute value inside the
integrals where, similarly to (\ref{eq:oneoverab}), we apply the
inequality 
\begin{align}
& \frac{1}{|z+\ci \kappa_{j,2 i}-\sigma_{j,i} \omega(k^+_j)|}
\frac{1}{|z+\ci \kappa_{0,0}-\sigma_{0,0} \omega(k^+_0)|}
 \nonumber \\ & \quad 
\le \frac{1}{2 \ommin} \Bigl(
\frac{1}{|z+\ci \kappa_{j,2 i}-\sigma_{j,i} \omega(k^+_j)|} +
\frac{1}{|z+\ci \kappa_{0,0}-\sigma_{0,0} \omega(k^+_0)|}
\Bigr) .
\end{align}
Since $(j,i)\ne (0,0)$, we must have $\bar{N}_1>0$, and we can 
apply Lemma \ref{th:missingpropag}.  This yields an upper bound
$\frac{2}{\ommin}\beta \sabs{\ln\beta}$ times
\begin{align}\label{eq:finalupperb}
   c''  \norm{f}_\infty    \Emax
(c \sabs{\vep t})^{\bar{N}/2} \sabs{\ln \frac{\sabs{\vep t}}{\vep}}^{2}
\end{align}
where $c=\sabs{3 \comega \ommax^2} \sabs{\bar{g}}$ and 
$c''= \rme^2 4 (2 \ommax+5)^2$ depend only on $\omega$.
The same estimate is valid also whenever there is
an index pair $(j,i)$ such that  $\sigma'_{j,i}=-\sigma'_{0,0}$.

Therefore, the sum over all those sign combinations which do not have
constant $\sigma$ and $\sigma'$ is bounded by (\ref{eq:finalupperb}) 
times
\begin{align}\label{eq:firstextra}
2^{\bar{N}}\frac{2^3}{\ommin} \beta \sabs{\ln \beta}.
\end{align}
Thus we have proven that up to such an error, $\Kpart$ is equal to 
\begin{align}\label{eq:goingtoKmain}
&
 \sum_{\sigma',\sigma\in \set{\pm 1}}
  \vep^{m} \int_{(\T^3)^{I'_m}} \!\!\rmd k\,
f_{\sigma'\sigma}(k_m)
 \FT{\psi}^\vep_{\sigma'}(k_0^-)^*  
 \FT{\psi}^\vep_{\sigma}(k_0^+)
 \oint_\gpath \frac{\rmd z}{2\pi} 
 \oint_\gpath \frac{\rmd z'}{2\pi} \, \rme^{-\ci t (z+z')}
\nonumber \\ & \qquad \times
\prod_{j=0}^m \Bigl[ 
  \frac{\ci}{z'+\ci\kappa'_{j,0}+\sigma'\omega(k_j^-)} 
\prod_{i=1}^{n'_j}
\Bigl(
\frac{\ci \vep g_{\sigma'\sigma'}(k_j^- ;-z'-\ci\kappa'_{j,2 i-1}) 
 }{z'+\ci\kappa'_{j,2 i}+\sigma'\omega(k_j^-)} \Bigr)
\nonumber \\ & \qquad
 \times 
  \frac{\ci}{z+\ci\kappa_{j,0}-\sigma\omega(k_j^+)}
\prod_{i=1}^{n_j}
\Bigl(\frac{-\ci \vep g_{\sigma\sigma}(k_j^+ ;z+\ci\kappa_{j,2 i-1})}{
z+\ci\kappa_{j,2 i}-\sigma\omega(k_j^+)} \Bigr)
\Bigr]
\nonumber \\ & \qquad
\times
\prod_{j=0}^{m-1} \Bigl[ 
 v_{\sigma'\sigma'}(k_j^-,k_{j+1}^-)
v_{\sigma\sigma}(k_{j+1}^+,k_j^+)
\Bigr] .
\end{align}
If $\bar{N}=0$, then this formula is equal to (\ref{eq:defKmain}). 
Otherwise, we can 
express the two $K$ factors in (\ref{eq:defKmain}) as integrals over 
$z'$ and $z$.  This yields a formula which would be equal to
(\ref{eq:goingtoKmain}) if we could change each $v_{\sigma\sigma}$ to 
$\sigma \omega$, and each $g_{\sigma\sigma}$ to $\Theta_\sigma$ there.
However, we can do these changes one by one and compute an upper bound for 
$|\mbox{(\ref{eq:goingtoKmain})}-\Kpmain|$ using the
following estimates:
\begin{align}
& \left|v_{\sigma\sigma}(k_{j+1}^+,k_{j}^+)-\sigma \omega(k_{j+1})\right|
\le |\omega(k^+_{j+1})-\omega(k_{j+1})| 
 + \frac{1}{2} |\kappa_{j,0}-\kappa_{j+1,0}|
\nonumber \\ & \qquad 
 + \frac{1}{2}|z+\ci\kappa_{j,0}-\sigma \omega(k^+_{j})|
 + \frac{1}{2}|z+\ci\kappa_{j+1,0}-\sigma \omega(k^+_{j+1})|
\nonumber \\ & \quad 
\le \frac{\norm{\nabla \omega}_\infty}{2} |p| + 
\frac{1}{2} \kappa
 + \frac{1}{2}|z+\ci\kappa_{j,0}-\sigma \omega(k^+_{j})|
 + \frac{1}{2}|z+\ci\kappa_{j+1,0}-\sigma \omega(k^+_{j+1})|,
\end{align}
and a similar estimate for 
$|v_{\sigma'\sigma'}(k_{j}^-,k_{j+1}^-)-\sigma'\omega(k_{j+1})|$.
By Lemma \ref{th:gateint},
\begin{align}
& \left|g_{\sigma\sigma}(k_j^+ ;z+\ci\kappa_{j,2 i-1})-\Theta_{\sigma}(k_0)\right|
\le c'_1 |\omega(k_0^+)-\omega(k_0)|
\nonumber \\ & \qquad 
+ c'_1 |z+\ci \kappa_{0,0}-\sigma \omega(k^+_0)|
+ (c'_1+c'_2/\beta^{1/2}) |z+\ci \kappa_{j,2 i}-\sigma \omega(k^+_j)|
\nonumber \\ & \qquad
+ c'_1 |\kappa_{j,2 i}-\kappa_{0,0}|
+ c'_2 |\kappa_{j,2 i}-\kappa_{j,2 i-1}|/\beta^{1/2}
+c'_3\sqrt{\beta+\kappa} 
\nonumber \\ & \quad 
\le c''_1 |z+\ci \kappa_{0,0}-\sigma \omega(k^+_0)|
+ \frac{c''_2}{\beta^{1/2}} |z+\ci \kappa_{j,2 i}-\sigma \omega(k^+_j)|
\nonumber \\ & \qquad
+  \frac{c''_2}{\beta^{1/2}} \kappa + c''_3 \sqrt{\beta+\kappa} 
+ c''_4 |p|  
\end{align}
where all the constants $c''_i$ depend only on $\omega$.
Since $-g_{\sigma\sigma}(k;-w)=g_{-\sigma,-\sigma}(k;w)$, a similar bound
is valid also for $|-g_{\sigma'\sigma'}(k_j^- ;-z-\ci\kappa'_{j,2 i-1})
  -\Theta_{-\sigma'}(k_0)|$.

We have to iterate $2 m$ times
the change of $v$ and $\bar{N}/2- m$ times the change of $g$.
Collecting the estimates, and applying Lemma \ref{th:missingpropag} 
when needed, shows that $|\mbox{(\ref{eq:goingtoKmain})}-\Kpmain|$
is bounded by
(\ref{eq:finalupperb}) times 
\begin{align}\label{eq:finalextra}
\bar{N} \sqrt{\beta} \Bigl(c_1 \frac{|p|}{\sqrt{\beta}}
  + c_2 \sabs{\ln \beta}
  + c_3 \frac{\kappa}{\beta} + c_4 \sqrt{1+\kappa/\beta}\Bigr)
\end{align}
where the constants $c_i$ depend only on $\omega$.  Then the terms
containing $\beta$ can be bounded from above as before and,
together with (\ref{eq:firstextra}) and after a
redefinition of the constants,
this proves (\ref{eq:simplebound4}).
\end{proofof}

\subsection{Convergence of the main term}
\label{sec:maintermconv}

It will be enough to study the limit of a sum of functions
$\Kpmain$ defined in (\ref{eq:kpmainbnd}), more precisely, the limit of 
\begin{align}\label{eq:Kpmainbegin}
 & \sum_{N_1,N_2=0}^{N_0(\vep)-1} \sum_{m=0}^{N_0(\vep)-1}\!\! 
\sum_{n,n'\in \N^{I'_m}} 
 \1\Bigl(N_1=m+2\sum_{j=0}^{m} n_j\Bigr)
 \1\Bigl(N_2=m+2\sum_{j=0}^{m} n'_j\Bigr) 
\nonumber \\ & \qquad \times
\Kpmain(\tmacro/\vep, S_m(n,n'),0,\vep,0,\vep\pmacro,e_{\nmacro}P_{++}) .
\end{align}
To see this, first note that the difference between this and
\begin{align}\label{eq:sumsimplie}
\sum_{N_1,N_2=0}^{N_0-1}  \sum_{\substack{S\in\pi(I_{N_1,N_2}),\\ S \text{ simple}}}
  \Kpart (\tmacro /\vep,S;(0,N_2),(0,N_1),\vep,0,\vep\pmacro,e_{\nmacro}P_{++}) 
\end{align}
is by Lemma \ref{th:maindiff} bounded by 
\begin{align}
& N_0^2 c'  \Emax (c \sabs{\tmacro})^{\!\frac{\bar{N}}{2}}
\Bigsabs{\ln \frac{\sabs{\tmacro}}{\vep}}^{3}
(\bar{N}+1) (1 + c_3+ c_1 \sqrt{\vep}|\pmacro|)
 \frac{ \sabs{\tmacro}}{\sabs{\tmacro/\vep}^{1/2}}
\end{align}
for some constants $c,c'$ and $\bar{N}=2(N_0-1)$. 
The bound goes to zero as $\vep\to 0$.
Secondly, by Corollaries \ref{th:EFF} and \ref{th:nonsimple},
for all $0<\vep\le \vep'$ and $N_1+N_2>0$, the difference between
(\ref{eq:sumsimplie}) 
and $\Fmain(\pmacro,\nmacro,\tmacro) $ is bounded by 
\begin{align}\label{eq:nonsimplebound2a}
& \sum_{N_1,N_2=0}^{N_0-1}  
\sum_{\substack{S\in\pi(I_{N_1,N_2}),\\ S \text{ not simple}}}
 \prod_{A\in S} |C_{|A|}| \, 
  |\Kpart (\tmacro/\vep,S;
  (0,N_2),(0,N_1),\vep,0,\vep\pmacro,e_{\nmacro}P_{++}) |
\nonumber \\ & \quad
\le  N_0^2 c' \Emax
  \left(c\sabs{\tmacro}\right)^{\!\frac{\bar{N}}{2}}
  \bar{N}!
  \Bigsabs{\ln \frac{\sabs{\tmacro}}{\vep}}^{\bar{N}+\max(2,d_2)}
  \bar{N}^{d_1} 
  \Bigl(\frac{\sabs{\tmacro}}{\sabs{\tmacro /\vep}}\Bigr)^{\gamma'}.
\end{align}
for some constants $c,c'$ and $\bar{N}=2(N_0-1)$.   Also this bound
vanishes as $\vep\to 0$, and it is thus sufficient to study the limit of 
(\ref{eq:Kpmainbegin}).

For any $n\in \N$,
\begin{align}
 1 =  \sum_{R\in \N} \1(R=n) = \sum_{R\in \N}
\int_0^1\rmd \varphi\, \rme^{\ci 2 \pi \varphi (R-n)}.
\end{align}
We insert this identity twice
into  (\ref{eq:Kpmainbegin}),
with $n=\sum_j n_j$ and with 
$n=\sum_j n'_j$.  Then we can perform the 
sums over $N_1$ and $N_2$.
We express the two $K$-factors again as
integrals over $z$ and $z'$, but this time choosing
$\beta = 2 \vep \bar{g}$, with $\bar{g}$ defined in (\ref{eq:defbarg}). 
This shows that (\ref{eq:Kpmainbegin}) equals
\begin{align}\label{eq:onlyafewmore}
& \sum_{m=0}^{N_0-1}
 \sum_{n,n'\in \N^{m+1}}  \sum_{R \in \N^2} 
\1(m+2 R_1\le N_0-1)\1(m+2 R_2\le N_0-1)
\nonumber \\ & \quad \times 
\int_{[0,1]^2} \rmd \varphi \, \rme^{\ci 2 \pi R\cdot \varphi}
  \vep^{m} \int_{(\T^3)^{I'_m}} \!\!\rmd k\,
 \FT{\psi}^\vep_{+}(k_0^-)^*  
 \FT{\psi}^\vep_{+}(k_0^+)
\rme^{\ci 2\pi \nmacro\cdot k_m} \prod_{j=1}^{m} \omega(k_{j})^2
\nonumber \\ & \quad \times
(-\vep \Theta_{+}(k_0)\rme^{-\ci 2\pi \varphi_1})^{\sum_{j=0}^m n_j}
(-\vep \Theta_{-}(k_0)\rme^{-\ci 2\pi \varphi_2})^{\sum_{j=0}^m n'_j}
\nonumber \\ & \quad \times
 \oint_\gpath \frac{\rmd z}{2\pi} 
 \oint_\gpath \frac{\rmd z'}{2\pi} \, \rme^{-\ci \frac{\tmacro}{\vep} (z+z')}
 \prod_{j=0}^{m}  \Bigl[ \Bigl( \frac{\ci}{z-\omega(k^+_j)} \Bigr)^{n_j+1}
  \Bigl(\frac{\ci}{z'+\omega(k^-_j)} \Bigr)^{n'_j+1}\Bigr] 
\end{align}
where $k^{\pm}_j = k_j \pm \frac{1}{2}\vep \pmacro $.
By our choice of $\beta$, we have for all $\vep\le (2\bar{g})^{-1}$ and
$z\in\gpath$, $|\vep\Theta(k_0)|/|z\pm\omega(k)|\le 1/2$.  
This implies that for such $\vep$ the sums over $n$ and $n'$ are
absolutely summable, and we can use Fubini's theorem to perform them first.
As for any $a,b\in \C$ such that $|b|<|a|$,
\begin{align}
  \sum_{n\in \N} \frac{1}{a} \left(\frac{b}{a}\right)^n
  = \frac{1}{a-b},
\end{align}
we find that the last two lines of (\ref{eq:onlyafewmore}), summed over $n$
and $n'$,  become
\begin{align}
 & \oint_\gpath\! \frac{\rmd z}{2\pi} 
 \oint_\gpath\! \frac{\rmd z'}{2\pi} \,  \rme^{-\ci \frac{\tmacro}{\vep} (z+z')}
 \prod_{j=0}^{m} \Bigl( 
 \frac{\ci}{z-\omega(k^+_j)+\ci\vep \Theta_+(k_0)\rme^{-\ci 2\pi \varphi_1}}
\nonumber \\ & \qquad \times
 \frac{\ci}{z'+\omega(k^-_j)+\ci\vep \Theta_-(k_0)\rme^{-\ci 2\pi \varphi_2}}
 \Bigr) .
\end{align}
We use here Theorem \ref{th:Knprop} to evaluate the
$z$ and $z'$ integrals and insert the result in 
(\ref{eq:onlyafewmore}).  
For any $a\in \C$, and $R\in \N$,
\begin{align}
 \int_0^1\rmd \varphi\, \rme^{\ci 2 \pi \varphi R}
 \exp\!\left( a\rme^{-\ci 2\pi \varphi}\right)
  = \frac{a^R}{R!},
\end{align}
which can be proven, e.g., by a series expansion.  Using this to evaluate
the $\varphi_1$ and $\varphi_2$ integrals, we 
arrive at the following expression for 
(\ref{eq:onlyafewmore})
\begin{align}\label{eq:nearly2}
 & \sum_{m=0}^{N_0-1}
 \sum_{R \in \N^2} \1(m+2 R_1\le N_0-1)\1(m+2 R_2\le N_0-1)
\nonumber \\ &  \times 
 \int_{(\T^3)^{I'_m}} \!\!\rmd k\,
 \FT{\psi}^\vep_{+}(k_0^-)^*  
 \FT{\psi}^\vep_{+}(k_0^+)
 \prod_{j=1}^{m} \omega(k_{j})^2  \rme^{\ci 2\pi \nmacro\cdot k_m}
  \frac{(-\tmacro\Theta_+(k_0))^{R_1}}{R_1!}
  \frac{(-\tmacro\Theta_-(k_0))^{R_2}}{R_2!}
\nonumber \\ & \times
   \vep^{m} \int_{\R_+^{I'_m}}\! \rmd s'  \int_{\R_+^{I'_m}}\! \rmd s \,
  \delta\Bigl(\frac{\tmacro}{\vep}-\sum_{j=0}^{m} s_j\Bigr) 
  \delta\Bigl(\frac{\tmacro}{\vep}-\sum_{j=0}^{m} s'_j\Bigr) 
\rme^{-\ci \sum\limits_{j=0}^{m}
    \left[ s_j \omega(k^+_j) - s'_j \omega(k^-_j) \right]} .
\end{align}

The $m=0$ term in the sum is equal to 
\begin{align}
& \sum_{R \in \N^2} 
\1(2 R_1\le N_0-1)\1(2 R_2\le N_0-1)
 \int_{\T^3} \!\!\rmd k_0\,
 \FT{\psi}^\vep_{+}(k_0^-)^*
 \FT{\psi}^\vep_{+}(k_0^+)
\nonumber \\ & \qquad \times 
  \frac{(-\tmacro\Theta_+(k_0))^{R_1}}{R_1!}
  \frac{(-\tmacro\Theta_-(k_0))^{R_2}}{R_2!}
  \rme^{\ci 2\pi \nmacro\cdot k_0} 
\rme^{-\ci \tmacro ( \omega(k^+_0) - \omega(k^-_0) )/\vep}.
\end{align}
For all $k_0$,
\begin{align}\label{eq:eme}
& 
\Bigl| \rme^{-\ci  \tmacro \bigl(
       \frac{\omega(k^+_0) - \omega(k^-_0) }{\vep} 
       - \pmacro\cdot \nabla \omega(k_0)\bigr)} -1 \Bigr| \le
\vep\frac{1}{4}\tmacro \bar{p}^2 \norm{D^2\omega}_\infty
\end{align}
which allows
replacing the last exponential by $\rme^{-\ci \tmacro p\cdot \nabla \omega(k_0)}$
with an error which vanishes in the $\vep \to 0$ limit.
The remaining integrand is $\vep$-independent, apart from the
$\FT{\psi}^\vep$ factors.  Dominated convergence can be applied to
take the $\vep\to 0$ limit inside the $R$-sum, where 
$N_0(\vep)\to \infty$ and, as $\Theta$ is continuous by Lemma 
\ref{th:gateint}, we can apply Lemma \ref{th:winitlim} and obtain the limit 
\begin{align}
& \sum_{R \in \N^2}  \int_{\T^3}\! \mu_{0}(\rmd x\, \rmd k)
\rme^{-\ci 2 \pi (\pmacro\cdot x-\nmacro\cdot k)}
\rme^{-\ci \tmacro \pmacro\cdot \nabla \omega(k)}
  \frac{(-\tmacro  \Theta_+(k))^{R_1}}{R_1!}
  \frac{(-\tmacro  \Theta_-(k))^{R_2}}{R_2!}
\nonumber \\ \quad & =
 \int_{\T^3}\! \mu_{0}(\rmd x\, \rmd k)
\rme^{-\ci 2 \pi (\pmacro\cdot x-\nmacro\cdot k)}
\rme^{-\ci \tmacro \pmacro\cdot \nabla \omega(k)}
\rme^{-2 \tmacro \re \Theta_+\!(k)}.
\end{align}
The equality follows from Fubini's theorem, which allows swapping the 
$R$-sum and the $k$-integral, and then using $\Theta_-=\Theta_+^*$.

Consider then the remaining case $m\ge 1$.  We make the same change of
variables as in (\ref{eq:stoab}),
\begin{align}
  r_j = \vep\frac{s_j+s'_j}{2}\qand b_j = s_j - s'_j
\end{align}
when $s_j=r_j/\vep +\frac{1}{2} b_j$, 
$s'_j = r_j/\vep  -\frac{1}{2} b_j$. The Jacobian is now 
$\vep^{-m}$, which cancels the remaining $\vep$-factors, and the
last line of (\ref{eq:nearly2}) becomes
\begin{align}
& \int_{\R_+^{I'_m}}\!\!\! \rmd r\, 
\delta\Bigl(\tmacro-\sum_{j=0}^{m} r_j\Bigr) 
  \int_{\R^{I_m}}\!\!\! \rmd b \,
  \1\Bigl(\Bigl|\sum_{j=1}^{m}b_j \Bigr| \le 2\frac{r_0}{\vep} \Bigr)
  \prod_{j=1}^{m} \1\!\left(|b_j|\le 2 \frac{r_j}{\vep}\right)
\nonumber \\ & \qquad \times 
\rme^{\ci \frac{\omega(k^+_0) + \omega(k^-_0)}{2} \sum_{j=1}^m b_j }
\prod_{j=1}^{m}\rme^{-\ci b_j \frac{\omega(k^+_j) + \omega(k^-_j)}{2}}
\prod_{j=0}^{m}\rme^{-\ci r_j \frac{ \omega(k^+_j) - \omega(k^-_j)}{\vep}} .
\end{align}
For any $\vep>0$ the
integration region over $(r,b)$ is bounded which allows using Fubini's
theorem and
performing first the $k$ integrals.  Therefore, for $m\ge 1$
the summand in (\ref{eq:nearly2}) is equal to
\begin{align}\label{eq:almostthere}
& \sum_{R \in \N^2} 
\1(m+2 R_1\le N_0-1)\1(m+2 R_2\le N_0-1)
\nonumber \\ & \quad \times 
 \int_{\R_+^{I'_m}}\!\!\! \rmd r\, 
\delta\Bigl(\tmacro-\sum_{j=0}^{m} r_j\Bigr) 
  \int_{\R^{I_m}}\!\!\! \rmd b \,
  \1\Bigl(\Bigl|\sum_{j=1}^{m}b_j \Bigr| \le 2\frac{r_0}{\vep} \Bigr)
  \prod_{j=1}^{m} \1\!\left(|b_j|\le 2 \frac{r_j}{\vep}\right)
\nonumber \\ & \quad \times 
 \int_{\T^3} \!\rmd k_0\,
 \FT{\psi}^\vep_{+}(k_0^-)^*  
 \FT{\psi}^\vep_{+}(k_0^+)
  \frac{(-\tmacro\Theta_+(k_0))^{R_1}}{R_1!}
  \frac{(-\tmacro\Theta_-(k_0))^{R_2}}{R_2!}
\nonumber \\ & \quad\times
\rme^{\ci \frac{\omega(k^+_0) + \omega(k^-_0)}{2} \sum_{j=1}^m b_j }
\rme^{-\ci r_0 \frac{ \omega(k^+_0) - \omega(k^-_0)}{\vep}}
 \int_{(\T^3)^{I_m}} \!\!\rmd k\,
 \rme^{\ci 2\pi \nmacro\cdot k_m}
\nonumber \\ & \quad\times
 \prod_{j=1}^{m} \Bigl( \omega(k_{j})^2  
\rme^{-\ci b_j \frac{\omega(k^+_j) + \omega(k^-_j)-2 \omega(k_j)}{2}}
\rme^{-\ci r_j \frac{ \omega(k^+_j) - \omega(k^-_j)}{\vep}} 
\rme^{-\ci b_j \omega(k_j)} \Bigr).
\end{align}

For any multi-index $\alpha$, differentiation with respect to $k$
satisfies
\begin{align}
  D^{\alpha}\! \left(\omega(k^+) - \omega(k^-)\right)
 = D^{\alpha} \omega(k^+) - D^{\alpha}\omega(k^-)
\end{align}
which, by the smoothness of $\omega$, is bounded by 
$\vep \bar{p}\norm{\omega}_{\infty,|\alpha|+1}$. When
multiplied with $r_j/\vep$, the bound remains uniformly bounded in $\vep$.
Similarly, 
\begin{align}
 & D^{\alpha} \!\left(\omega(k^+) - \omega(k)
    + \omega(k^-) - \omega(k) \right)
\nonumber \\ &\quad = 
 D^{\alpha} \omega(k^+) -  D^{\alpha} \omega(k)
    +  D^{\alpha}\omega(k^-) -  D^{\alpha}\omega(k) 
\end{align}
is bounded by $\frac{1}{4}\vep^2\pmacro^2 \norm{\omega}_{\infty,|\alpha|+2}$,
and thus, when multiplied by $b_j$, it is bounded by 
$c_2\tmacro \vep \pmacro^2$ where $c_2$ is a constant independent of
$\vep$. Applying (DR\ref{it:suffdisp}) we thus find that there is
a constant $c'$ such that the $k_j$
integral in (\ref{eq:almostthere}) 
is for any $j=1,\ldots,m$ bounded by $c'/\sabs{b_j}^{3/2}$ which is
integrable over $b_j$.  
Therefore, by a similar argument as in the $m=0$ case, we can now use
\begin{align}
&\rme^{\ci \frac{\omega(k^+_0) + \omega(k^-_0)}{2} \sum_{j=1}^m b_j }
\rme^{-\ci r_0 \frac{ \omega(k^+_0) - \omega(k^-_0)}{\vep}}
= \rme^{\ci \omega(k_0) \sum_j b_j }
\rme^{-\ci r_0 \pmacro \cdot\nabla \omega(k_0)} 
+\order{\vep}
\end{align}
to remove the $\vep$-dependence from this term.

Let us then consider the sum over all $m=1,\ldots,N_0-1$.
We apply the above bounds to justify using dominated convergence 
to take the $\vep\to 0$ limit up to inside the $b$-integrals (for the
sum over $m$, note that due to the $r$-integral each term has
an upper bound of the type $(c \tmacro)^m/m!$).  Applying Lemma
\ref{th:winitlim}, we then find that the sum over
these $m$ converges to 
\begin{align}
& \sum_{m=1}^\infty \sum_{R \in \N^2} 
 \int_{\R_+^{I'_m}}\!\!\! \rmd r\, 
\delta\Bigl(\tmacro-\sum_{j=0}^{m} r_j\Bigr) 
  \int_{\R^{I_m}}\!\!\! \rmd b \,
\prod_{j=1}^{m-1}
 \int_{\T^3} \!\!\rmd k_j\, \omega(k_{j})^2 
 \rme^{-\ci b_j \omega(k_j)}
 \rme^{-\ci r_j \pmacro\cdot \nabla \omega(k_j)}
\nonumber \\ & \quad \times 
 \int_{\T^3} \!\!\rmd k_m\,
\rme^{\ci 2\pi \nmacro\cdot k_m}  \omega(k_{m})^2 
 \rme^{-\ci b_m \omega(k_m)}
 \rme^{-\ci r_m \pmacro\cdot \nabla \omega(k_m)}
\int \mu_{0}(\rmd x\, \rmd k_0)
\rme^{-\ci 2 \pi \pmacro\cdot x}
\nonumber \\ & \quad\times
 \rme^{\ci \omega(k_0) \sum_j b_j }
 \rme^{-\ci r_0 \pmacro \cdot\nabla \omega(k_0)} 
  \frac{(-\tmacro\Theta_+(k_0))^{R_1}}{R_1!}
  \frac{(-\tmacro\Theta_-(k_0))^{R_2}}{R_2!}. 
\end{align}
Since $\mu_0$ is a bounded Borel measure, we can apply Fubini's theorem here to
reorder the integrals so that we first perform the $k_j$ integrals for
$j=1,\ldots,m$, then the sum over $R$,
then all $b$-integrals, and 
finally the $\mu_0$ integral.  This shows that the above sum is equal to 
\begin{align}
& \sum_{m=1}^\infty \int_{\R_+^{I'_m}}\!\!\! \rmd r\, 
\delta\Bigl(\tmacro-\sum_{j=0}^{m} r_j\Bigr) 
\int \mu_{0}(\rmd x\, \rmd k_0)
\rme^{-\ci 2 \pi \pmacro\cdot x}
 \rme^{-\ci r_0 \pmacro \cdot\nabla \omega(k_0)} 
 \rme^{-\tmacro 2 \re \Theta_+(k_0)}
\nonumber \\ & \quad \times 
\prod_{j=1}^{m-1}  \int_{-\infty}^\infty\! \rmd b_j \,
 \int_{\T^3} \!\!\rmd k_j\, \omega(k_{j})^2 
 \rme^{-\ci b_j (\omega(k_j)-\omega(k_0))}
 \rme^{-\ci r_j \pmacro\cdot \nabla \omega(k_j)}
\nonumber \\ & \quad \times 
 \int_{-\infty}^\infty\!  \rmd b_m \,
  \int_{\T^3} \!\!\rmd k_m\,
  \rme^{\ci 2\pi \nmacro\cdot k_m}  \omega(k_{m})^2 
  \rme^{-\ci b_m (\omega(k_m)-\omega(k_0))}
  \rme^{-\ci r_m \pmacro\cdot \nabla \omega(k_m)} .
\end{align}
Using the equation (\ref{eq:smoothdelta}) in Proposition
\ref{th:defcrosssect}, and (\ref{eq:reTheta}) in Lemma \ref{th:gateint},
we obtain, by collecting all the results proven in this section,
\begin{align}\label{eq:simplelim}
& \lim_{\vep\to 0}\Fmain(\pmacro,\nmacro,\tmacro) 
\nonumber \\ & \quad 
  = \sum_{m=0}^\infty
\int_{\R_+^{I'_m}}\!\!\! \rmd r\, 
\delta\Bigl(\tmacro-\sum_{j=0}^{m} r_j\Bigr) 
\int \mu_{0}(\rmd x\, \rmd k_0)
 \rme^{-\ci r_0 \pmacro \cdot\nabla \omega(k_0)} 
 \rme^{-\tmacro \sigma(k_0)}
\nonumber \\ & \qquad \times 
\prod_{j=1}^{m-1}  \int_{\T^3} \!\nu_{k_0}(\rmd k_j)
\rme^{-\ci r_j \pmacro\cdot \nabla \omega(k_j)}
  \int_{\T^3} \!\nu_{k_0}(\rmd k_m)
  \rme^{-\ci r_m \pmacro\cdot \nabla \omega(k_m)}
  \rme^{-\ci 2 \pi (\pmacro\cdot x- \nmacro\cdot k_m)}
\nonumber \\ & \quad 
  = \sum_{m=0}^\infty
\int_{\R_+^{I'_m}}\!\!\! \rmd r\, 
\delta\Bigl(\tmacro-\sum_{j=0}^{m} r_j\Bigr) 
\int \mu_{0}(\rmd x\, \rmd k_0)
\int_{\T^3} \!\nu_{k_0}(\rmd k_1)\cdots
\int_{\T^3} \!\nu_{k_m-1}(\rmd k_m)\,
\nonumber \\ & \qquad \times 
\prod_{j=0}^m \rme^{-r_j (\sigma(k_j)+\ci \pmacro\cdot \nabla \omega(k_j))}
  \rme^{-\ci 2 \pi (\pmacro\cdot x- \nmacro\cdot k_m)}
\end{align}
where $\sigma(k)=\nu_k(\T^3)$ is the total collision rate, and
we used Proposition \ref{th:crosssectprop} to derive the second
equality.
The final form is a Dyson series solution to the characteristic function
of the  Boltzmann equation (\ref{eq:Btransporteq}) 
at time $\tmacro$ with the required initial conditions.
This proves that (\ref{eq:Fveptomut2}) holds
and concludes the proof of the main theorem.

\section{Dispersion relation}
\label{sec:dispersion}

To make the main theorem, Theorem \ref{th:main},
a nonempty statement, we still have
to  discuss how the assumptions
(DR\ref{it:DC1}) -- (DR\ref{it:crossing}) 
could be verified for a given dispersion
relation $\omega$.  We will also 
give two explicit examples of elastic couplings
which satisfy the conditions.

The bound (\ref{eq:suffdisp}) follows immediately by standard stationary
phase methods in case $\omega$ is a Morse
function, i.e., if $\omega$ has only isolated,
non-degenerate critical points. For instance, one can then
use a partition of unity to isolate the critical points and then apply
Theorem 7.7.5.\ in \cite{horm:PDE1} which proves the validity of the bound
with $d_1=4$.
The suppression of crossings, (DR\ref{it:crossing}), is
much harder to verify.  It has been shown to be valid 
for the function 
$\sum_{\nu=1}^3 2(1-\cos(2 \pi k^{\nu}))$ in \cite{chen03} with 
$\gamma=1/5$ and $d_2=2$
and, independently, in \cite{erdyau04} with $\gamma=1/4$ and $d_2=6$.  
Therefore, 
$k\mapsto \omega_0^2+\sum_{\nu=1}^3 2(1-\cos(2 \pi k^{\nu}))$ is 
a Morse function satisfying (DR\ref{it:DC1}) -- (DR\ref{it:crossing})  
for any $\omega_0>0$.
We prove in Proposition \ref{th:crossing1},
that the taking of the square root, which is necessary to get the dispersion
relation from the Fourier transform of the elastic couplings, very
generally preserves the Assumptions \ref{th:disprelass}.
In particular, this is then true for
\begin{align}\label{eq:nndisp}
 \omega_{\text{nn}}(k) = \Bigl[\omega_0^2 + \sum_{\nu=1}^3
    2(1-\cos(2 \pi k^{\nu}))\Bigr]^{\frac{1}{2}}
\end{align}
whenever $\omega_0>0$.
Both $\omega_{\text{nn}}(k)$, and $\omega_{\text{nn}}(k)^2$ are
dispersion relations of simple lattice systems.  
$\omega_{\text{nn}}$ corresponds 
to the nearest neighbour elastic couplings, $\alpha(0)=\omega_0^2+6$,
$\alpha(y)=-1$ for $|y|=1$, and $\alpha(y)=0$ otherwise, while
$\omega_{\text{nn}}^2$ corresponds to $\alpha(0)=(\omega_0^2+6)^2+6$,
$\alpha(y)=-2(\omega_0^2+6)$ for $|y|=1$, 
$\alpha(y)=2$ for $|y|=\sqrt{2}$, 
$\alpha(y)=1$ for $|y|=2$, 
and $\alpha(y)=0$ otherwise.

\begin{proposition}\label{th:crossing1}
If $\omega$ is a Morse function which satisfies all of the
Assumptions \ref{th:disprelass}, then $\sqrt{\omega}$
is also a Morse function satisfying them 
with the same value for the parameter $\gamma$.
\end{proposition}
\begin{proof}
Since $\omega\ge \ommin>0$, the function $g(k)=\sqrt{\omega(k)}$ is
well-defined and smooth.  The assumptions also immediately imply that 
$g$ is symmetric and $g\ge\sqrt{\ommin}>0$, and thus $g$ satisfies
(DR\ref{it:DC1}) and (DR\ref{it:DC2}).  As also
\begin{align}
  D g(k) = \frac{1}{2 g(k)} D \omega(k),
\end{align}
the critical points of $g$ and $\omega$ coincide, and if $k_0$ is a critical
point, 
\begin{align}
D^2 g(k_0) = \frac{1}{2 g(k_0)} D^2 \omega(k_0) -
\frac{1}{2 g(k_0)^2} D\omega(k_0)\otimes D\omega(k_0) = 
\frac{1}{2 g(k_0)} D^2 \omega(k_0)
\end{align}
which is non-degenerate since $\omega$ is a Morse function.  This proves
that $g$ is a Morse function, which implies that assumption
(DR\ref{it:suffdisp}) holds. 

Then we only need to check the crossing estimate.  
If $|\alpha_1|\le \sqrt{\ommin}/2$, we can prove (\ref{eq:crossingest})
for the function $g$ using the trivial bound 
\begin{align}
 |\alpha_1-\sigma_1 g(k_1)+\ci\beta|\ge 
 g(k_1)-|\alpha_1|\ge \sqrt{\ommin}/2 
\end{align}
and evaluation of the remaining integrals
by Lemma \ref{th:morseprop}:\ref{it:mz1}. This yields a bound
$\order{\sabs{\ln \beta}^2}$.  
If $|\alpha_1|\ge 2\sqrt{\bar{\omega}}$, we get the same result using
the bound $ |\alpha_1-\sigma_1 g(k_1)+\ci\beta|\ge 
\sqrt{\bar{\omega}}$.  If either 
$|\alpha_i|\le \sqrt{\ommin}/2$ or
$|\alpha_i|\ge 2\sqrt{\bar{\omega}}$, for $i=2$, or $i=3$, 
we get similarly a bound $\order{\sabs{\ln \beta}^2}$.  

Let us then assume that 
$\sqrt{\ommin}/2\le |\alpha_i|\le 2\sqrt{\bar{\omega}}$ for all $i=1,2,3$.
Then we can apply the following bound to all of the three fractions in the
integrand,
\begin{align}
& \frac{1}{|\alpha-\sigma g(k)+\ci\beta|}\le 
\left|\frac{\alpha+\ci\beta+\sigma g(k)}{
    (\alpha+\ci\beta)^2-g(k)^2}\right|\le 
\frac{3\sqrt{\bar{\omega}}+1}{
   \left| \alpha^2-\beta^2-\omega(k)+\ci 2\alpha\beta\right|}
\nonumber  \\ & \quad \le 
\frac{3\sqrt{\bar{\omega}}+1}{
   \left| \alpha^2-\beta^2-\omega(k)+\ci \sqrt{\ommin}\beta\right|} .
\end{align}
This allows using the crossing bound of $\omega$ to prove that of $g$. 
\end{proof}

Finally, let us give a result which could become useful if one needs to check
whether a given dispersion relation satisfies the crossing condition.
We will show that Assumption
(DR\ref{it:crossing}) 
can also be replaced by the following one which should be more accessible
to stationary phase methods.
\begin{assumption}[DR\ref{it:crossing}']
\label{th:crossing2a}
Assume that
there are constants $c'_2>0$, $0<\gamma\le 1$ and $d'_2\in\N$ such that
for all $0< \beta \le 1$, using $k_3=k_1-k_2+u$,
\begin{align}\label{eq:crossingcond2}
 \sup_{u\in \T^3}
\int_{\R^3} \!\rmd s\, \rme^{-\beta |s|}
\left|  \int_{(\T^3)^2} \rmd k_1\rmd k_2\, 
\rme^{-\ci \sum_{i=1}^3 s_i \omega(k_i)}\right| 
\le c_2' \beta^{\gamma-1} \sabs{\ln \beta}^{d'_2} .
\end{align}
\end{assumption}
\begin{proposition}\label{th:crossing2}
Let the assumption \ref{th:crossing2a} be satisfied.  Then 
there is constant $c_2$ such
the assumption (DR\ref{it:crossing})   holds for this 
$\gamma$ and for $d_2=3+d_2'$.
\end{proposition}
\begin{proof}
By Lemma \ref{th:absest}, one has for any $a\in\R$,
\begin{align}
 \frac{1}{|a+\ci \beta|} =
 \int_{-\infty}^\infty\!\! \rmd s \, \rme^{\ci s a}
 f(\beta |s|)
\end{align}
with 
\begin{align}
 0\le f(\beta |s|) \le  
\sabs{\ln\beta} \rme^{-\beta |s|} + \1(|s|\le 1)\ln |s|^{-1} .
\end{align}
We use this to evaluate the left hand side of (\ref{eq:crossingest})
and then Fubini's theorem to swap the order of the $s$- and
$k$-integrals.  We then split the integration region $\R^3$ over the
$s$-variables into two parts: $\norm{s}_\infty \le 1$ and 
$\norm{s}_\infty > 1$.  
The first integration region yields a value bounded by a constant times
$\sabs{\ln \beta}^3$.  For the second region we use
$\sum_{i=1}^3 |s_j| \ge |s|$ combined with the estimate
(\ref{eq:crossingcond2}),
and obtain a bound which proves the stated result.
\end{proof}

\section{Energy transport for harmonic lattice dynamics}
\label{sec:classical}

We return to the lattice dynamics
in Section \ref{sec:model} with the goal of reading off from the main theorem
the implications on energy transport
in the kinetic scaling limit.   Let us first 
consider a fixed realization of the random masses 
and a state $(q,v)$ with a finite energy: $E(q,v;\xi)<\infty$ with 
$E$ defined in (\ref{eq:defHam}).  An energy density is a function
$E(x;q,v,\xi)$ such that 
$\int_{\R^3}\! \rmd x \, E(x;q,v,\xi) = E(q,v;\xi)$.  In general, 
there are many ways to divide up the energy into local pieces. 
However, there is
one particularly convenient choice in our case: we define the energy
density at a scale $\vep^{-1}>0$
as the random distribution $\Edens^\vep[q,v]$ with
\begin{align}
\mean{f,\Edens^\vep[q,v]} = \sum_{y\in\Z^3} f(\vep y)^*
 \frac{1}{2}  \Bigl( (1+\sqrt{\vep}\, \xi_y)^{-2} 
 v^2_y + |(\Omega q)_{y}|^2 \Bigr)
\end{align}
for any test function $f\in \cals(\R^3)$.  Here $\Omega$ is the convolution
operator defined in (\ref{eq:defOm}).  Since it is assumed that
$E(q,v;\xi)<\infty$, one has $\Omega q\in \ell_2$.  This implies that
the above formula makes sense for any 
$f\in C^\infty(\R^3)\cap L^\infty(\R^3)$ and that 
$\mean{1,\Edens^\vep[q,v]} = E(q,v)$.  This particular choice for energy
density is appealing since it 
is a positive distribution for any choice of $(q,v)$ -- in fact, when
divided by the total energy, it defines a probability measure on $\R^3$.

Let us then consider some initial conditions $q(0)=q^\vep_0$ and 
$v(0)=v^\vep_0$ with a bounded unperturbed energy, i.e., with 
\begin{align}
\sup_\vep \left. E(q_0^\vep,v_0^\vep)\right|_{\xi=0} <\infty.
\end{align}
We define further $\psi^\vep\in\hilb$ by
\begin{align}\label{eq:defPsiinit}
  \psi^\vep_{\sigma,y} =  \frac{1}{2} \left( (\Omega q_0^\vep)_y +  
    \ci \sigma (v_0^\vep)_y  \right),
\end{align} 
which differs from $\psi(0)$ defined in (\ref{eq:defPsi})
by omission of the random perturbations.  This omission will lead to errors
which are uniformly of order $\sqrt{\vep}$:
The mechanical energy density and the Wigner function of $\psi^\vep$ evolved
according to (\ref{eq:hilbevol}) are related by
\begin{align}
\Edens^\vep[q(t),v(t)](x) = 
 2\int_{\T^3}\!\rmd k\, 
 W_{++}^\vep[\rme^{-\ci t H_\vep} \psi^\vep](x,k) +
 \order{\smash{\vep^{\frac{1}{2}}}}. 
\end{align}
More precisely, if $f\in \cals(\R^3)$, we
define $J(x,k)=f(x)$ as a test-function in $\cals(\R^3\times\T^3)$,
and then for all $t\in \R$ and all sufficiently small $\vep$,
\begin{align}\label{eq:initcerror}
& \bigl|\mean{f,\Edens^\vep[q(t),v(t)]} -
 2\mean{J, W_{++}^\vep[\rme^{-\ci t H_\vep} \psi^\vep]} \bigr|
\le c \norm{f}_{4,\infty} \norm{\psi^\vep}^2 \sqrt{\vep}
\end{align}
where $c$  is a constant which depends only on $\bar{\xi}$.

To prove (\ref{eq:initcerror}), first note that, if $\psi(t)$ is defined
by (\ref{eq:defPsi}), then by unitarity 
\begin{align}
\norm{\psi(t)-\rme^{-\ci t H_\vep}\psi^\vep} =
\norm{\psi(0)-\psi^\vep} \le \sqrt{2\vep} 
\frac{\bar{\xi}}{1-\sqrt{\vep}\bar{\xi}} \norm{v^\vep_0}.
\end{align}
Therefore, using (\ref{eq:JW2}) and $\norm{v^\vep_0}^2\le 2\norm{\psi^\vep}^2$, 
there is a constant $c'$ such that
\begin{align}\label{eq:Jdiff2}
\bigl|\mean{J,W_{++}^\vep[\psi(t)]}-
  \mean{J,W_{++}^\vep[\rme^{-\ci t H_\vep}\psi^\vep]} \bigr|
\le c' \norm{f}_{4,\infty} \norm{\psi^\vep}^2 (2 \sqrt{\vep} b + \vep b^2),
\end{align}
where $b=2 \bar{\xi}/(1-\sqrt{\vep}\bar{\xi})$ which goes to 
$2 \bar{\xi}$ when $\vep\to 0$.  On the other hand, since $J$ does not
depend on $k$, we obtain directly from the definition (\ref{eq:defEW})
\begin{align}
\mean{J,W_{++}^\vep[\psi(t)]} =
\sum_{y\in\Z^3} f(\vep y)^* |\psi(t)_{+,y}|^2 =
\frac{1}{2}\mean{f,\Edens^\vep[q(t),v(t)]}.
\end{align}
Therefore, (\ref{eq:Jdiff2}) implies (\ref{eq:initcerror}) for
all sufficiently small $\vep$.

The following result establishes 
that the time-evolved, disorder-averaged energy
density of the harmonic lattice dynamics 
in the kinetic scaling limit can be obtained by solving the 
linear Boltzmann equation and then integrating out the $k$-variable.
\begin{corollary}
Consider the lattice dynamics (\ref{eq:defDyn}) with
initial conditions $q_0^\vep$, $v_0^\vep$, and let $\psi^\vep$
be defined as in (\ref{eq:defPsiinit}).  Assume that 
the initial conditions are
independent of $\xi$, and the family $(\psi^\vep)$ satisfies the
assumptions (IC\ref{it:I1}) -- (IC\ref{it:I3}), and suppose
that the elastic couplings satisfy (E\ref{it:EC0}) -- (E\ref{it:EC3})
and have a dispersion relation which satisfies
(DR\ref{it:suffdisp}) and (DR\ref{it:crossing}).

Then there is a family of bounded positive measures $\mu_t$, $t\ge 0$, on
$\R^3\times \T^3$ which satisfy the Boltzmann equation
(\ref{eq:Btransporteq}), such that for any $f\in \cals(\R^3)$ and $t\ge 0$
\begin{align}
\lim_{\vep\to 0}\E[\mean{f,\Edens^\vep[q(t/\vep),v(t/\vep)]}]=
 2 \int_{\R^3\times \T^3} \!\!\mu_t(\rmd x \,\rmd k)\, f(x)^*.
\end{align}
\end{corollary}
\begin{proof}
Since (E\ref{it:EC0}) -- (E\ref{it:EC3}) imply 
(DR\ref{it:DC1}) and (DR\ref{it:DC2}) we can now apply Theorem 
\ref{th:main} to compute the limit of
$\mean{J,W_{++}^\vep[\rme^{-\ci t H_\vep}\psi^\vep]}$ for all $J$ of the
form $J(x,k)=f(x)$, $f\in\cals$.  Together with 
(\ref{eq:initcerror}) this proves the corollary.
\end{proof}

In the previous section 
we have already given examples of elastic couplings which satisfy the
assumptions of the Corollary. The assumptions on the 
initial conditions can be satisfied, for
instance, by using the following two 
standard examples of Wigner functions in the semi-classical
limit:
\begin{enumerate}
\item $\vep$-independent $\psi\in\hilb$: Then we have the weak-$*$ limit
\begin{align}
\lim_{\vep\to 0^+} W^\vep[\psi](x,k)
=\delta(x)\, \FT{\psi}(k)^*\otimes \FT{\psi}(k).
\end{align}
\item WKB-type $\psi^\vep\in\hilb$:
For some given $h,S\in\cals(\R^3)$, $S$ real, define
\begin{align}
\psi^\vep_{+,y} = \vep^{3/2} h(\vep y) \rme^{\ci S(\vep y)/\vep},
\end{align}
and $\psi^\vep_{-,y} = (\psi^\vep_{+,y})^*$.  Then
for both $\sigma= \pm 1$,
\begin{align}
\lim_{\vep\to 0^+} W_{\sigma\sigma}^\vep(x,k) = 
|h(x)|^2 \delta\!\left(k-\left[\frac{\sigma}{2 \pi}\nabla S(x)\right]\right).
\end{align}
where $[\cdot]$ denotes the 
natural injection of $\R^3$ to $\T^3$ defined by removal of the integer part.
The off-diagonal components $W^\vep_{+-}$ and $W^\vep_{-+}$
do not necessarily have a weak-$*$ limit as $\vep\to 0$.
Note that the normalization has been
chosen so that $\sup_\vep\norm{\psi^\vep} < \infty$.  
\end{enumerate}
Given such $\psi^\vep$, initial positions and velocities of the particles
are obtained from
\begin{align}
q_0^\vep = \Omega^{-1}(2\, \re \psi^\vep_+),
\quad\text{and}\quad v_0^\vep = 2\,\im \psi^\vep_+ .
\end{align}

\appendix

\section{Definition of the collision operator}
\label{sec:appBoltzmann}

For writing down the collision term in the Boltzmann equation, we
need to know that our assumptions yield
``energy-level'' measures which are sufficiently regular.  This is
the content of the following two propositions:
\begin{proposition}\label{th:defcrosssect}
Let $\omega:\T^3\to \R$ be measurable and assume it satisfies 
(DR\ref{it:suffdisp}). Then
for all $\alpha\in \R$, the mapping
\begin{align}\label{eq:deltadef}
& C(\T^3)\ni f \mapsto
 \lim_{\beta\to 0^+} \int_{\T^3}\!\rmd k\,
  \frac{\beta}{\pi}
 \frac{1}{(\alpha-\omega(k))^2+\beta^2}  f(k)
\end{align}
defines a positive bounded Borel measure which
we denote by $\rmd k\, \delta(\alpha-\omega(k))$.  In addition, for all 
$f\in C^{\infty}(\T^3)$,
\begin{align}\label{eq:smoothdelta}
 \int_{-\infty}^\infty \rmd s
 \left( \int_{\T^3} \! \rmd k\, f(k) 
   \rme^{\ci s (\alpha-\omega(k))} \right)
 = 2\pi \int_{\T^3} \! \rmd k\,  \delta(\alpha-\omega(k)) f(k) .
\end{align}
\end{proposition}
\begin{proof}
Let us consider the family of linear mappings 
$\Lambda_{\alpha,\beta}\in C(\T^3)^*$ 
defined by the formula
\begin{align}\label{eq:defLambda}
\Lambda_{\alpha,\beta}[f] =  \int_{\T^3}\!\rmd k\,
  \frac{\beta}{\pi}
 \frac{1}{(\alpha-\omega(k))^2+\beta^2}  f(k)
\end{align}
for all $0<\beta\le 1$.  Then
\begin{align} 
\norm{\Lambda_{\alpha,\beta}} \le \int_{\T^3}\!\rmd k\,
  \frac{\beta}{\pi} \frac{1}{(\alpha-\omega(k))^2+\beta^2}
\end{align}
and we shall soon prove that the integral has an upper bound $\comega$.
Therefore, the family is equicontinuous.
The set of smooth functions is dense in $C(\T^3)$, and if we can prove that
the limit $\beta\to 0^+$ exists for all smooth functions, it follows that the
limit in fact exists in all of $C(\T^3)$, and the limit functional belongs to 
$C(\T^3)^*$ with a norm bounded by $\comega$.  The limit is
positive for any positive $f$, implying
that the limit functional is positive, and thus 
is determined by a unique regular positive Borel measure on $\T^3$,
bounded by $\comega$.

Suppose thus that $f\in C^{\infty}(\T^3)$.
By (\ref{eq:deltaappr}),
\begin{align}
\Lambda_{\alpha,\beta}[f]
= \int_{-\infty}^\infty \frac{ \rmd s}{2\pi}
   \rme^{\ci s \alpha-\beta |s|}
\Bigl( \int_{\T^3} \! \rmd k\, f(k) 
   \rme^{-\ci s \omega(k)} \Bigr).
\end{align}
However, the dispersion bound 
then also provides a bound for dominated
convergence theorem, which implies that the limit in 
(\ref{eq:deltadef}) exists and is equal to
\begin{align}
 \int_{-\infty}^\infty  \frac{\rmd s}{2\pi}
 \int_{\T^3} \! \rmd k\, f(k)  \rme^{\ci s(\alpha-\omega(k))}.
\end{align}
In addition, we have then also
\begin{align}\label{eq:betaintbound}
 \int_{\T^3}\!\rmd k\,  \frac{\beta}{\pi}
 \frac{1}{(\alpha-\omega(k))^2+\beta^2} \le
 \int_{-\infty}^\infty  \frac{\rmd s}{2\pi} \frac{\comega}{\sabs{s}^{3/2}}
 \le \comega.
\end{align}

Therefore, we can conclude that the limit defines a bounded
positive Borel measure, 
such that for any smooth function
$f$ equation (\ref{eq:smoothdelta}) holds.
\end{proof}

For the next result, we also need to require continuity of $\omega$.
\begin{proposition}\label{th:crosssectprop}
Let $\omega$ satisfy the assumptions of Proposition \ref{th:defcrosssect}
and let
\begin{align}\label{eq:defnuk2}
\nu_k(\rmd k') = \rmd k' \delta(\omega(k)-\omega(k')) 2\pi \omega(k')^2,
\quad k\in\T^3.
\end{align}
If $\omega$ is continuous, then all of the
following statements are true:
\begin{enumerate}
\item\label{it:contin}  For any $g\in C(\T^3)$, the functions 
$\R\ni \alpha \mapsto  \int \rmd k' \delta(\alpha-\omega(k')) g(k')$
and
$\T^3\ni k \mapsto  \int \nu_k(\rmd k') g(k')$
are continuous.
\item\label{it:gsymm} For any $g\in C(\T^3\times \T^3)$
\begin{align}
  \int_{\T^3}\!\rmd k \left(   \int_{\T^3} \nu_k(\rmd k') g(k,k')\right)
= \int_{\T^3}\!\rmd k' \left(   \int_{\T^3} \nu_{k'}(\rmd k) g(k,k')\right).
\end{align}
\item\label{it:movecontel}  If $f\in C(\R)$, then 
for all $k\in \T^3$,
\begin{align}\label{eq:fswap}
&  \rmd k'\delta(\omega(k)-\omega(k'))f(\omega(k)) =
 \rmd k'\delta(\omega(k)-\omega(k')) f(\omega(k')) .
\end{align}
\end{enumerate}
\end{proposition}
\begin{proof}
For $g\in C(\T^3)$ and $0<\beta\le 1$, let
$h_\beta(\alpha;g)=\Lambda_{\alpha,\beta}[g]$ with $\Lambda$ defined in
(\ref{eq:defLambda}).  Then by Proposition \ref{th:defcrosssect} the limit
$h_0(\alpha;g)=\lim_{\beta\to 0^+} h_\beta(\alpha;g)$ exists.  
To prove item \ref{it:contin},
we only need to prove that $h_0$ is continuous: this is
equal to the first statement and also implies the second, as then
$k\mapsto h_0(\omega(k);\omega^2 g)\in C(\T^3)$.

Since for any $x\in \R$,
\begin{align}\label{eq:bxunif}
  \frac{\beta |x|}{x^2+\beta^2} \le \frac{1}{2}.
\end{align}
we get from (\ref{eq:betaintbound}),
\begin{align}
  |h_\beta(\alpha')-h_\beta(\alpha)| \le
\frac{|\alpha'-\alpha|}{\beta} \comega \norm{g}_\infty.
\end{align}
Taking $\beta$ sufficiently small then allows us 
to conclude that $h_0$ is continuous.

The proof of \ref{it:gsymm} is a straightforward 
application of the dominated convergence
and Fubini's theorems, with the necessary bounds provided by 
(\ref{eq:betaintbound}).  To prove 
item \ref{it:movecontel}, let us first assume that $f$ is smooth.  Then
\begin{align}\label{eq:dfbound}
 |f(\omega(k'))-f(\omega(k))|\le 
 |\omega(k')-\omega(k)| \sup_{|x|\le \ommax} |f'(x)|,
\end{align}
where $\ommax=\sup_k|\omega(k)|<\infty$ since $\omega$ is continuous.
Thus for all $g\in C(\T^3)$,
\begin{align}\label{eq:vara1}
& \int \rmd k'\delta(\omega(k)-\omega(k')) ( f(\omega(k)) -
 f(\omega(k')) ) g(k')
\nonumber \\ & \quad = 
\lim_{\beta\to 0^+} 
 \int_{\T^3}\!\rmd k'\,
  \frac{\beta}{\pi}
 \frac{1}{(\omega(k)-\omega(k'))^2+\beta^2}  ( f(\omega(k)) -
 f(\omega(k')) ) g(k') =0,
\end{align}
as (\ref{eq:bxunif}) and (\ref{eq:dfbound}) allow applying
dominated convergence theorem to take the limit inside the $k'$-integral.
However, since the left hand side of (\ref{eq:vara1}) is continuous in
$f$ in the $\sup$-norm,
this implies that (\ref{eq:vara1}) holds also for all continuous $f$.
This proves (\ref{eq:fswap}).
\end{proof}

\section{Lattice Wigner transform}
\label{sec:appWigner}

It will be convenient for us to generalize the definition of the Wigner
transform slightly, and consider also Wigner transforms of probability
measures.  Since the following results do not depend on the specific model
of our study and can be of use in later work, we state the results in
greater generality than what was assumed for the main theorem.  
In particular, we
consider here the Wigner transform in any dimension $d\in \N_+$ and with
arbitrary number of components $N\in \N_+$.

Using our conventions for Fourier transform, 
the Wigner transform of $\psi\in L^2(\R^d)$ would be defined as
the function
\begin{align}\label{eq:L2Wigner}
 \R^{3}\times\R^{3} \ni (x,k) \mapsto
  \int_{\R^d}\! \!\rmd p\, \rme^{-\ci 2\pi x\cdot p}
  \FT{\psi}\Bigl(k+\frac{1}{2}\vep p\Bigr)^*
  \FT{\psi}\Bigl(k-\frac{1}{2}\vep p\Bigr)
\end{align}
where $\FT{\psi}$ is the Fourier transform of $\psi$ --
this is often also called the Wigner function.  Most of the
properties listed below have then been proven in \cite{gerard97}, 
but Wigner transforms of lattice
systems have not been so widely discussed. We are aware only  of
\cite{macia04,mielke05}.  In \cite{mielke05} the approach is to consider
$\FT{\psi}$ as a function in $L^2(\R^d)$ by setting $\FT{\psi}(k)=0$ for
$k$ not in the fundamental Brillouin zone.  One can then apply the standard
results valid for wave functions on $\R^d$.
In \cite{macia04}, the discrete Wigner transform
is defined as a distribution, similarly to what we have done here.
Similar proposals have been made in the context of studying
semi-classical limits of the  Schr\"{o}dinger equation in a periodic potential,
see for instance \cite{bal99,gerard97,SP04}.

We find it convenient to define the  Wigner transform as a
distribution which, for $\psi\in L^2(\R^d)$, would correspond to using
(\ref{eq:L2Wigner}) as an integral kernel.
\begin{definition}\label{th:genWigner}
Let $\nu$ be a Borel probability measure on $\ell_2(\Z^d,\C^N)$
equipped with its weak topology, and
let $\E_\nu$ denote the expectation value with respect to  $\nu(\rmd \psi)$. 
Whenever $\E_\nu\!\left[\norm{\psi}^2\right] <\infty$, we define
for any $\vep>0$ the Wigner transform $W_\nu^\vep$
of $\nu$ at the scale $\vep^{-1}$ via
\begin{align}\label{eq:defEW}
  \mean{J,\wvep_\nu}
= \sum_{y',y\in\Z^d} \sum_{i',i=1}^N
  \E_\nu\!\!\left[\psi_{i',y'}^* \psi_{i,y} \right]
 \int_{\T^d}\! \rmd k\, \rme^{\ci 2\pi k\cdot (y'-y)} 
 J_{i',i}\Bigl(\vep\frac{y'+y}{2},k\Bigr)^*.
\end{align}
where $J\in \cals(\R^d\times \T^d,\M_N)$.
\end{definition}
This definition includes the deterministic case, where $\nu$ is
the Dirac measure 
$\nu=\delta_{\phi}$ for any $\phi\in \ell_2(\Z^d,\C^N)$.  In this case
$\wvep_\nu $ is the Wigner transform of the vector $\phi$, 
denoted by $\wvep[\phi]$.

The topology of 
$\cals(\R^d\times \T^d, \M_N)$ is defined as usual, via a countable family
of seminorms (see e.g.\ \cite{gelfandII}).
The next theorem proves that, under the above assumptions, 
$W_\nu^\vep$ is a tempered distribution, and it lists
some of their general properties.  In
particular, item (b) establishes that 
this definition coincides with the one given
in Eq.~(\ref{eq:Wdef}).
\begin{theorem}\label{th:cwvep}
Under the assumptions of Definition \ref{th:genWigner}, 
$W_\nu^\vep\in \cals'(\R^d\times \T^d,\M_N)$, and 
for every test-function $J$, 
\begin{align}\label{eq:meanlast}
\mean{J,W_\nu^\vep} = \E_\nu\!\left[\mean{J,W^\vep[\psi]} \right].
\end{align}
Furthermore, denoting $\hilb=\ell_2(\Z^d,\C^N)$ and for arbitrary
$J\in \cals(\R^d\times \T^d,\M_N)$, the following properties hold:
\begin{jlist}[(\alph{jlisti})]
\item There is a bounded operator $\cwvep [J]$ and 
a constant $c$, depending only on the dimensions $d$ and $N$, such that for
all  $\psi\in \hilb$
\begin{gather}\label{eq:JW2}
 \mean{J,\wvep [\psi]} = \braket{\psi}{\cwvep [J]\psi}
 \qquad \text{with}\qquad
 \norm{\cwvep[J]} \le c \norm{J}_{d+1,\infty} . 
\end{gather}
In addition, for the same constant $c$ as above,
\begin{gather}\label{eq:Wnubound}
\left|\mean{J,W_\nu^\vep}\right| \le
c \norm{J}_{d+1,\infty}  \E_\nu\!\!\left[\norm{\psi}^2\right].
\end{gather}
\item\label{it:GW2} For any $\psi\in \hilb$,
\begin{align}\label{eq:WJFourier}
&   \mean{J,\wvep [\psi]} =
  \int_{\R^d}\! \!\rmd p 
  \int_{\T^d}\! \!\rmd k\,
  \FT{\psi}\Bigl(k-\frac{1}{2}\vep p\Bigr) \cdot
  \FT{J}(p,k)^*
  \FT{\psi}\Bigl(k+\frac{1}{2}\vep p\Bigr)
\end{align}
where $\FT{J}$ is the Fourier transform of $J$ in the first variable,
as in (\ref{eq:defFT1J}).
\end{jlist}
\end{theorem}
\begin{proof}
Consider $\vep>0$ and an arbitrary test-function $J$.  Define
component-wise the operator $\cwvep [J]$ by
\begin{align}
\cwvep [J](i',y';i,y) =
  \int_{\T^d}\! \rmd k\, \rme^{\ci 2\pi k\cdot (y'-y)} 
 J_{i',i}\Bigl(\vep\frac{y'+y}{2},k\Bigr)^*.
\end{align}
By partial integration in $k$ we find that there is a constant $c'<\infty$,
depending only on $d$, such that
\begin{align}\label{eq:Jschbound}
\left| \cwvep [J](i',y';i,y)\right| \le \frac{c'}{\sabs{y'-y}^{d+1}} 
\norm{J}_{d+1,\infty} .
\end{align}
Therefore, for all $\phi,\psi\in \hilb$,
\begin{align}
& \sum_{y',y\in\Z^d} \sum_{i',i=1}^N
  \left|\phi_{i',y'}^* \psi_{i,y} \cwvep [J](i',y';i,y)\right| \
\le  \norm{J}_{d+1,\infty} \norm{\phi} \norm{\psi}
  \sum_{n\in\Z^d} \frac{c' N}{\sabs{n}^{d+1}}.
\end{align}
Let us denote the result from the sum over $n$ by $c$. Then $c$ is finite and
depends only on $d$ and $N$, and we have proven that
$\norm{\cwvep[J]} \le c \norm{J}_{d+1,\infty}$. 
By the definition (\ref{eq:defEW}), 
$\mean{J,\wvep[\psi]} = \braket{\psi}{\cwvep [J]\psi}$, and
(\ref{eq:JW2}) is valid.

Under the assumptions made on $\nu$, $\psi_{i',y'}^* \psi_{i,y}$
is measurable, and the mean of its absolute value 
is bounded by
$\E_\nu[|\psi_{i',y'}|^2]^{1/2}\E_\nu[|\psi_{i,y}|^2]^{1/2}$
by the Schwarz inequality.  An application of (\ref{eq:JW2}) shows that
the sum in (\ref{eq:defEW}) is absolutely summable, and it is
bounded by 
$c \norm{J}_{d+1,\infty} \E_\nu\!\!\left[\norm{\psi}^2\right]<\infty$.
Therefore,
$\mean{J,\wvep_\nu}$ is well-defined, the inequality 
(\ref{eq:Wnubound}) is satisfied, and by Fubini's
theorem (\ref{eq:meanlast}) holds.
The mapping $\wvep_\nu$ is linear and, as
$\norm{J}_{d+1,\infty}$ is bounded from above by one of the semi-norms 
defining
the topology of $\cals$, (\ref{eq:Wnubound}) implies that $\wvep_\nu$ is a
tempered distribution.

Now we only need to prove the item (b). By (\ref{eq:JW2}) both
sides of the equality (\ref{eq:WJFourier}) are continuous in $\psi$, and
thus it is 
enough to prove it for $\psi\in\hilb$ which have a compact support.
However, then we can first use
\begin{align}
 J\Bigl(\vep\frac{y'+y}{2},k\Bigr)^*
= \int_{\R^d} \!\rmd p\, \rme^{-\ci 2\pi (\vep p/2)\cdot (y'+y)}  
 \FT{J}(p,k)^*
\end{align}
in the definition (\ref{eq:defEW})
and perform the finite sums over $y$ and $y'$.
This yields (\ref{eq:WJFourier}) after changing the order of $p$ and $k$
integral, which is possible by the integrability of
$p\mapsto\sup_k\norm{\FT{J}(p,k)}$.
\end{proof}

Let us next investigate properties of
limit points of a sequence of Wigner
transforms when $\vep\to 0$.  
For simplicity, we shall do this only in the
case $N=1$, but for arbitrary $d$.
In most cases, it is sufficient to study the limit of the
Fourier transforms of the Wigner distributions.  Explicitly, 
let $\ell_2=\ell_2(\Z^d)$,  
let $\nu$ be a probability measure on $\ell_2$ 
satisfying the assumptions of Definition \ref{th:genWigner},
let $W_\nu^\vep$ denote its Wigner transform for some $\vep>0$, 
and define the function $F^\vep_\nu:\R^d\times \Z^d\to \C$ by the formula
\begin{align}\label{eq:defFvep}
F^\vep_\nu(p,n) =  
\sum_{y\in \Z^d} 
  \E_\nu\!\!\left[\psi_{y-n}^* \psi_{y} \right]
  \rme^{-\ci 2\pi \vep p \cdot (y-n/2)} .
\end{align}
\begin{proposition}\label{th:Fnuprop}
For $\nu$, $F^\vep_\nu$ and $W^\vep_\nu$ defined as above, all of the
following hold: 
\begin{jlist}[(\alph{jlisti})]
\item $|F^\vep_\nu(p,n)|\le F^\vep_\nu(0,0)=\E_\nu[\norm{\psi}^2]$.
\item $F^\vep_\nu$ is the Fourier transform of $W^\vep_\nu$: for
all $J\in \cals(\R^d\times \T^d)$,
\begin{align}\label{eq:FTWnu}
\mean{J,W_\nu^\vep} = \sum_{n\in\Z^d}  \int_{\R^d}\! \!\rmd p \,
\mathcal{J}(p,n)^* F^\vep_\nu(p,n) 
\end{align}
where $\mathcal{J}$ is the Fourier transform of $J$ in both variables.
\item For all $p\in \R^d$ and $n\in \Z^d$, 
\begin{align}\label{eq:FvepFourier}
&  F^\vep_\nu(p,n) =
  \E_{\nu}\!\!\left[ \int_{\T^d}\! \!\rmd k\, \rme^{\ci 2 \pi n\cdot k}
  \FT{\psi}\Bigl(k-\frac{1}{2}\vep p\Bigr)^*
  \FT{\psi}\Bigl(k+\frac{1}{2}\vep p\Bigr)\right] .
\end{align}
\end{jlist}
\end{proposition}
\begin{proof}
That $F^\vep_\nu(p,n)$ is well-defined, and the bounds in (a) follow as in
the proof of Theorem \ref{th:cwvep}.  (b) follows directly from Fubini's
theorem, since performing the integral over $k$ yields $\1(y'=y-n)$.  
The proof of (c) is similar to that of (b) in Theorem
\ref{th:cwvep}. First prove the result for $\nu=\delta_\phi$ with $\phi$
having a compact support, then extend it to all $\phi\in \ell_2$ by
continuity, and finally use Fubini's theorem to change the order of the sum
and the expectation value in (\ref{eq:defFvep}).  
\end{proof}

In both of the following theorems, 
let $I=(\vep_k)$, $k=1,2,\ldots$, be a sequence in 
$(0,\infty)$ such that $\vep_k\to 0$ when $k\to\infty$. For notational 
simplicity, we will again denote the limits of the type
$\lim_{k\to\infty} f(\vep_k)$ by $\lim_{\vep\to 0} f(\vep)$.
Let $(\nu^\vep)_{\vep\in I}$ be a family of probability
measures on $\ell_2$ 
satisfying the assumptions of Definition \ref{th:genWigner},
and denote also $\E^\vep = \E_{\nu^\vep}$,
$\owvep = W_{\nu^\vep}^\vep$  and $F^\vep =F_{\nu^\vep}^\vep$.  
\begin{theorem}\label{th:weakimpliesborel}
If $\owvep \to \owl$ in the weak-$*$
topology and
\begin{align}\label{eq:Emaxass}
 \sup_{\vep \in I} \E^{\vep}\!\left[\norm{\psi}^2\right] <\infty,
\end{align}
then there is a unique bounded positive Borel measure $\mu$ on 
$\R^d\times \T^d$ such that for all test functions $J$
\begin{align}\label{eq:Wismeas}
  \mean{J,\owl} = 
 \int_{\R^d\times \T^d} \!\!\mu(\rmd x \,\rmd k)\, J(x,k)^*
\end{align}
and $\mu$ is bounded by
$\sup_{\vep \in I}\E^{\vep}\!\left[\norm{\psi}^2\right]$.

If, in addition, the family $(\nu_\vep)_\vep$ is tight on the scale
$\vep^{-1}$, in the sense that
\begin{align}\label{eq:mthightness}
\limsup_{\vep\to 0} \sum_{|y|> R/\vep} 
\E^\vep\!\!\left[|\psi_{y}|^2\right] \to 0 \text{, when }R\to \infty,
\end{align}
then $F^\vep$ converges to the characteristic function of $\mu$:
For all $p\in \R^d$ and $n\in \Z^d$, 
\begin{align}\label{eq:Fvepchar}
  \lim_{\vep\to 0} F^\vep(p,n) = 
 \int_{\R^d\times \T^d} \!\!\mu(\rmd x \,\rmd k)\, 
\rme^{-\ci 2 \pi (p\cdot x- n\cdot k)}.
\end{align}
\end{theorem}
\begin{proof}
We start by proving that
\begin{align}\label{eq:owlposit}
\mean{|J|^2,\owl}\ge 0,\qquad \text{for all }J\in \cals .
\end{align}
Since 
$\R^d\times \Z^d$ is a locally compact Abelian group,
it then follows from  the Bochner-Schwartz theorem \cite{warz68}
that there is a unique tempered positive Borel measure $\mu$ such
that (\ref{eq:Wismeas}) holds.  Let $J\in \cals$. Then for all 
$k\in \T^d$, and $y,n\in \Z^d$,
\begin{align}
 J\Bigl(\vep y + \vep \frac{n}{2},k\Bigr) -
 J(\vep y,k)
 = \frac{\vep}{2} \int_0^1\! \rmd s \, 
n\cdot \nabla_{\! 1} J\Bigl(\vep y + s \vep \frac{n}{2},k\Bigr) .
\end{align}
Therefore, for any $y',y\in \Z^d$,
there is $c$, depending only on $d$, such that
\begin{align}
& \Bigl| \int_{\T^d}\! \rmd k\, \rme^{\ci 2\pi k\cdot (y'-y)} 
 \Bigl( \Bigl|J\Bigl(\vep\frac{y'+y}{2},k\Bigr)\Bigr|^2
 -  J(\vep y',k)^*  J(\vep y,k) \Bigr) \Bigr| 
\nonumber \\ & \quad
\le \vep \frac{c}{\sabs{y'-y}^{d+1}} \norm{J}_{d+2,\infty}^2,
\end{align}
which can be proven, e.g., by $d+2$ partial integrations over $k_i$
with $i$ chosen so that $|y'_i-y_i|$ is at maximum.
Proceeding as in the proof of Theorem \ref{th:cwvep}, we find
\begin{align}
& \mean{|J|^2,\owvep} =
\E^\vep\Bigl[ \int_{\T^d}\!\rmd k \Bigl| \sum_{y\in\Z^d} J(\vep y,k) 
\rme^{-\ci  2\pi k\cdot y} \psi_y \Bigr|^2 \Bigr] + \order{\vep}.
\end{align}
This implies that $\mean{|J|^2,\owl} = 
\lim_{\vep\to 0}\mean{|J|^2,\owvep} \ge 0$, and proves (\ref{eq:owlposit}).

We still need to prove that $\mu$ is bounded.  For this, let
\begin{align}\label{eq:defJlambda}
J_{\lambda,p,n}(x,k) = 
\rme^{-\lambda^2 x^2+\ci 2\pi (p\cdot x-n\cdot k)}\text{ for }\lambda>0.
\end{align}
Then
\begin{align}\label{eq:Jlambda}
& \mean{J_{\lambda,p,n},\owvep} 
= \sum_{y\in \Z^d} \rme^{-\lambda^2 \vep^2 (y-n/2)^2}
  \E^\vep\!\!\left[\psi_{y-n}^* \psi_{y} \right]
  \rme^{-\ci 2\pi \vep p \cdot (y-n/2)} ,
\end{align}
and thus $|\mean{J_{\lambda,p,n},\owvep} |\le\E^\vep[\norm{\psi}^2]$.
On the other hand, by monotone convergence
$\mu(\R^d\times \T^d) = \lim_{\lambda\to 0}  \int\mu(\rmd x \,\rmd k)\,
 \rme^{-\lambda^2 x^2} =  \lim_{\lambda\to 0} \mean{J_{\lambda,0,0},\owl}$, 
and we can infer that $\mu$ is bounded  
by $\sup_{\vep} \E^\vep[\norm{\psi}^2]<\infty$.

Let us then assume that also (\ref{eq:mthightness}) holds, and consider any
fixed $p\in \R^d$ and $n\in \Z^d$.  By
(\ref{eq:Jlambda}),
\begin{align}\label{eq:deltaFv}
& \left| \mean{J_{\lambda,p,n},\owvep}  - F^\vep(p,n)\right|  
\le \sum_{y\in \Z^d} \left|1-\rme^{-\lambda^2 \vep^2 (y-n/2)^2}\right|
  \E^\vep\!\left[|\psi_{y-n}||\psi_{y}| \right].
\end{align}
Then for any $R>0$ and $\vep \le 2/|n|$ we have
\begin{align}
& \sum_{y\in \Z^d} \left(1-\rme^{-\lambda^2 \vep^2 (y\pm n/2)^2}\right)
  \E^\vep\!\!\left[|\psi_{y}|^2 \right]
\le \lambda^2 (R+1)^2  \E^\vep[\norm{\psi}^2]
+ \sum_{|y|>R/\vep}  \!\!\! \E^\vep\!\!\left[|\psi_{y}|^2 \right],
\end{align}
which can be applied in (\ref{eq:deltaFv}) yielding
\begin{align}
& \left| \mean{J_{\lambda,p,n},\owvep}  - F^\vep(p,n)\right|  
\le 
 \lambda^2 (R+1)^2  \E^\vep[\norm{\psi}^2]
+ \sum_{|y|>R/\vep}   \E^\vep\!\!\left[|\psi_{y}|^2 \right].
\end{align}
Let $F$ denote the characteristic function of $\mu$, defined by the right hand
side of (\ref{eq:Fvepchar}).  By dominated convergence, 
$F(p,n) = \lim_{\lambda\to 0}(\lim_{\vep\to 0}
\mean{J_{\lambda,p,n},\owvep})$, and thus for all $R>0$,
\begin{align}
& \limsup_{\vep\to 0}
\left| F(p,n) - F^\vep(p,n)\right|  
\le  \limsup_{\vep\to 0}
\sum_{|y|>R/\vep}   \E^\vep\!\!\left[|\psi_{y}|^2 \right].
\end{align}
We take here $R\to\infty$, when the tightness assumption implies that 
(\ref{eq:Fvepchar}) holds.
\end{proof}
For the converse of this theorem, we do not even need to require tightness.
The main part of the statement can be summarized as follows: If the
Fourier transforms $F^\vep$ converge pointwise almost everywhere, then the
corresponding Wigner transforms converge to a measure whose characteristic
function coincides  almost everywhere with $\lim_{\vep\to 0}F^\vep$.
\begin{theorem}\label{th:FimpliesWweak}
Let the family $(\nu^\vep)$ satisfy Eq.~(\ref{eq:Emaxass}), and 
assume that for all $n\in \Z^d$ and almost every $p\in \R^d$, the limit
$\lim_{\vep\to 0} F^\vep(p,n)$ exists.  Then 
$\owvep \to \owl$ in the weak-$*$ topology  when $\vep\to 0$, and
$\owl$ is given by a bounded
positive Borel measure $\mu$ such that almost everywhere
\begin{align}\label{eq:Fveptomuch}
  \lim_{\vep\to 0} F^\vep(p,n) = 
 \int_{\R^d\times \T^d} \!\!\mu(\rmd x \,\rmd k)\, 
\rme^{-\ci 2 \pi (p\cdot x- n\cdot k)}.
\end{align}
In addition, for almost every $p\in \R^d$, in particular for every $p$
for which (\ref{eq:Fveptomuch}) holds
for all $n$, we have for any $f\in C(\T^d)$
\begin{align}\label{eq:Fveptomucont}
\lim_{\vep\to 0} \E^\vep\!\!\left[ \int_{\T^d} \!\!\rmd k f(k)\,
  \FT{\psi}\Bigl(k-\frac{1}{2}\vep p\Bigr)^*
  \FT{\psi}\Bigl(k+\frac{1}{2}\vep p\Bigr) \right]
 =  \int_{\R^d\times \T^d}\!\! \!\!\mu(\rmd x \,\rmd k)\, 
\rme^{-\ci 2 \pi p\cdot x} f(k) .
\end{align}
\end{theorem}
\begin{proof}
Let us define $F^0(p,n)=\lim_{\vep\to 0} F^\vep(p,n)$ for every $p$, $n$
for which the limit exists and $0$ elsewhere.  By Proposition
\ref{th:Fnuprop}, items (a) and (b), we can apply dominated convergence to
the equation (\ref{eq:FTWnu}) which proves that
for any test-function $J$, with Fourier transform $\mathcal{J}$,
\begin{align}\label{eq:F0eq}
\lim_{\vep\to 0}
\mean{J,\owvep} = \sum_{n\in\Z^d}  \int_{\R^d}\! \!\rmd p \,
\mathcal{J}(p,n)^* F^0(p,n) .
\end{align}
By the Banach-Steinhaus theorem,  then there is $\owl\in \cals'$ such that
$\lim_{\vep\to 0}\owvep = \owl$ in the weak-$*$ topology.  

Therefore,
we can apply Theorem \ref{th:weakimpliesborel} and conclude that there
is a bounded positive Borel measure such that (\ref{eq:Wismeas}) holds.
Let $F$ be the characteristic function of $\mu$.  Then by Fubini's theorem,
we have for all test-functions $J$
\begin{align}
\mean{J,\owl} = \sum_{n\in\Z^d}  \int_{\R^d}\! \!\rmd p \,
\mathcal{J}(p,n)^* F(p,n) .
\end{align}
As this needs to be equal to (\ref{eq:F0eq}), for all $n$
there is $A_n\subset \R^d$ with Lebesgue measure zero such that
$F(p,n)=F^0(p,n)$ for $p\not\in A_n$.  Then also $A=\cup_n A_n$ has
Lebesgue measure zero, and we have proven (\ref{eq:Fveptomuch}).

For the final result, consider any $n\in \Z^d$ and $p\not\in A$.
As for any $\vep>0$, $|\E^\vep[ \int \rmd k f(k)
 \FT{\psi}(k-\vep p/2)^* \FT{\psi}(k+\vep p/2)]| \le \norm{f}_\infty
 \E^\vep[\norm{\psi}^2]$, and smooth functions are dense in
 $C(\T^d)$, it is enough to prove 
(\ref{eq:Fveptomucont}) for all smooth $f$. Let thus 
$f\in C^{\infty}(\T^d)$ and let  
$\IFT{f}$ denote its Fourier transform.  Then
by Fubini's theorem and Proposition \ref{th:Fnuprop}, item (c),
\begin{align}
 \E^\vep\!\!\left[ \int_{\T^d} \!\!\rmd k f(k)\,
  \FT{\psi}\Bigl(k-\frac{1}{2}\vep p\Bigr)^*
  \FT{\psi}\Bigl(k+\frac{1}{2}\vep p\Bigr) \right]
 =  \sum_{n\in Z^d} \IFT{f}(n) F^\vep(p,-n).
\end{align}
Using  the dominated convergence theorem and the assumption $p\not\in A$, we
find that, when $\vep\to 0$, this converges to 
\begin{align}
\sum_{n\in Z^d} \IFT{f}(n) F(p,-n)=
 \int_{\R^d\times \T^d}\!\! \!\!\mu(\rmd x \,\rmd k)\, 
\rme^{-\ci 2 \pi p\cdot x} f(k).
\end{align}
This finishes the proof of the theorem.
\end{proof}

\section{Cumulant bounds}
\label{sec:appComb}

Let $\nu$ denote the distribution of $\xi_0$, which, by assumption, has zero
mean, unit variance, and whose support is bounded by $\bar{\xi}$.  Let
$g_m(z) = \int \nu(\rmd x) \exp(\ci x z)$ and $g_c(z)=\ln g_m(z)$
denote its moment and cumulant generating functions, respectively.
The cumulants $C_n$ of $\nu$ are then defined by the formula
$C_n = (-\ci)^n g_c^{(n)}(0)$, for $n\in \N_+$.
\begin{lemma}\label{th:cumulb}
$C_1=0$ and $C_2=1$, and for all $n> 2$,
\begin{equation}
  \label{eq:bcumul}
  |C_n| \le 3 \bar{\xi}^n n!. 
\end{equation}
\end{lemma}
\begin{proof}
Since $\nu$ has a zero mean and unit variance,
$C_1=0$ and $C_2=1$.  Due to the compact range,
the generating function $g_m$ is analytic near origin, and since
$g_m(0)=1$, so is then the cumulant generating function $g_c=\ln g_m$.  
On the other hand, for $|z|\le 1/\bar{\xi}$ we have by
Jensen's inequality
\begin{align}
 &   \re g_m(z) = \int_{[-\bar{\xi},\bar{\xi}]} 
  \nu(\rmd x)\, \rme^{-x \im z}\cos\left(x\re z\right)  
\nonumber \\ &  \quad  
\ge  \cos(1) \int
  \nu(\rmd x)\, \rme^{-x \im z} \ge
  \cos(1) \exp\Bigl[-\im z \int \nu(\rmd x)\, x \Bigr]
    =  \cos(1)>0.
\end{align}
Thus $g_c(z)$ is analytic in an open set containing the closed disc
$|z|\le 1/\bar{\xi}$, and inside the disc
$|g_c(z)|\le |\ln|g_m(z)||+|\arg g_m(z)| \le 1 + \pi/2<3$, 
since now $\cos(1)\le |g_m(z)| \le \rme^1$.  Then the
Cauchy estimates for derivatives yield (\ref{eq:bcumul}).
\end{proof}

\begin{definition}\label{th:defPiI}
For any $N\in \N_+$, let $I_N=\set{1,\ldots,N}$, and define
$I_0=\emptyset$.  
For any finite, non-empty set $I$, let $\pi(I)$ denote  
the set of all its partitions:
$S\in \pi(I)$ if and only if $S \subset \mathcal{P}(I)$ such that
each $A\in S$ is non-empty, $\cup_{A\in S} A = I$, and
if $A,A'\in S$ with $A'\ne A$ then $A'\cap A=\emptyset$.
In addition, we define $\pi(\emptyset)=\set{\emptyset}$.
\end{definition}

\begin{lemma}[Moments to cumulants formula]\label{th:momtocum}
Let $N\in\N_+$ and $I=I_N$.  Then for any mapping $i:I\to \Z^d$
\begin{align}\label{eq:momtocum}
  \E\Bigl[\prod_{\ell=1}^N \xi_{i_\ell}\Bigr] & = 
 \sum_{S\in\pi(I)} \prod_{A\in S} \Bigl[ C_{|A|} 
 \sum_{y\in \Z^{3}} \prod_{\ell\in A} \delta_{i_\ell,y} \Bigr] .
\end{align}
\end{lemma}
\begin{proof}
Proof is by induction in $N$.  The formula is clearly true for $N=1$.  
For the induction step, let us assume it is true for all values less than a
given $N>1$.

Let us first define the finite set $X=i(I) \subset \Z^3$, when
\begin{align}
  \E\Bigl[\prod_{\ell=1}^N \xi_{i_\ell}\Bigr] & = 
 \prod_{\ell=1}^N 
 \Bigl(-\ci \frac{\partial}{\partial z_{i_\ell}}\Bigr)
  \E\Bigl[\rme^{\ci \sum_{x\in X}\! \xi_x z_x}\Bigr]_{z=0} .
\end{align}
By the assumed independence of $(\xi_y)$, we get 
for all $z\in \R^X$ in a sufficiently small neighbourhood of zero,
\begin{align}
& \prod_{\ell=1}^N 
 \Bigl(-\ci \frac{\partial}{\partial z_{i_\ell}}\Bigr)
  \E\Bigl[\rme^{\ci \sum_{x\in X}\! \xi_x z_x}\Bigr] 
= \prod_{\ell=1}^N 
 \Bigl(-\ci \frac{\partial}{\partial z_{i_\ell}}\Bigr)
  \Bigl[\prod_{x\in X} g_m(z_{x})\Bigr] 
\nonumber \\ & \quad =
 \prod_{\ell=1}^{N-1} 
 \Bigl(-\ci \frac{\partial}{\partial z_{i_\ell}}\Bigr)
 \Bigl[-\ci \partial_{z_{i_N}} \exp\Bigl(\sum_{x\in X}
  g_c(z_{x})\Bigr)\Bigr]
\nonumber \\ & \quad 
= \prod_{\ell=1}^{N-1} 
 \Bigl(-\ci \frac{\partial}{\partial z_{i_\ell}}\Bigr)
 \Bigl\{\bigl(-\ci g'_c(z_{i_N})\bigr) 
 \E\Bigl[\rme^{\ci \sum_{x\in X}\! \xi_x z_x}\Bigr]\Bigr\} .
\end{align}
By induction, we then can prove that this is equal to 
\begin{align}
   \sum_{B\subset\set{1,\ldots,N-1}}
 \Bigl\{  \Bigl[\prod_{\ell\in B\cup\set{N}} \!\!
 \bigl(-\ci \partial_{z_{i_\ell}}\bigr)\Bigr] g_c(z_{i_N})
 \Bigl[\prod_{\ell\in B^c} 
 \bigl(-\ci \partial_{z_{i_\ell}}\bigr)\Bigr]
 \E\Bigl[\rme^{\ci \sum_{x\in X}\! \xi_x z_x}\Bigr]\Bigr\}.
\end{align}
We evaluate this at $z=0$, showing
\begin{align}
&  \E\Bigl[\prod_{\ell=1}^N \xi_{i_\ell}\Bigr] 
 =  \sum_{B\subset\set{1,\ldots,N-1}}
 \Bigl\{ C_{|B|+1} \sum_{y\in\Z^3}
 \prod_{\ell\in B\cup\set{N}} \delta_{i_\ell,y} \Bigr\}
 \E\Bigl[ \prod_{\ell\in B^c} \xi_{i_\ell}\Bigr]
\end{align}
where an application of the induction assumption yields (\ref{eq:momtocum}).
\end{proof}

Finally, we need the following bound:
\begin{lemma}\label{th:highosum}
Let $I$ be an index set with $|I|=N$, let $M$ be an integer such that
$2\le M< N$, and define
$a=2\bar{\xi}(3\bar{\xi}^2+1)$. Then for any $0\le r \le 1/a$,
\begin{align}
 \sum_{\substack{S\in\pi(I),\\ |A|>M\text{ for some }A\in S}}
 r^{N-2 |S|} \prod_{A\in S} \left|C_{|A|} \right| \le 
  N!\, (a r)^{M-\1(N-M\text{ is odd})}.
\end{align}
\end{lemma}
\begin{proof}
We begin by
\begin{align}
 \sum_{\substack{S\in\pi(I),\\
     |A|>M\text{ for some }A\in S}} r^{N-2 |S|}
  \prod_{A\in S} \left|C_{|A|} \right| =
  \sum_{m=1}^N  r^{N-2 m}
 \sum_{\substack{S\in\pi(I): |S|=m,\\
     |A|>M\text{ for some }A\in S}}
  \prod_{A\in S} \left|C_{|A|} \right|
\end{align}
and focus on the second sum. For any finite, non-empty
set $S$, let $\text{Ind}_S$ denote the set of indexings of $S$,
i.e., it is the collection of all bijections 
$\set{1,\ldots,|S|} \to S$. As this set always has $|S|!$ elements
and the summand is indexing invariant, we get
\begin{align}
 & \sum_{\substack{S\in\pi(I): |S|=m,\\
     |A|>M\text{ for some }A\in S}}
  \prod_{A\in S} \left|C_{|A|} \right|
  = \frac{1}{m!} 
 \sum_{\substack{S\in\pi(I): |S|=m,\\
     |A|>M\text{ for some }A\in S}} \sum_{P\in \text{Ind}_S}
  \prod_{j=1}^m \left|C_{|P(j)|} \right|
\nonumber \\ & \quad 
=  \frac{1}{m!} 
 \sum_{\substack{S\in\pi(I): |S|=m,\\
     |A|>M\text{ for some }A\in S}} \sum_{P\in \text{Ind}_S}
 \sum_{n\in \N_+^m} \prod_{j=1}^m \1(|P(j)|=n_j)
 \prod_{j=1}^m \left|C_{n_j} \right|
\nonumber \\ & \quad 
=  \frac{1}{m!} 
 \sum_{n\in \N_+^m} \1\Bigl(\sum_j n_j = N\Bigr) 
\1(\exists j: n_j>M)
\prod_{j=1}^m \left|C_{n_j} \right|
\nonumber \\ & \qquad \times
 \sum_{S\in\pi(I): |S|=m} \sum_{P\in \text{Ind}_S}
 \1(\forall j: |P(j)|=n_j) \label{eq:sumSAM}
\end{align} 
Every pair $S$, $P$ defines, by the formula $S_j=P(j)$,
a sequence $(S_1,\ldots,S_m)$ which is a collection of non-empty, disjoint
sets which partition $I$. Conversely, to every such sequence corresponds a
unique pair $S$ and $P\in \text{Ind}_S$.  On the other hand, since the
number of such sequences, which also satisfy $n_j=|S_j|$ for all $j$, is
exactly
\begin{align}
\prod_{j=1}^m \binom{N-\sum_{j'=1}^{j-1} n_{j'}}{n_j} =
\frac{N!}{\prod_{j=1}^m n_j!},
\end{align}
we have now proven that (\ref{eq:sumSAM}) is equal to
\begin{align}
&\frac{N!}{m!} \sum_{n\in \N_+^m} \1\Bigl(\sum_j n_j = N\Bigr) 
\1(\exists j:n_j>M)
\prod_{j=1}^m \frac{\left|C_{n_j} \right|}{n_j!} .
\end{align}
Since $C_1=0$, the summand in 
is zero, unless $n_j \ge 2$ for all $j$.  Taking also into account that 
$C_2=1$, we get from Lemma \ref{th:cumulb}
\begin{align}\label{eq:onlycumul}
 & \sum_{\substack{S\in\pi(I): |S|=m,\\
     |A|>M\text{ for some }A\in S}}
  \prod_{A\in S} \left|C_{|A|} \right|  
\nonumber \\ &  \quad 
\le \frac{N!}{m!} \sum_{\substack{n\in \N_+^m\\n_j\ge 2}} 
\1\Bigl(\sum_j n_j = N\Bigr)  \1(\exists j: n_j>M)
\prod_{j:n_j>2} \left(3 \bar{\xi}^{n_j}\right) .
\end{align}
Let us denote $k=|\set{j:n_j=2}|$, when the second condition implies
$1\le k< m$.  Then, by shifting each $n_j$ by $2$
and by using the permutation invariance of the summand,
the right hand side of (\ref{eq:onlycumul}) can also be written as
\begin{align}
& \frac{N!}{m!} \sum_{k=0}^{m-1} \binom{m}{k}
 \sum_{n\in \N_+^{m-k}} \1\Bigl(\sum_j n_j = N-2 m\Bigr) 
  \1(\exists j: n_j>M-2)
3^{m-k} \bar{\xi}^{N-2 k}
\nonumber \\ & \  \le
\frac{N!}{m!} \sum_{k=0}^{m-1} \binom{m}{k} (m-k)
3^{m-k} \bar{\xi}^{N-2 k}
 \sum_{n'\in \N_+^{m-k}} \!\!\1\Bigl(\sum_j n'_j = N-2 m-M+2\Bigr) 
\end{align}
where we have estimated $\1(\exists j: n_j>M-2) \le
\sum_{j=1}^{m-k} \1(n_j>M-2)$. 

Now for any $m'\ge k'\ge 1$,
\begin{align}
 \sum_{n\in \N_+^{k'}} \1\Bigl(\sum_j n_j = m'\Bigr) = \binom{m'-1}{k'-1},
\end{align}
and, if $m'<k'$, the sum obviously yields zero.
Therefore, we have now proven
\begin{align}
 & \sum_{\substack{S\in\pi(I),\\
     |A|>M\text{ for some }A\in S}} r^{N-2 |S|}
  \prod_{A\in S} \left|C_{|A|} \right|  
 \le N! \sum_{m=1}^{\bar{n}}  r^{N-2 m}
\nonumber \\  &  \quad \times
 \sum_{k=\max(0,3m-N+M-2)}^{m-1}
 \frac{m-k}{m!}\binom{m}{k} 3^{m-k} \bar{\xi}^{N-2 k}
  \binom{N-2m-M+1}{m-k-1}.
\end{align}
where $\bar{n}=\lfloor \frac{N-M+1}{2} \rfloor$, so that 
$N=2 \bar{n} +M -\sigma$ with $\sigma=\1(N-M\text{ is odd})$.
Using the new summation variable $k'=m-1-k$ and estimating 
$\frac{m-k}{m!} \binom{m}{k} \le 1$, we obtain a new upper bound
\begin{align}
 &     N! \sum_{m=1}^{\bar{n}}  (\bar{\xi} r)^{N-2 m} 
 \sum_{k'=0}^{\min(m-1,N-2m-M+1)} \binom{N-2m-M+1}{k'}
 (3 \bar{\xi}^{2})^{k'+1}
\nonumber \\ & \quad
\le   N! \sum_{m=1}^{\bar{n}}  (\bar{\xi} r)^{N-2 m} 
 3 \bar{\xi}^{2}  (3 \bar{\xi}^{2}+1)^{N-2m-M+1}
\nonumber \\ & \quad =
  N! \frac{3 \bar{\xi}^{2}}{(3 \bar{\xi}^{2}+1)^{M-1}}
\sum_{m'=0}^{\bar{n}-1}  (a r/2)^{2m'+M-\sigma} 
 \le  N!   (a r/2)^{M-\sigma} \frac{1}{1-(ar/2)^2}
\nonumber \\ & \quad 
 \le  N!  (a r)^{M-\sigma}
\end{align}
where we have used the assumption $a r \le 1$. 
\end{proof}


\end{document}